\documentclass[a4paper,11pt]{article}
\pdfoutput=1 

\usepackage{jheppub} 

\usepackage{amsfonts,mathrsfs}

\usepackage[T1]{fontenc} 

\usepackage{tikz}
\usepackage{physics}
\usepackage{cancel}
\usepackage{float,graphicx}
\usepackage{comment}

\usetikzlibrary{arrows,arrows.meta,shapes,shapes.misc}

\newcommand{\PhiSG}{\varPhi}

\newcommand{\phiSG}{\varphi}
\newcommand{\phiMON}{\widehat{\varphi}}

\newcommand{\AdSST}{\text{AdS}_3 \times\text{S}^3 \times \text{T}^4}

\newcommand{\su}{\mathfrak{su}}

\newcommand{\so}{\mathfrak{so}}

\def\({\left(}
\def\){\right)}
\def\abs#1{\left| #1\right|}
\def\alg#1{\mathfrak{#1}}
\def\bb#1{\mathbb{#1}}

\newcommand{\scr}[1]{\mathscr{#1}}
\def\pv{\mathrm{p.v.}}

\def\AdSxT{{AdS${}_3 \times {}$S${}^3 \times T^4$}}

\newcommand{\BES}{\sigma_{\text{\tiny BES}}}
\newcommand{\bes}{{\text{\tiny BES}}}
\newcommand{\de}{\text{d}}

\usepackage{color}

\def\bQ{\bar{Q}}

\def\cE{{\mathcal E}}
\def\cK{{\mathcal K}}
\def\cO{{\mathcal O}}

\def\tE{\widetilde \cE}
\def\tK{\tilde K}

\def\hstar{\,\hat{\star}\,}
\def\cstar{\,\breve{\star}\,}

\title{More on the tensionless limit of pure-Ramond-Ramond AdS3/CFT2}

\author[a,b]{Alberto Brollo,}
\author[c]{Dennis le Plat,}
\author[a,d,e]{Alessandro Sfondrini,}
\author[f]{Ryo Suzuki}

\affiliation[a]{Dipartimento di Fisica e Astronomia, Universit\`a degli Studi di Padova,\\
via Marzolo 8, 35131 Padova, Italy}
\affiliation[b]{Fakult\"at f\"ur Mathematik, Technische Universit\"at M\"unchen, \\ Boltzmannstra\ss e 3, 85748 Garching, Germany}
\affiliation[c]{Institut f\"ur Mathematik und Physik, Humboldt-Universit\"at zu Berlin,\\
Zum gro{\ss}en Windkanal 2, 12489 Berlin, Germany}
\affiliation[d]{
Istituto Nazionale di Fisica Nucleare, Sezione di Padova,\\
via Marzolo 8, 35131 Padova, Italy}
\affiliation[e]{Institute for Advanced Study, School of Natural Sciences,\\
1 Einstein Drive,
Princeton, NJ 08540, USA}
\affiliation[f]{Shing-Tung Yau Center of Southeast University,\\
No.2 Sipailou, Xuanwu district, Nanjing, Jiangsu, 210096, China}

\emailAdd{alberto.brollo@tum.de}
\emailAdd{diplat@physik.hu-berlin.de}
\emailAdd{alessandro.sfondrini@unipd.it}
\emailAdd{rsuzuki.mp@gmail.com}

\abstract{
In a recent letter we presented the equations which describe tensionless limit of the excited-state spectrum for strings on  $AdS_3\times S^3\times T^4$ supported by Ramond-Ramond flux, and their numerical solution.
In this paper, we give a detailed account of the derivation of these equations from the mirror TBA equations proposed by Frolov and Sfondrini, discussing the contour-deformation trick which we used to obtain excited-state equations and the tensionless limit. We also comment at length on the algorithm for the numerical solution of the equations in the tensionless limit, and present a number of explicit numerical results, as well as comment on their interpretation.
}

\begin{document} 
\bibliographystyle{JHEP}
\maketitle
\flushbottom

\newpage
\section{Introduction}
An important family of string backgrounds in the AdS3/CFT2 holographic correspondence~\cite{Maldacena:1997re} involves the~$\AdSST$ geometry and supports  16 Killing spinors --- the maximal amount for an $\text{AdS}_3$ background~\cite{Haupt:2018gap}.
This is a \textit{family} of backgrounds as it depends on several moduli, see~\cite{Larsen:1999uk,OhlssonSax:2018hgc} for a detailed description. A very intriguing one-parameter family of supergravity backgrounds is the one interpolating between the case with \textit{no Kalb-Ramond $B$-field} (but with Ramond-Ramond background fluxes) and the one with \textit{no RR fluxes} (but with a $B$-field). In supergravity this is a continuous interpolation, but in string theory the coefficient~$k$ of the $B$-field has to be quantised. In perturbative string theory, which will be our focus here, the string coupling~$g_s$ is vanishingly small, while the string tension is sourced by both~$k$ and by the RR coupling~$g$,
\begin{equation}
\label{eq:stringtension}
    T=\frac{R^2}{2\pi\,\alpha'}=\sqrt{\frac{k^2}{4\pi^2}+g^2}\,,\qquad  k\in\mathbb{N}_0\,,\qquad g\geq0\,.
\end{equation}
Here $R$ is the radius of the of the three-sphere, which is equal to the radius of~$\text{AdS}_3$.

By far the best understood setup in this duality is the case in which $g=0$ and $k=1,2,3,\dots$. Then, the model can be described as worldsheet CFT, namely a supersymmetric Wess-Zumino-Novikov-Witten model~\cite{Giveon:1998ns}, and it can be solved~\cite{Maldacena:2000hw}.%
\footnote{To be precise, the WZNW description can be used without issue for~$k\geq2$, while the special case~$k=1$ is best understood by the ``hybrid'' worldsheet approach~\cite{Berkovits:1999im}, see~\cite{Eberhardt:2018ouy}.}
In particular, the free-string spectrum can be worked out explicitly and it features a discrete set of (highly degenerate) states, as well as a continuum.%
\footnote{
The continuum does not exist for~$k=1$~\cite{Giribet:2018ada,Eberhardt:2018ouy}.
}
This knowledge of the spectrum allowed to conjecture the dual of these string backgrounds, which are believed to take the form of symmetric-product orbifold CFTs~\cite{Giribet:2018ada,Eberhardt:2018ouy,Eberhardt:2021vsx}. Of these, the simplest and best understood is the symmetric-product orbifold of~$T^4$, (meaning, of four free bosons with $\mathcal{N}=(4,4)$ supersymmetry), which is dual to the $g=0$, $k=1$ string theory.

Things are significantly more involved for $g>0$, \textit{i.e.} when turning on RR-background fields (which can be done by tuning the RR scalar~$C_0$, see~\cite{OhlssonSax:2018hgc}).%
\footnote{While the backgrounds obtained in this way are in principle related by stringy dualities, such maps are non-perturbative in~$g_s$.}
In this case, the worldsheet CFT becomes nonlocal~\cite{Berenstein:1999jq,Cho:2018nfn}, and it is hard to decouple its ghost sector~\cite{Berkovits:1999im,Eberhardt:2018exh}. Qualitatively, we expect the continuum part of the spectrum to disappear and the degeneracies in the discrete spectrum to lift --- aside of course from the ones due to superconformal symmetry.
The polar opposite of the WZNW setup is the case where $k=0$ (\textit{i.e.}, there is no $B$-field at all) and the tension~\eqref{eq:stringtension} is entirely sourced by the RR fluxes. This case is qualitatively similar to that of strings on~$\text{AdS}_5\times \text{S}^5$ or on~$\text{AdS}_4\times \text{CP}^3$, and indeed we expect the spectrum to be as nontrivial as the one of a planar gauge theory. Even in such complicated cases, not all hope is lost as \textit{integrability} can be exploited to understand the planar spectrum, see~\cite{Arutyunov:2009ga,Beisert:2010jr} for reviews. One may hope that the same is true for $\text{AdS}_3$ backgrounds too, and this turns out to be true, see~\cite{Sfondrini:2014via}.

More specifically, the Green-Schwarz action $\AdSST$ with generic $k$ and $g$ is classically integrable~\cite{Cagnazzo:2012se}, and there is good evidence that integrability persists at the quantum level too, in terms of factorised scattering on the worldsheet.\footnote{%
See~\cite{Frolov:2023lwd} for recent work in this direction and further references.}
Moreover, the integrability construction is particularly robust in the case of~$g>0$ and~$k=0$. In this case, the construction of the integrable S~matrix has been recently completeted by a proposal for the dressing factors~\cite{Frolov:2021fmj}. This allowed to derive the mirror TBA equations~\cite{Frolov:2021bwp} which describe the whole planar spectrum of the theory (for arbitrary values of the tension~$g$).%
\footnote{%
An independent set of equations describing the spectrum, the so-called quantum spectral curve~\cite{Cavaglia:2021eqr} has also been recently proposed based on symmetry considerations. It is however presently unclear if and how these equations encode certain sectors of the theory, and in particular the part which is leading at low-tension (which will be the focus of this paper). 
}
This provides us with the possibility of making quantitative predictions for pure-RR $\AdSST$ strings. This is the main aim of this paper.

It is generally the case that the TBA equations cannot be solved in closed-form for generic unprotected states, and that each family of states (differing by the number of asymptotic worldsheet excitations) require a separate analysis. For this reason, here we focus on a subset of states, \textit{i.e.}\ those that acquire an anomalous dimension at low tension~$T\ll1$.
This is the limit where the dual-CFT description, whatever it may be, should simplify significantly. For instance, in this limit in $\text{AdS}_5\times \text{S}^5$ one recovers a nearest-neighbour integrable spin-chain~\cite{Minahan:2002ve}. It is natural to ask whether in our case we may find some similar spin chain, or a free CFT (like the symmetric-product orbifold of~$\text{T}^4$), or something else entirely. In this paper we will not be able to construct the dynamics (\textit{i.e.}, the Hamiltonian) in the $T\ll1$ regime, but only to read off the spectrum. This will still be sufficient to rule out some otherwise believable scenarios.

The main results of this paper were presented in short form elsewhere~\cite{Brollo:2023pkl}. As outlined there, our strategy is to start from the full, nonperturbative equations of~\cite{Frolov:2021bwp}, appropriately modified to described excited states, and expand them at small tension. The equations depend on a parameter~$h(T)$ which at small tension can be identified with~$T$ itself,
\begin{equation}
    h(T)\sim T= g\,,\qquad T\ll 1\,.
\end{equation}
In what follows, in analogy with the notation of the existing literature, we will indicate the parameter as~$h$. From the small-$h$ expansion we will find a set of equations which we will solve numerically to high precision.

This paper is structured as follows.
In section~\ref{sec:kinematics} we briefly review the model and in section~\ref{sec:groundstateTBA} we recap the mirror TBA equations for the ground state. In section~\ref{sec:excitedstatetba} we derive the equations for excited states through the contour-deformation trick; we focus on massless excitations, which are leading at small tension. In section~\ref{sec:smalltension} we take the small-tension limit and show that several of the mirror TBA equations decouple. In section~\ref{sec:smalltensionTBA} we further simplify and summarise the small-tension TBA equations. Finally, in section~\ref{sec:tensionlessspectrum} we discuss their numerical solution and present the results, and we conclude in section~\ref{sec:conclusions}.
We also include some further details in appendices. Appendix~\ref{app:notation} summarises our notation. Appendix~\ref{app:Kren} discusses some identities of the integration kernels which we need to simplify the TBA equations. Appendix~\ref{app:weak-coupling exp} proves some of the identities that we use in the weak-tension expansion of the TBA equations, while appendix~\ref{app:BES phase} contains a detailed discussion of the Beisert-Eden-Staudacher phase in various regimes (in the massive and massless kinematics). In appendix~\ref{app:largeL} we discuss in more detail the large-volume behaviour of the mirror TBA equations. Finally, appendix~\ref{app:numerical} contains several tables containing numeric results.

\section{Particle Content and Kinematics}
\label{sec:kinematics}
Let us begin by briefly reviewing the model of interest. Strings on $\AdSST$ with RR flux feature infinitely many worldsheet excitations, labelled by a number $m\in\mathbb{Z}$, with periodic momentum~$p$ and energy~\cite{Lloyd:2014bsa}
\begin{equation}
\label{eq:string dispersion}
\mathcal{E}(p)= \sqrt{m^2+4h^2\sin^2(p/2)}\,,
\end{equation}
where $h$ is the coupling constant. In the mirror model, the dispersion relation becomes
\begin{equation}
\label{eq:mirrordispersion}
    \widetilde{\mathcal{E}}(\tilde{p}) = 2\,\text{arcsinh}\left(\frac{\sqrt{m^2+\tilde{p}^2}}{2h}\right)\,.
\end{equation}
Notice that, for $m=0$, both dispersion relations have no mass gap. Strictly speaking, this dispersion relation applies to a whole supermultiplet of particles, which takes a different form in the string and mirror model, see~\cite{Seibold:2022mgg}. Notably, in either case, the dimension of the representation does not grow with~$|m|$, but it is fixed to being four-dimensional (this is unlike what happens, for instance, in~$\text{AdS}_5\times \text{S}^5$).

When analysing the mirror model in the thermodynamic limit, we identify the following type of particles:
\begin{itemize}
    \item \textit{$Q$-particles.} These are essentially bound states with $Q=m=1,2,\dots$.
    \item \textit{$\overline{Q}$-particles.} These are essentially bound states with $\overline{Q}=-m=1,2,\dots$.
    \item \textit{Massless particles,} which we also indicate with $Q=0$. These are the modes with $m=0$, and come in two copies, labelled by an index~$\dot{\alpha}=1,2$. In the literature, this index is associated with the so call $\su(2)_\circ$ which is part of the $\so(4)$ acting geometrically on the four flat directions.%
    \footnote{%
    Like in~\cite{Frolov:2021bwp}, we only consider states with no winding nor momentum along~$\text{T}^4$.}
    We will slightly generalise this construction, and allow the index~$\dot{\alpha}$ to run from~$1$ to~$N_0$, without specifying that $N_0=2$ until the very end. This will allow us, later on, to discuss a very recent proposal for the TBA equations~\cite{Frolov:2023wji}.%
    \footnote{The observation of~\cite{Frolov:2023wji} is that the mirror TBA in their original form do not seem to yield the correct energy for a twisted ground state, and that this can be seemingly fixed by tuning ``by hand''~$N_0=1$.
    It is worth remarking that while the $N_0=2$ equations are derived from a string hypothesis, there is a priori no justification for tuning~$N_0$ to different values. 
    }
    \item \textit{Auxiliary particles,} which account for the supermultiplet structure of the various bound-state representations and do not contribute to the energy. They are labelled by an index $\alpha=1,2$ (without dot), related to the $\su(2)_\bullet$ which also comes from the $\so(4)$ acting geometrically on the four flat directions, $\so(4)\cong\su(2)_\bullet\oplus\su(2)_\circ$. Later on, we will attach a further index~$\pm$ to these auxiliary roots, which will make it easier to parametrise them (just like it is done in $\text{AdS}_5\times \text{S}^5$~\cite{Arutyunov:2009zu}).
\end{itemize}
It is worth emphasising that, unlike what happens in $\text{AdS}_5\times \text{S}^5$, there are no Bethe strings involving fundamental particles and auxiliary roots. This is because the dimension of the bound state representation does not grow with $Q$ or~$\overline{Q}$.

Finally, we stress that this construction, which is the one of~\cite{Frolov:2021bwp}, relies on the scattering between massless representations with different~$\su(2)_\circ$ quantum number being trivial. This is what seems to appear from comparison with perturbative computations, see the discussions of~\cite{Lloyd:2014bsa}. Were we to allow for a non-trivial~$\su(2)_\circ$ S~matrix, we would obtain a different string hypothesis. This is discussed in~\cite{toappear}.

\subsection{String-region parametrisation}
\label{sec:kinematics:string}
It is convenient to introduce suitable variables to parametrise the momentum and energy in the string and mirror models.

In the string region we can introduce Zhukovsky variables satisfying, for $m\neq 0$,
\begin{equation}
    \frac{x^+}{x^-}=e^{ip},\qquad
    \frac{h}{2i}\left(x^+ -\frac{1}{x^+}-x^-+\frac{1}{x^-}\right)=\mathcal{E}(p)\,,
\end{equation}
subject to the constraint
\begin{equation}
\label{eq:Zhukovsky-constraint}
    \frac{h}{2i}\left(x^+ +\frac{1}{x^+}-x^--\frac{1}{x^-}\right) = |m|\,.
\end{equation}
For the special case $m=0$, we have
\begin{equation}
\label{eq:massless-shortening}
    x_s\equiv x^+ =\frac{1}{x^-}\,,\qquad (m=0)\,.
\end{equation}
Hence $(x_s)^2=e^{ip}$ and we take $x$ to lie \textit{on the upper half-circle} for particles of real momentum.
It is convenient to solve the constraint~\eqref{eq:Zhukovsky-constraint} in terms of an unconstrained variable~$u$
\begin{equation}
    u=x^+ +\frac{1}{x^+}-\frac{i|m|}{h}
    =x^- +\frac{1}{x^-}+\frac{i|m|}{h} = x+\frac{1}{x}\,.
\end{equation}
In the string region we may solve this relation by defining
\begin{equation}
    x_s(u) = \frac{u}{2}\left(1+\sqrt{1-\frac{4}{u^2}}\right)\,,
\end{equation}
which has cuts for $-2\leq u\leq +2$, and letting
\begin{equation}
    x^\pm(u) = x_s(u\pm \tfrac{i}{h}|m|)\,.
\end{equation}
For massive particles with real momenta, the rapidities take values on the real line.
In the massless case $m=0$, this formula should be understood as a ``$\pm i0$'' prescription at the cut $-2\leq u\leq +2$, which fits with~\eqref{eq:massless-shortening}. In particular, in eq.~\eqref{eq:massless-shortening}, $x=x_s(u+i0)$ with $-2\leq u\leq +2$.
It is also useful to introduce the following $\gamma$-parametrisation~\cite{Frolov:2021fmj}, see also~\cite{Fontanella:2019baq},
\begin{equation}
    x^\pm = \frac{i\mp e^{\gamma^\pm}}{i\pm e^{\gamma^\pm}}\,. 
\end{equation}
To invert this relation we need to pick a branch,%
\footnote{In what follows, $\ln(z)$ denotes the principal branch of the logarithm, with the branch cut running on the negative real axis.}
and we choose in the string region~\cite{Frolov:2021fmj}
\begin{equation}
    \gamma^\pm  = \ln\left(\mp i \frac{x^\pm -1}{x^\pm + 1}\right)\,.
\end{equation}
Note that with these choices we have the following reality conditions for real-momentum particles in the string region:
\begin{equation}
    (x^+)^* = x^-\,,\qquad 
    (\gamma^+)^* = \gamma^-\,.
\end{equation}
Analogously, for massless particles we set
\begin{equation}
\label{eq:string-gamma-x}
    x_s(\gamma_s) = \frac{i- e^{\gamma_s}}{i+ e^{\gamma_s}}\,,\qquad
    \gamma_s(x_s) = \ln\left(-i\frac{x_s-1}{x_s+1}\right)\,,
\end{equation}
with the reality conditions, for real-momentum particles in the string model
\begin{equation}
\label{eq:massless-conjugation-string}
    (x_s)^* = \frac{1}{x_s}\,,\qquad  (\gamma_s)^*= \gamma_s\,,
\end{equation}
where the subscript ``$s$'' emphasises that we are discussing the string model (this is to avoid clashes with later notation).

\subsection{Mirror-region parametrisation}
\label{sec:kinematics:mirror}

In this paper we will be chiefly interested in the mirror model, which is related to the string model by analytic continuation~\cite{Arutyunov:2007tc,Frolov:2021zyc}.
In this case we introduce the (mirror) Zhukovsky parametrisation
\begin{equation}
\label{eq:Zhukovsky}
    x(u) = \frac{1}{2}\left(u - i \sqrt{4-u^2} \right) 
\end{equation}
which has cuts on the real $u$-line for $|u|>2$. For massive particles we write
\begin{equation}
    x^{\pm}(u) = x(u \pm \tfrac{i}{h}Q)\,,\qquad
    x^{\pm}(u) = x(u \pm \tfrac{i}{h}\overline{Q})\,,
\end{equation}
For massless particles, like before, this means
\begin{equation}
\label{eq:mirrorx-Qis0}
    x(u) = x(u + i0 ) = \frac{1}{x(u-i0)}\,.
\end{equation}
Now $u$ is on the real $u$-line with $|u|>2$, 
and $x(u+i0)$ lies on the interval $-1 \le x \le 1$ on the real axis.
In the mirror region the momentum for a $Q$-particle is given by
\begin{equation}
    \tilde{p}^Q(u) = h \left( x(u - \tfrac{i}{h}Q) - x(u+ \tfrac{i}{h}Q) \right) +i Q\,,
\end{equation}
while the mirror energy for a $Q$-particle is
\begin{equation}
\label{def:tilde EQ}
\tilde{\mathcal{E}}^Q(u) = \ln \frac{x(u-\frac{i}{h}Q)}{x(u+\frac{i}{h}Q)}\,,
\end{equation}
where we made the $Q$-dependence explicit for later convenience. The formulae for $\overline{Q}$-particles are identical.
For $Q=0$ we have
\begin{equation}
\tilde{\mathcal{E}}^0(u)=-\ln \left(x(u+i0)\right)^2\,.
\end{equation}
Note that this function is not analytic, see also~\cite{Frolov:2021zyc}. We will discuss its branches in Section \ref{sec:kinematics:massless}.
It can be checked that, with these definitions, we reproduce the mirror dispersion relation~\eqref{eq:mirrordispersion} with $|m|=Q$ or~$ |m|=\overline{Q}$.
Let us emphasise that \textit{real-momentum particles} are defined for
\begin{itemize}
    \item $u\in\mathbb{R}$ for $Q$ and $\overline{Q}$ particles,
    \item $u\in(-\infty,-2)\cup (+2,+\infty)$ for massless particles.
\end{itemize}
The relation between the $\gamma$-parameter in the mirror theory satisfies, for massless mirror particles,
\begin{equation}
    x(\gamma)=-\tanh\frac{\gamma}{2}\,,\qquad
    \gamma(x)=-2\,\text{atanh}(x)\,,
\end{equation}
so that
\begin{equation}
\label{eq:MasslessGammaRapidity}
    \gamma(u)=\gamma(x(u))=\frac{1}{2} \ln\left(- \frac{u-2}{u+2}\right) +\frac{i \pi}{2} \,. 
\end{equation}
Using this formula we can define for $Q$-particles (or $\overline{Q}$-particles),
\begin{equation}
    \gamma^-(u) = \gamma(u-\tfrac{i}{h}Q), \qquad \gamma^+(u)=\gamma(u+\tfrac{i}{h}Q) - i \pi \,.
\end{equation}
The reality conditions in the mirror theory are, for real-mirror-momentum particles,
\begin{equation}
    (x^\pm)^*=\frac{1}{x^{\mp}}\,,\qquad
    (\gamma^\pm)^* = \gamma^{\mp} \,,
\end{equation}
and
\begin{equation}
    x^*=x\,,\qquad \gamma^* = \gamma\,,
\end{equation}
where in these two last formulae it is necessary to correctly account for the branch cut on real $u$, $|u|>2$.
Finally, by directly comparing this definition of $\gamma$ with the one given in the string region we see that
\begin{equation}
    \gamma(u)\Big|_{\text{string}} = 
    \gamma(u)\Big|_{\text{mirror}} - \frac{i\pi}{2}\,,
\end{equation}
where $\gamma(u)_{\text{mirror}}$ is defined by~\eqref{eq:MasslessGammaRapidity} and $\gamma(u)_{\text{string}} = \gamma(x_s(u))$. In what follows, we will indicate the latter by $\gamma_s$, so that when treating $\gamma$ as a free variable we will write
\begin{equation}
\label{eq:string-mirror-gamma}
    \gamma_s = \gamma - \frac{i\pi}{2}\,.
\end{equation}
The inverse of $\gamma(u)$ is given by
\begin{equation}
\begin{aligned}
u (\gamma) &= - 2 \coth (\gamma), &\qquad 
&u \in (-\infty, -2) \cup (2, \infty)
\\[1mm]
u_s (\gamma_s) &= -2 \tanh (\gamma_s), &\qquad 
&u_s \in (-2,2).
\end{aligned}
\label{def:u to gamma's}
\end{equation}
If $u$ is slightly above the real axis, $\gamma$ is found at:
\begin{equation}
u+i0 \ \ \leftrightarrow \ \ \gamma +i0, \qquad
u_s + i0 \ \ \leftrightarrow \ \ \gamma_s -i0.
\end{equation}
Neither $u(\gamma)$ nor $u_s(\gamma_s)$ covers the whole real line.

\subsection{Explicit formulae for massless particles}
\label{sec:kinematics:massless}
In what follows the kinematics of the massless particles will be particularly important. Let us spell out explicitly the parametrisation of various physical quantities in terms of the $\gamma$ and $\gamma_s$ rapidities.
In the string theory we have
\begin{equation}
    \mathcal{E}_0=\frac{2h}{\cosh\gamma_s}\,,\qquad
    p_0=-i\ln
    \left(\frac{i-e^{\gamma_s}}{i+e^{\gamma_s}}\right)^2,
\end{equation}
where the branches of $\log z^2$ can be resolved as
\begin{equation}
    p^0=\begin{cases}
    -2i\ln\left(+\frac{i-e^{\gamma_s}}{i+e^{\gamma_s}}\right),\qquad&\gamma_s>0\quad (-\pi\leq p\leq 0),\\
    -2i\ln\left(-\frac{i-e^{\gamma_s}}{i+e^{\gamma_s}}\right),\qquad&\gamma_s<0\quad (0\leq p\leq +\pi).
    \end{cases}
\end{equation}
The necessity of picking different branches for positive and negative momentum is a consequence of the gapless dispersion relation and it was discussed at length in~\cite{Frolov:2021zyc}.
For mirror particles we have, for $\gamma$ slightly above the real axis,
\begin{equation}
    \tilde{p}_0=-\frac{2h}{\sinh\gamma}\,,\qquad
    \widetilde{\mathcal{E}}_0=-\ln\left(\frac{1-e^\gamma}{1+e^\gamma}\right)^2\,.
\end{equation}
Again we can resolve the branches of the logarithm as~\cite{Frolov:2021zyc}
\begin{equation}
    \widetilde{\mathcal{E}}_0=\begin{cases}
        -2\ln(\frac{1-e^\gamma}{1+e^\gamma})+2\pi i,\qquad
        &\gamma>0\quad (\tilde{p}^0<0),\\
        -2\ln(\frac{1-e^\gamma}{1+e^\gamma}),\qquad
        &\gamma<0\quad (\tilde{p}^0>0),
    \end{cases}
\end{equation}
where the variables take values just above the real-$\gamma$ line.

\subsection{Parametrisation of (mirror) auxiliary particles}
\label{sec:kinematics:auxiliary}
Let us finally briefly discuss the parametrisation of the auxiliary particles. In this work, we will only need their kinematics in the mirror theory. In~\cite{Frolov:2021bwp} it was argued that, for the Bethe-Yang equations of the mirror model to admit a solution, the rapidity variable of the auxiliary particles must lie on the unit circle.
Hence, we have that quite curiously \textit{the mirror kinematics of auxiliary particles is the same as the string kinematics of massless particles} (see~\eqref{eq:massless-conjugation-string} for the string massless kinematics).
One important difference is that auxiliary particles can take values on either the upper or lower half-circle --- not just the upper, like for string massless particles. (Another difference is  that there is no momentum or energy associated to the auxiliary particles.)
Therefore we will define two types of auxiliary ``particles'':
\begin{itemize}
    \item \textit{$y^-$ particles}, whose rapidity is parametrised by $x(u)$ with $-2<u<+2$, which parametrise the lower half-circle; equivalently, we can use $x_s(u-i0)$.
    \item \textit{$y^+$ particles}, whose rapidity is parametrised by $1/x(u)$ with $-2<u<+2$, which parametrise the upper half-circle; equivalently, we can use $x_s(u+i0)$.
\end{itemize}
In terms of the $\gamma$ variable therefore we can use the string parametrisation~\eqref{eq:string-gamma-x}. 

\section{Ground-state mirror TBA equations}
\label{sec:groundstateTBA}
Let us collect here the ground-state TBA equations which were derived in~\cite{Frolov:2021bwp} for the mirror model. The equations are written in the $u$-variables introduced above, and all quantities are in the mirror kinematics. The domain of $u$ is chosen to cover all real values of momentum --- or, in case of the auxiliary particles, the unit circle of the Zhukovsky plane.
The various kernels appearing in the equations below are defined as the logarithmic derivatives of appropriate S~matrices in the mirror-mirror kinematics. Schematically,
\begin{equation}
    K(u,u^\prime) = \frac{1}{2 \pi i} \frac{\de}{\de u} \log{S(u,u^\prime)}\,.
\end{equation}
The precise form of the various S~matrices and kernels is collected in appendix~\ref{app:notation}.
We will use the short-hand notation 
\begin{equation}
    \rho_j \star K_{ji}(u) = \sum_j \int \de u^\prime \rho_j(u^\prime) K_{ji}(u^\prime,u) \,.
\end{equation}
When writing the convolutions, we use
\begin{equation}
    \star \leftrightarrow \int\limits_{-\infty}^{+\infty} \de u, \qquad \hat{\star} \leftrightarrow \int\limits_{-2}^{+2} \de u, \qquad \check{\star} \leftrightarrow \int\limits_{-\infty}^{-2}\de u + \int\limits_{+2}^{+\infty}\de u \,.
\end{equation}
In particular, in view of the discussion in the previous section, we will use the convolutions
\begin{itemize}
    \item     ``$\star$'' for $Q$ and $\overline{Q}$ particles,
    \item     ``$\hat{\star}$'' for auxiliary particles,
    \item     ``$\check{\star}$'' for massless particles.
\end{itemize}
We will later rewrite the kernels and TBA equations in terms of the $\gamma$ variables, and introduce ad hoc symbols for kernels and convolutions.

It is worth emphasising a potential ambiguity in the derivation of the mirror TBA equations --- whether certain expressions should be understood as \textit{sum of logarithms}, or \textit{logarithms of products}. For the time being, we will not specify the branch of the logarithm. We will address this ambiguity later on.

\subsection{Mirror TBA equations}
The equations are written in terms of $Y$-functions. We have~\cite{Frolov:2021bwp}:

\paragraph{Equations for $Q$-particles.}
\begin{equation} 
\begin{aligned}
    -\log Y_Q &= L \tilde{\mathcal{E}}_Q - \log{(1+Y_{Q^\prime})} \star K_{\textit{sl}}^{Q^\prime Q} - \log{(1+ \bar{Y}_{Q^\prime})} \star \tilde{K}_{\textit{su}}^{Q^\prime Q} \\
    &- \sum_{\dot{\alpha}=1}^{N_0} \log{(1+Y_0^{(\dot{\alpha})})} \check{\star} K^{0Q} \\
    &- \sum_{\alpha=1,2} \log{\left(1-\frac{1}{Y_+^{(\alpha)}}\right)} \hat{\star} K_+^{yQ} - \sum_{\alpha=1,2} \log{\left(1-\frac{1}{Y_-^{(\alpha)}}\right)} \hat{\star} K_-^{yQ} \,.
\end{aligned}
\end{equation}

\paragraph{Equations for $\bar{Q}$-particles.}
\begin{equation} 
\begin{aligned}
    -\log \bar{Y}_Q &= L \tilde{\mathcal{E}}_Q - \log{(1+\bar{Y}_{Q^\prime})} \star K_{\textit{su}}^{Q^\prime Q} - \log{(1+ Y_{Q^\prime})} \star \tilde{K}_{\textit{sl}}^{Q^\prime Q} \\
    &- \sum_{\dot{\alpha}=1}^{N_0} \log{(1+Y_0^{(\dot{\alpha})})} \check{\star} \tilde{K}^{0Q} \\
    &- \sum_{\alpha=1,2} \log{\left(1-\frac{1}{Y_+^{(\alpha)}}\right)} \hat{\star} K_-^{yQ} - \sum_{\alpha=1,2} \log{\left(1-\frac{1}{Y_-^{(\alpha)}}\right)} \hat{\star} K_+^{yQ} \,.
\end{aligned}
\end{equation}

\paragraph{Equations for Massless particles.}
\!\!\!\!\!\!\footnote{In both $K^{y0}$ and $K^{0y}$, the $y$-particle lies on the \textit{upper} half-circle. In this way, $K^{y0} = - K^{0y}$ is positive when the arguments are in the appropriate ranges.}
\begin{equation}
\begin{aligned}
    -\log Y_0^{(\dot{\alpha})} &= L \tilde{\mathcal{E}}_0 - \sum_{\dot{\beta}=1}^{N_0} \log{(1+Y_0^{(\dot{\beta})})} \check{\star} K^{00} - \log{(1+Y_Q)} \star K^{Q0} - \log{(1+\bar{{Y}_Q})} \star \tilde{K}^{Q0} \\
    &- \sum_{\alpha=1,2} \log{\left(1-\frac{1}{Y_+^{(\alpha)}}\right)} \hat{\star} K^{y0} - \sum_{\alpha=1,2} \log{\left(1-\frac{1}{Y_-^{(\alpha)}}\right)} \hat{\star} K^{y0} \,.
\end{aligned}    
\end{equation}

\paragraph{Equation for auxiliary $y^-$-particles.}
\begin{equation}
    \log{Y_-^{(\alpha)}} = -\log{(1+Y_Q)} \star K_-^{Qy} + \log{(1+\bar{Y}_Q)} \star K_+^{Qy} + \sum_{\dot{\alpha}=1}^{N_0} \log{(1+Y_0^{(\dot{\alpha})})} \check{\star} K^{0y} \,.
\label{TBA ym gnd}
\end{equation}

\paragraph{Equation for auxiliary $y^+$-particles.}
\begin{equation}
    \log{Y_+^{(\alpha)}} = -\log{(1+Y_Q)} \star K_+^{Qy} + \log{(1+\bar{Y}_Q)} \star K_-^{Qy} - \sum_{\dot{\alpha}=1}^{N_0} \log{(1+Y_0^{(\dot{\alpha})})} \check{\star} K^{0y} \,.
\label{TBA yp gnd}
\end{equation}

\subsection{Energy and momentum}
Once the Y-functions are determined, it is possible to compute the (ground-state) energy by the following formula:
\begin{equation}
\label{eq:groundstateen}
\begin{aligned}
    E(L) =& - \int\limits_{-\infty}^\infty \frac{\de u}{2 \pi} \frac{\de \tilde{p}^Q}{\de u} \log{\left((1+Y_Q)(1+\bar{Y}_Q)\right)}\\
    &- \int\limits_{|u|>2} \frac{\de u}{2 \pi} \frac{\de \tilde{p}^0}{\de u} \sum_{\dot{\alpha}=1}^{N_0} \log{\left(1+Y_0^{(\dot{\alpha})}\right)} \,.
\end{aligned}
\end{equation}
Note that auxiliary Y-functions do not contribute to the formula. It is also useful to write a similar formula to impose that the total-momentum of the (ground) state vanishes, namely
\begin{equation}
\begin{aligned}
    0 =& - \int\limits_{-\infty}^\infty \frac{\de u}{2 \pi} \frac{\de \tilde{\mathcal{E}}_Q}{\de u} \log{\left((1+Y_Q)(1+\bar{Y}_Q)\right)}\\
    &- \int\limits_{|u|>2} \frac{\de u}{2 \pi} \frac{\de \tilde{\mathcal{E}}_0}{\de u} \sum_{\dot{\alpha}=1}^{N_0} \log{\left(1+Y_0^{(\dot{\alpha})}\right)} \,.
\end{aligned}
\end{equation}
This is the level-matching condition in string theory (in a sector without winding around the lightcone, see~\cite{Arutyunov:2009ga}).

\section{Exciting the massless modes}
\label{sec:excitedstatetba}
Let us now discuss how the ground-state equations change if we consider states containing massless excitations. This can be done by the contour-deformation trick of Dorey and Tateo~\cite{Dorey:1996re}. For simplicity, we consider only states involving \textit{an even number of massless modes with real momenta coming in pairs}, $(p_j,-p_j)$, and without any auxiliary excitations.\footnote{%
In other words, we consider excited states which  do not contain auxiliary Bethe roots. However, even for this simpler set of states, it is absolutely necessary to take into account the role of auxiliary Y-functions: in a large-volume picture, our state consists only of highest-weight particles of momentum $(p_j,-p_j)$ in the massless representations, but it is subject to finite-volume effects due to \textit{all} types of virtual mirror particles, including auxiliary ones.} 
As it will become clear in a moment, this will ensure that the $Y$-functions are symmetric under $u\to-u$ and, in turn, that the level-matching condition is satisfied.

\subsection{Contour-deformation trick}
We expect the equations to be modified by ``driving terms'' which arise out of the contour-deformation trick. They come from picking up residues in the various convolution in places where
\begin{equation}
\label{eq:exactBethe-schematic}
    Y_{0,\text{string}}^{(\dot{\alpha}_j)}\big(u_j^{\dot{\alpha}_j}\big)=-1\,,\qquad
    j=1,\dots 2M\,.
\end{equation}
The subscript ``string'' reminds us that the values $u_j^{\dot{\alpha}_j}$ where this may happen lie in the string, rather than mirror, region. In fact, for the case at hand, this will happen when $u_j^{\dot{\alpha}_j}$ is on the real-string line for massless particles. We indicate with an asterisk the S-matrix elements which have one leg in the string-region. Furthermore, we indicated by a dot the free argument of each functional equation (which takes value on the real mirror line).
Formally, we find a simple modification of the equations, written in blue.

\paragraph{Equations for $Q$-particles.}
\begin{equation} 
\begin{aligned}
    -\log Y_Q &= L \tilde{\mathcal{E}}_Q - \log{(1+Y_{Q^\prime})} \star K_{\textit{sl}}^{Q^\prime Q} - \log{(1+ \bar{Y}_{Q^\prime})} \star \tilde{K}_{\textit{su}}^{Q^\prime Q} \\
    &- \sum_{\dot{\alpha}=1}^{N_0} \log{(1+Y_0^{(\dot{\alpha})})} \check{\star} K^{0Q}
    {\color{blue}+ \sum_{j=1}^{2M} \log S^{0_*Q}(u_j^{\dot{\alpha}_j},\,\cdot\,)}\\
    &- \sum_{\alpha=1,2} \log{\left(1-\frac{1}{Y_+^{(\alpha)}}\right)} \hat{\star} K_+^{yQ} - \sum_{\alpha=1,2} \log{\left(1-\frac{1}{Y_-^{(\alpha)}}\right)} \hat{\star} K_-^{yQ} \,.
\end{aligned}
\end{equation}

\paragraph{Equations for $\bar{Q}$-particles.}
\begin{equation} 
\begin{aligned}
    -\log \bar{Y}_Q &= L \tilde{\mathcal{E}}_Q - \log{(1+\bar{Y}_{Q^\prime})} \star K_{\textit{su}}^{Q^\prime Q} - \log{(1+ Y_{Q^\prime})} \star \tilde{K}_{\textit{sl}}^{Q^\prime Q} \\
    &- \sum_{\dot{\alpha}=1}^{N_0} \log{(1+Y_0^{(\dot{\alpha})})} \check{\star} \tilde{K}^{0Q}
    {\color{blue}+ \sum_{j=1}^{2M} \log \tilde{S}^{0_*Q}(u_j^{\dot{\alpha}_j},\,\cdot\,)} \\
    &- \sum_{\alpha=1,2} \log{\left(1-\frac{1}{Y_+^{(\alpha)}}\right)} \hat{\star} K_-^{yQ} - \sum_{\alpha=1,2} \log{\left(1-\frac{1}{Y_-^{(\alpha)}}\right)} \hat{\star} K_+^{yQ} \,.
\end{aligned}
\end{equation}

\paragraph{Equations for  Massless particles.}
\begin{equation}
\label{eq:TBA0-excited}
\begin{aligned}
    -\log Y_0^{(\dot{\alpha})} &= L \tilde{\mathcal{E}}_0 - \sum_{\dot{\beta}=1}^{N_0} \log{(1+Y_0^{(\dot{\beta})})} \check{\star} K^{00}
    {\color{blue}+ \sum_{j=1}^{2M} \log S^{0_*0}(u_j^{\dot{\alpha}_j},\,\cdot\,)}\\
    &- \log{(1+Y_Q)} \star K^{Q0} - \log{(1+\bar{{Y}}_Q)} \star \tilde{K}^{Q0} \\
    &- \sum_{\alpha=1,2} \log{\left(1-\frac{1}{Y_+^{(\alpha)}}\right)} \hat{\star} K^{y0} - \sum_{\alpha=1,2} \log{\left(1-\frac{1}{Y_-^{(\alpha)}}\right)} \hat{\star} K^{y0} \,.
\end{aligned}    
\end{equation}

\paragraph{Equation for auxiliary $y^-$-particles.}
\begin{equation}
\begin{aligned}
    \log{Y_-^{(\alpha)}} =& -\log{(1+Y_Q)} \star K_-^{Qy} + \log{(1+\bar{Y}_Q)} \star K_+^{Qy}\\
    &+ \sum_{\dot{\alpha}=1}^{N_0} \log{(1+Y_0^{(\dot{\alpha})})} \check{\star} K^{0y} 
    {\color{blue}- \sum_{j=1}^{2M} \log S^{0_*y}(u_j^{\dot{\alpha}_j},\,\cdot\,)}\,.
\end{aligned}
\end{equation}

\paragraph{Equation for auxiliary $y^+$-particles.}
\begin{equation}
\begin{aligned}
    \log{Y_+^{(\alpha)}} =& -\log{(1+Y_Q)} \star K_+^{Qy} + \log{(1+\bar{Y}_Q)} \star K_-^{Qy}\\
    &- \sum_{\dot{\alpha}=1}^{N_0} \log{(1+Y_0^{(\dot{\alpha})})} \check{\star} K^{0y}
    {\color{blue}+ \sum_{j=1}^{2M} \log S^{0_*y}(u_j^{\dot{\alpha}_j},\,\cdot\,)}\,.
\end{aligned}
\end{equation}

\subsection{Exact Bethe equations}
The rapidities $u_j^{\dot{\alpha}_j}$ which appear in the equations above are far from arbitrary, as they have to satisfy~\eqref{eq:exactBethe-schematic}. We can rewrite that constraint by analytically continuing equation~\eqref{eq:TBA0-excited} to the string region. Because that equation is for the logarithm of the Y-function, we can choose any branch labeled by $\nu_k^{\dot{\alpha}}\in\mathbb{Z}$ and get
\begin{equation}
\begin{aligned}
    {\color{blue}i\pi(2\nu_k^{\dot{\alpha}_k}+1)}=&
    {\color{blue}-i L p_k^{\dot{\alpha}_k}}- \sum_{\dot{\beta}=1}^{N_0} \log{(1+Y_0^{(\dot{\beta})})} \check{\star} K^{00_*}
    {\color{blue}+ \sum_{j=1}^{2M} \log S^{0_*0_*}(u_j^{\dot{\alpha}_j},\,u_k^{\dot{\alpha}_k})}\\
    &- \log{(1+Y_Q)} \star K^{Q0_*} - \log{(1+\bar{{Y}}_Q)} \star \tilde{K}^{Q0_*} \\
    &- \sum_{\alpha=1,2} \log{\left(1-\frac{1}{Y_+^{(\alpha)}}\right)} \hat{\star} K^{y0_*} - \sum_{\alpha=1,2} \log{\left(1-\frac{1}{Y_-^{(\alpha)}}\right)} \hat{\star} K^{y0_*}\,,
\end{aligned}
\end{equation}
for $k=1,\dots, 2M$. 
Note that we have added an asterisk to the second index of the various kernels, to recall that they are analytically continued to $u_k^{\dot{\alpha}_k}$ in the string region (and avoid writing down all arguments explicitly).
Let us focus on the blue terms. On the left-hand side we have the mode number~$\nu_k^{\dot{\alpha}_k}$; on the right-hand side we have \textit{the momentum in the string region} (which comes from analytically continuing the mirror energy $\tilde{E}_0$) and a sum of $\log S$ terms, with both arguments in the string region. It is clear that the blue terms alone give the asymptotic Bethe equations (see~\cite{Frolov:2021fmj}) and the other terms are finite-size corrections to those equations. This justifies the name ``exact Bethe equations''.

\paragraph{Repeated roots.}
Whether or not it is allowed to consider more than one number with the same quantum number depends on the model, and namely if the particles behave effectively like Fermions or Bosons. In the case at hand we are dealing with Fermions, and the exclusion principle imposes that we may have
\begin{equation}
    \nu_j^{\dot{\alpha}_j}=\nu_k^{\dot{\alpha}_k}\qquad
    \text{only if}\qquad \dot{\alpha}_j\neq \dot{\alpha}_k\,.
\end{equation}
This is necessary to reproduce the expected number of states, as it can be seen already from looking at protected states~\cite{Baggio:2017kza}.

\paragraph{Dependence on  the $\su(2)$ labels.}
Looking more closely at the above equations we find that, on the real mirror line
\begin{equation}
    Y_0^{(\dot{\alpha})}(u)=Y_0^{(\dot{\beta})}(u) \equiv Y_0 (u),
    \qquad
    Y_\pm^{(1)}(u)=Y_\pm^{(2)}(u) \equiv Y_\pm (u).
\end{equation}
This can be seen by taking the difference of the equations.
In other words, the Y functions do not depend on the $\su(2)$ labels. However, this does not mean that such labels can be completely dropped. In fact, as discussed just above the exact Bethe roots must satisfy the Pauli exclusion principle (that would also be true for the auxiliary roots, which we do not consider here). In other words, two Bethe roots may take the same value only if they correspond to different labels.

\subsection{Slightly simplified form of the equations}
Bearing in mind the observations above, we can simplify the equations by eliminating the $\su(2)$ labels from auxiliary and massless Y~functions (but keeping it on the roots).
Note that for the moment we do not specify the branch choice of the logarithm, which we will fix later by comparing with the asymptotic results.
We obtain the following equations.

\paragraph{Equations for $Q$-particles.}
\begin{equation} 
\begin{aligned}
    -\log Y_Q &= L \tilde{\mathcal{E}}_Q - \log{(1+Y_{Q^\prime})} \star K_{\textit{sl}}^{Q^\prime Q} - \log{(1+ \bar{Y}_{Q^\prime})} \star \tilde{K}_{\textit{su}}^{Q^\prime Q} \\
    &- \log{(1+Y_0)^{N_0}} \, \check{\star} \, K^{0Q}
    {\color{blue}+ \sum_{j=1}^{2M} \log S^{0_*Q}(u_j^{\dot{\alpha}_j},\,\cdot\,)}\\
    &-  \log{\left(1-\frac{1}{Y_+}\right)^2} \hat{\star} K_+^{yQ} -  \log{\left(1-\frac{1}{Y_-}\right)^2} \hat{\star} K_-^{yQ} \,.
\end{aligned}
\label{ex-TBA for YQ}
\end{equation}

\paragraph{Equations for $\bar{Q}$-particles.}
\begin{equation}
\begin{aligned}
    -\log \bar{Y}_Q &= L \tilde{\mathcal{E}}_Q - \log{(1+\bar{Y}_{Q^\prime})} \star K_{\textit{su}}^{Q^\prime Q} - \log{(1+ Y_{Q^\prime})} \star \tilde{K}_{\textit{sl}}^{Q^\prime Q} \\
    &- \log{(1+Y_0)^{N_0}} \, \check{\star} \,  \tilde{K}^{0Q}
    {\color{blue}+ \sum_{j=1}^{2M} \log \tilde{S}^{0_*Q}(u_j^{\dot{\alpha}_j},\,\cdot\,)} \\
    &-  \log{\left(1-\frac{1}{Y_+}\right)^2} \hat{\star} K_-^{yQ} - \log{\left(1-\frac{1}{Y_-}\right)^2} \hat{\star} K_+^{yQ} \,.
\end{aligned}
\label{ex-TBA for YbQ}
\end{equation}

\paragraph{Equations for Massless particles.}
\begin{equation}
\label{eq:TBA0-excited-label}
\begin{aligned}
    -\log Y_0 &= L \tilde{\mathcal{E}}_0 -  \log{(1+Y_0)^{N_0}} \check{\star} K^{00}
    {\color{blue}+ \sum_{j=1}^{2M} \log S^{0_*0}(u_j^{\dot{\alpha}_j},\,\cdot\,)}\\
    &- \log{(1+Y_Q)} \star K^{Q0} - \log{(1+\bar{{Y}}_Q)} \star \tilde{K}^{Q0} \\
    &- \log{\left(1-\frac{1}{Y_+}\right)^2} \hat{\star} K^{y0} - \log{\left(1-\frac{1}{Y_-}\right)^2} \hat{\star} K^{y0} \,.
\end{aligned}    
\end{equation}

\paragraph{Equation for auxiliary $y^-$-particles.}
\begin{equation}
\label{eq:auxiliary-ym-TBA}
\begin{aligned}
    \log{Y_-} =& -\log{(1+Y_Q)} \star K_-^{Qy} + \log{(1+\bar{Y}_Q)} \star K_+^{Qy}\\
    &+ \log{(1+Y_0)^{N_0}} \check{\star} K^{0y} 
    {\color{blue}- \sum_{j=1}^{2M} \log S^{0_*y}(u_j^{\dot{\alpha}_j},\,\cdot\,)}\,.
\end{aligned}
\end{equation}

\paragraph{Equation for auxiliary $y^+$-particles.}
\begin{equation}
\label{eq:auxiliary-yp-TBA-naive} 
\begin{aligned}
    \log{Y_+} =& -\log{(1+Y_Q)} \star K_+^{Qy} + \log{(1+\bar{Y}_Q)} \star K_-^{Qy}\\
    &-  \log{(1+Y_0)^{N_0}} \check{\star} K^{0y}
    {\color{blue}+ \sum_{j=1}^{2M} \log S^{0_*y}(u_j^{\dot{\alpha}_j},\,\cdot\,)}\,.
\end{aligned}
\end{equation}

\subsection{Regularisation of the TBA equation and ``renormalisation'' of the kernel}
The above TBA equations are potentially problematic because, in the mirror-mirror region,
\begin{equation}
    S^{0y}(u,u')=\frac{1}{\sqrt{x(u)^2}}\frac{x(u)-x_s(u')}{\frac{1}{x(u)}-x_s(u')}\,,\qquad
    S^{y0}(u',u)=\frac{1}{S^{0y}(u,u')}\,.
\label{def:S0y Sy0 uu'}
\end{equation}
with $u$ just above the long cut, and $u'$ just above the short cut.%
\footnote{
Note that with this definition, the kernel $K^{y0}(u',u)$ is positive as it should be.
} 
Hence $ S^{0_*y}(u_j^{\dot{\alpha}_j},u)$ has a zero when $u'$ is on the real mirror line, because $u_j^{\dot{\alpha}_j}$ is in the string region, \textit{i.e.} $x_s(u_j^{\dot{\alpha}_j})$ is on the upper half-circle, just like $x(u)$.

Because of this behaviour of $ S^{0_*y}(u_j^{\dot{\alpha}_j},u)$, at the special points~$u_j^{\dot{\alpha}_j}$ we have that $Y_+(u)\sim (u-u_j^{\dot{\alpha}_j})$.
In particular, this results in a logarithmic singularity in the convolution involving $\log(1-1/Y_+)$ in~\eqref{eq:TBA0-excited-label} (as well as in the massive TBA equations), and in a possible change of the sign in the argument of the logarithm. It is convenient to rewrite the convolution as 
\begin{equation}
    -\log{\left(1-\frac{1}{Y_+}\right)^2} \hat{\star} K^{y0}=
    \left(-\log{\left(1-Y_+\right)^2} +\log(Y_+)^2\right)\hat{\star} K^{y0}\,.
\label{rewrite 1-1/Yp}
\end{equation}
It is worth noting that, depending on how precisely we represent the logarithm (that is, whether we write $\log z^2$ or $2\log z$), we might obtain an additional $2i\pi$ term which needs to be convoluted with the kernel~$K^{y0}$. Observing that the convolution with a constant gives $1\hat{\star}K^{y0}=+\tfrac{1}{2}$, see appendix~\ref{app:notation}, we could obtain an additional $i\pi$ term in the TBA equations. As we will see, we can fix this ambiguity by requiring that the exact Bethe equations are compatible with the asymptotic Bethe equations.

After this rewriting, the first convolution of \eqref{rewrite 1-1/Yp} is regular at $u\approx u_j^{\dot{\alpha}_j}$. The second is not, but it can be written quite explicitly by plugging in the explicit form of $\log Y_+$ from~\eqref{eq:auxiliary-yp-TBA-naive}. (The upshot of doing this is that it will make it manifest that the potential divergences cancel, yieldsing well-behave eqeuations.) It is then natural to define
\begin{equation}
\begin{aligned}
K_{\rm ren}^{00} &= K^{00} + 2 K^{0y} \hstar K^{y0},\\
K_{\rm ren}^{Q 0} &= K^{Q 0} + 2 K^{Qy}_+ \hstar K^{y0}, \\
\tK_{\rm ren}^{Q 0} &=  \tK^{Q 0} - 2 K^{Qy}_- \hstar K^{y0},
\end{aligned}
\label{def:kernels ren}
\end{equation}
as well as
\begin{equation}
    \log S^{0_*0}_{\text{ren}}(u_j,u')=\log S^{0_*0}(u_j,u')
    +2\int\limits_{-2}^{+2}\de v
    \log S^{0_*y}(u_j,v)\,K^{y0}(v,u')\,,
\end{equation}
we can rewrite the TBA for massless modes for future convenience:
\begin{equation}
\begin{aligned}
    -\log Y_0 &= L \tilde{\mathcal{E}}_0 -  \log{(1+Y_0)^{N_0}} \check{\star} K^{00}_{\text{ren}}
    {\color{blue}+ \sum_{j=1}^{2M} \log S^{0_*0}_{\text{ren}}(u_j^{\dot{\alpha}_j},\,\cdot\,)}\\
    &- \log{(1+Y_Q)} \star K^{Q0}_{\text{ren}} - \log{(1+\bar{{Y}}_Q)} \star \tilde{K}^{Q0}_{\text{ren}} \\
    &- \log{\left(1-Y_+\right)^2} \hat{\star} K^{y0} - \log{\left(1-\frac{1}{Y_-}\right)^2} \hat{\star} K^{y0} \,.
\end{aligned}    
\label{eqY0 renorm}
\end{equation}
After this rewriting, all the terms involving logarithm of the $Y$ functions are regular in the integration domain. Hence, as long as the various kernels are not singular on the integration domain, the equation is well defined.

In a similar way,  we define 
\begin{equation} \label{eq:RenormKernels}
\begin{aligned}
K_{\rm ren}^{00_*} &= K^{00_*} + 2 K^{0y} \hstar K^{y0_*},\\
K_{\rm ren}^{Q 0_*} &= K^{Q 0_*} + 2 K^{Qy}_+ \hstar K^{y0_*}, \\
\tK_{\rm ren}^{Q 0_*} &=  \tK^{Q 0_*} - 2 K^{Qy}_- \hstar K^{y0_*},
\end{aligned}
\end{equation}
as well as
\begin{equation}
    \log S^{0_*0_*}_{\text{ren}}(u_j,u_k)=\log S^{0_*0_*}(u_j,u_k)
    +2\int\limits_{-2}^{+2}\de v
    \log S^{0_*y}(u_j,v)\,K^{y0_*}(v,u_k)\,,
\end{equation}
so that the exact Bethe equations can be written as
\begin{equation}
\begin{aligned}
    {\color{blue}i\pi(2\nu_k^{\dot{\alpha}}+1)}=&
    {\color{blue}-i L p_k^{\dot{\alpha}_k}}- \ln{(1+Y_0)^{N_0}} \check{\star} K_{\text{ren}}^{00_*}
    {\color{blue}+ \sum_{j=1}^{2M} \ln S_{\text{ren}}^{0_*0_*}(u_j^{\dot{\alpha}_j},\,u_k^{\dot{\alpha}_k})}\\
    &- \ln{(1+Y_Q)} \star K_{\text{ren}}^{Q0_*} - \ln{(1+\bar{{Y}}_Q)} \star \tilde{K}_{\text{ren}}^{Q0_*} \\
    &-  \ln{\left(1-Y_+\right)^2}  \hat{\star} K^{y0_*} - \sum_{\alpha=1,2} \ln{\left(1-\frac{1}{Y_-}\right)^2} \hat{\star} K^{y0_*}\,.
\end{aligned}
\label{eqBethe renorm}
\end{equation}
The new kernels in \eqref{def:kernels ren} have a simple form, as explained in Appendix \ref{app:Kren}.
A similar rewriting could be done also for the equations of $Q$- and $\overline{Q}$-particles, but as we will see in the next subsection these are not important for our weak-coupling analysis.
It is also worth pointing out that in this equation it is important to pick the branch of the logarithm in an appropriate way, because this may result in a misidentification of~$\nu_k$. For this reason we write~$\ln(z)$ to identify the principal branch of the log; it is also important to write the driving term as a sum of logarithms, rather than the logarithm of a product.

\subsection{Exact energy and momentum}

The exact energy also receives a correction, namely
\begin{equation}
\begin{aligned}
    \mathcal{E}(L) =& - \int\limits_{-\infty}^\infty \frac{\de u}{2 \pi} \frac{\de \tilde{p}^Q}{\de u} \log{\left((1+Y_Q)(1+\bar{Y}_Q)\right)}\\
    &- \int\limits_{|u|>2} \frac{\de u}{2 \pi} \frac{\de \tilde{p}^0}{\de u}  \log{\left(1+Y_0\right)^{N_0}} {\color{blue}+\sum_{j=1}^{2M}\mathcal{E}_0(u_j^{\dot{\alpha}_j})}\,.
\end{aligned}
\end{equation}
Once again, this does not depend on $\dot{\alpha}_j$.
The level-matching condition becomes
\begin{equation}
\label{eq:levelmatching}
\begin{aligned}
    0 =& - \int\limits_{-\infty}^\infty \frac{\de u}{2 \pi} \frac{\de \tilde{\mathcal{E}}_Q}{\de u} \log{\left((1+Y_Q)(1+\bar{Y}_Q)\right)}\\
    &- \int\limits_{|u|>2} \frac{\de u}{2 \pi} \frac{\de \tilde{\mathcal{E}}_0}{\de u} \sum_{\dot{\alpha}=1}^{N_0} \log{\left(1+Y_0\right)^{N_0}}
    {\color{blue}+\sum_{j=1}^{2M}p_j}\,.
\end{aligned}
\end{equation}
It is now easy to see a possible way to satisfy this equation. If we choose the mode numbers to come in pairs with opposite sign (regardless of the value of $\dot{\alpha}_j$), we can arrange the momenta to come in pairs $p_{2j-1}=-p_{2j}$, $j=1,\dots M$. It can be checked then that the Y-functions are even under $u\to-u$, which makes the integrals in~\eqref{eq:levelmatching} vanish~too.

\section{Simplification in the small-tension limit}
\label{sec:smalltension}
We now want to find the limit of the excited-state mirror TBA equations, the exact Bethe equations and the exact energy as $h\to0$.

\subsection{Na\"ive scaling}
\label{sec:smalltension:energy}

We start by noting the scaling of the mirror energy $\tilde{\mathcal{E}}^Q(u)$ as $h\ll1$. The mirror energy is bounded from below by its value at $\tilde{p}=0$, that is at $u=0$. Hence
\begin{equation}
 \tilde{\mathcal{E}}^Q(u)\geq \log\frac{Q^2}{h^2}\,,\qquad Q\neq0\,,
\end{equation}
while for $Q=0$, $\tilde{\mathcal{E}}^0(u)$ does not explicitly depend on $h$ at all, and hence has a finite limit as~$h\to0$.
We conclude that in the equations for $Y_Q(u)$ and $\bar{Y}_Q$ with $Q\neq0$, at least one term in the right-hand-side is divergent (and negative). Let us assume that all remaining terms in the excited-state mirror TBA equations for $Y_Q(u)$ and $\bar{Y}_Q$ with $Q\neq0$ (i.e., the convolutions and the driving terms) admit a finite limit as $h\to0$ --- we will discuss this in detail in the next subsection. Then we would immediately have that
\begin{equation}
    Y_Q(u; h) = h^{2L} \, y_Q(u)\,,\qquad
    \bar{Y}_Q(u; h) = h^{2L} \, \bar{y}_Q(u)\,,
\label{YQYbQ naive scaling}
\end{equation}
where $y_Q(u)$ and $\bar{y}_Q(u)$ are uniformly convergent and finite as $h\to0$.
This is basically what happens for $\text{AdS}_5\times\text{S}^5$ in the small tension limit, where $h^2$ would be replaced by the 't Hooft coupling~$\lambda$. At weak coupling, the Y-functions are suppressed exponentially in the volume of the system~$L$.
However, this is not the case for $Y_0$ functions, whose mirror energy remains finite. This is a major difference to the case of $\text{AdS}_5\times\text{S}^5$, and a signature of the gapless dynamics of~$\AdSST$.

Let us now come to the equations for $Y_0(u)$. Here too we need to make some assumption about kernels and driving terms, to be proven later.  
Let us assume that the kernels $K^{Q0}$ and $\tilde{K}^{Q0}$, which couple massive and massless particles, do not diverge as $h\to0$. If that is the case, we have that
\begin{equation}
    - \log{(1+Y_Q)} \star K^{Q0} - \log{(1+\bar{{Y}}_Q)} \star \tilde{K}^{Q0}=\mathcal{O}(h^{2L})\,,
\end{equation}
which can be neglected in comparison to $\tilde{\mathcal{E}}_0(u)=\mathcal{O}(h^{0})$. At this order,  the massive Y-functions decouple from the TBA equations for the massless modes. The same needs to be checked for the coupling to the auxiliary modes, given by $K^{Qy}_{\pm}$, and for the exact Bethe equations, where we encounter the analytically continued kernels $K^{Q0_*}$ and $\tilde{K}^{Q0_*}$.

Now we are left with a system of equations involving only $Y_0(u)$ and $Y_\pm(u)$. Let us first consider~$Y_0(u)$. The mirror-energy term goes like $\mathcal{O}(h^{0})$. It remains to see whether the remaining terms admit a finite limit too as $h\to0$. 
This requires again an analysis of the various kernels and driving terms. The story is the same for the auxiliary Y-functions. We will argue below that indeed
\begin{equation}
    Y_0(u)=\mathcal{O}(h^0)\,,\qquad Y_\pm(u)=\mathcal{O}(h^0)\,,\qquad
    p_j=\mathcal{O}(h^0)\,,\quad j=1,\dots,2M\,.
\end{equation}
Let us turn to the formula for the energy of an excited state. The integrals involving $Y_{Q}$ and $\bar{Y}_Q$ functions are, as usual, suppressed as~$h^{2L}$. The remaining term is given only by
\begin{equation}
    \mathcal{E}(L) =- \int\limits_{|u|>2} \frac{\de u}{2 \pi} \frac{\de \tilde{p}^0}{\de u} \log{\left(1+Y_0\right)^{N_0}} +\sum_{j=1}^{2M}\mathcal{E}_0(u_j^{\dot{\alpha}_j})= \mathcal{O}(h^1)\,,
\end{equation}
where the dependence on~$h$ comes from $\tilde{p}^0(u)$ and $\mathcal{E}_0(u)$.
Therefore, under the assumption listed above we have that, when $h\ll 1$,
\begin{enumerate}
    \item For a state including only massless excitations, massive modes decouple at leading order. Note that, had we included massive excitations, we would have expected them to contribute at $\mathcal{O}(h^0)$ to the energy, by virtue of the dispersion relation~\eqref{eq:string dispersion} (in $\mathcal{N}=4$ SYM, this contribution comes from the engineering dimension of the fields).
    \item Massless modes and auxiliary modes have some nontrivial dynamics described by a set of mirror TBA equations.
    \item The energy of all massless excitations goes to zero as $h\to0$, and the first non-trivial contribution is at $\mathcal{O}(h^1)$. It is interesting to note that, unlike the case of $\mathcal{N}=4$ SYM, here \textit{odd powers} of $h$ appear at~$h\to0$, associated to the massless modes.
\end{enumerate}

\subsection{Equations for massive particles}
\label{sec:smalltension-massive}

We now argue that none of the convolutions or of the driving terms affect the na\"ive scaling of the massive Y-functions in \eqref{YQYbQ naive scaling}.
Below we examine the TBA equations for $Q$-particles. The equations for $\bQ$-particles, as well as details of the analytic properties of the kernels and S-matrices will be discussed in Appendix \ref{app:weak-coupling exp}.

Consider the equation for $Q$-particles \eqref{ex-TBA for YQ} in the limit $h\to0$.
First, the kernels $K_{\textit{sl}}^{Q^\prime Q}$ and $\tilde{K}_{\textit{su}}^{Q^\prime Q}$ simplify at $\cO(h^0)$ as \begin{equation}
\begin{aligned}
- \log{(1+Y_{Q^\prime})} \star K_{\textit{sl}}^{Q^\prime Q} 
&\simeq
2 \log{(1+Y_{Q^\prime})} \star K_{\Sigma}^{Q_1 Q_2} (u_1,u_2) ,
\\
- \log{(1+ \bar{Y}_{Q^\prime})} \star \tilde{K}_{\textit{su}}^{Q^\prime Q} 
&\simeq
2 \log{(1+ \bar{Y}_{Q^\prime})} \star \tilde K_{\Sigma}^{\bar{Q}_1 Q_2} (u_1,u_2) .
\end{aligned}
\end{equation}
The non-zero piece comes from non-BES terms in the massive dressing factor \eqref{log of massive dressing}, which are regular for real $u_1 \,, u_2$\,.

Second, the kernel $K^{0Q} (u_1, u_2)$ is a quantity of $\cO(h)$. However, the coefficient diverges at $u_1=\pm 2$ and $u_1 \to \pm \infty$. The divergences at $u_1 = \pm 2$ are $O(1/\sqrt{u_1 \mp 2})$, which is integrable as long as $Y_0 (u_1)$ remains finite. The divergences at $u_1 \to \pm \infty$ come from the BES phase, which behaves as $O(1/x_1^2)$ as $x_1 \to 0$ for $|x_1| \ll h \ll 1$.
This would prevent us from taking the limit $h\to0$ in the integrand, unless we are certain that the kernel is integrated against a function which vanishes in the vicinity of $x_1=0$. 
But this is the case for 
\begin{equation}
- \log{(1+Y_0)^2} \, \check{\star} \, K^{0Q}
\end{equation}
In fact, if this were not the case, the energy formula~\eqref{eq:groundstateen} would be ill-defined. The reason for this suppression around~$x_1=0$ is  the driving term~$L\tilde{\mathcal{E}}^0$ in the massless TBA equation, which ensures that
\begin{equation}
\label{eq:masslessYatxis0}
    \log(1+Y_0^{(\dot{\alpha})}(x))\approx Y_0^{(\dot{\alpha})}(x)\approx x^{2L}\,,\qquad |x|\ll1\,.
\end{equation}
Hence, while the kernel is singular for $|x_1|\ll1$, the whole convolution is regular and we can take the limit $h\to0$ without any issue. 
Strictly speaking, the S-matrix $\log S^{0_*Q}(u_1 \,, u_2)$ also contributes to the expansion \eqref{eq:masslessYatxis0}. 
We will refine our argument in Appendix \ref{app:K massless^2}.

Finally, the kernels $K^{yQ_2}_\pm$ are also potentially dangerous at $u_1 = \pm 2$. Again this singularity is integrable as long as $Y_\pm (u_1)$ remain finite as $u_1 \to \pm 2$.

Hence, we conclude that all terms in the equations for $Q$-particles are regular at small $h$, except for the driving term $L \tilde{\mathcal{E}}_Q$\,. This shows that $Y_Q = \cO ( h^{2L} )$ as expected.

\subsection{Equations for massless particles}
\label{sec:smalltension-massless}

Consider the equations for massless particles \eqref{eqY0 renorm}, with the renormalised kernels given in eq. \eqref{eq:RenormKernels} (see also eqs. \eqref{KQ0 ren-ev} and \eqref{tKQ0 ren-ev} in appendix \ref{app:Kren}).
The auxiliary kernel $K^{y0}$ is regular and is explicitly given in \eqref{weak Kaux}.
The behaviour of the BES kernel $K_\bes^{Q 0} (x_2^\pm, x_1)$ at small $h$ can be estimated from \eqref{psi-integrand-strongly-weak} and \eqref{log-gamma-strongly-weak}. In particular, in the region $|x_1|\ll h\ll1$ we find
\begin{equation}
- \log{(1+Y_Q)} \star K_{\rm ren}^{Q0} - \log{(1+\bar{{Y}}_Q)} \star \tilde{K}_{\rm ren}^{Q0} \ \sim \ 
h^{2L} \log (x_1^2) .
\label{massive contribution to massless TBA}
\end{equation}
Although singular at $x_1=0$, we can safely neglect this contribution from massive particles. 
As discussed in \eqref{eq:masslessYatxis0}, the function $Y_0$ is suppressed by the driving term $-L\, \tE_{0} \simeq - L \log (x_1^2)$ around $x_1=0$. Thus the terms \eqref{massive contribution to massless TBA} just renormalises $L$ at higher order in $h$.

Since the massive dynamics decouples from the massless one, it is convenient to work always in terms of the~$\gamma$ rapidity, and use ``calligraphic'' kernels defined through
\begin{equation}
    K^{AB}(u,u')=\frac{\de\gamma}{\de u}\,\mathcal{K}^{AB} (\gamma(u), \gamma(u'))\,, \quad \text{with} \quad \mathcal{K}^{AB}(\gamma, \gamma') = \frac{1}{2 \pi i} \frac{\de}{\de \gamma} \log{S^{AB}(\gamma, \gamma')}\,.
\end{equation}
We also introduce a distinguished symbol for the convolution on the real-$\gamma$ line,
\begin{equation}
    \big(F*\mathcal{K}\big)(\gamma) = \int\limits_{-\infty}^{+\infty}\de\gamma'\,F(\gamma') \mathcal{K}(\gamma-\gamma')\,.
\end{equation}
We are interested in the leading-order expression for the calligraphic kernels, which is~$\mathcal{O}(h^0)$.
At this order, a function which will play a distinguished role in the $\gamma$-variable is the Cauchy kernel
\begin{equation}
    s(\gamma)=\frac{1}{2\pi\cosh(\gamma)}\,,
\end{equation}
which is related to the S~matrix~$S(\gamma)$ as in appendix~\ref{app:notation}.

Most of the kernels are outright independent on~$h$, with the exception of~$\cK_{\rm ren}^{00}$ which contains the BES dressing factor of massless particles~\cite{Frolov:2021fmj}. This requires some study. 
Firstly, we observe that the  sine-Gordon function~$\PhiSG(\gamma)$ disappears in the renormalised kernel, see appendix~\ref{app:massless^2 Kren}, and we have\footnote{Recall that the kernel $\cK_{\rm ren}^{00}$ was derived by substituting one of the TBA equations into another. Thus the new equations are mathematically equivalent to the original mirror TBA, even if they have a simplified form; this simplification is quite common in TBA equations, see \textit{e.g.}\ section 2.5 in~\cite{vanTongeren:2016hhc}.}
\begin{equation}
\cK_{\rm ren}^{00} (\gamma_j, \gamma_k) 
= \frac{1}{2 \pi i} \, \frac{\de}{\de \gamma_j} \Bigl[ \log S (\gamma_{jk})
- 2 \log \Sigma_{\bes}^{0 0}(x (\gamma_j) , x (\gamma_k)) \Bigr] ,
\end{equation}
so that we are left with the Cauchy and BES kernels.
We furthermore argue that the massless-massless BES kernel $\cK_\bes^{00} (\gamma_1, \gamma_2)$ also decouples at $\mathcal{O}(h^0)$. This is not immediately obvious because  this kernel is singular around $\gamma_1=0$. 
Let us collect the driving terms in the massless TBA \eqref{eqY0 renorm} and define
\begin{equation}
\scr{Y}_0 (\gamma) \equiv  e^{- L\, \tE_{0} (\gamma)} \prod_{j=1}^{2M}  {S}_{\rm ren}^{00}( \gamma_{*j} , \gamma) , \qquad
{S}_{\rm ren}^{0_* 0}( \gamma_{*} , \gamma)  = S (\gamma_{*} , \gamma ) \Sigma_\bes^{0_*0}(\gamma_{*} , \gamma )^{-2} .
\label{def:scr Y0}
\end{equation}
where $\gamma$ is the $\gamma$-rapidity in the mirror region, $\gamma_j - \frac{i \pi}{2} = \gamma_{*j} \in \bb{R}$ as in \eqref{analytic continuation to string gamma}.
The function $\scr{Y}_0 (\gamma)$ behaves as
\begin{equation}
\lim \limits_{\gamma \to 0} \, \scr{Y}_0 (\gamma) \simeq \frac{\gamma^{2 E_L^{\rm (ex)} }}{h^{2 E_L^{\rm (ex)} - 2L  } }
\,, \qquad
E_L^{\rm (ex)} \equiv L + \sum_{j=1}^{2M} \cE_0 (\gamma_{*j}) \ge 0.
\end{equation}
We approximate $\log (1+Y_0)^2 * \cK_\bes^{00}$ by $\scr{Y}_0 * \cK_\bes^{00}$\,, assuming that other terms in the TBA equations \eqref{eqY0 renorm} do not significantly modify the behaviour of $Y_0 (\gamma)$ at small $\gamma$.
This is a natural assumption because the remaining terms are convolutions with the Cauchy kernel.
Then we find
\begin{equation}
\log (1+Y_0)^2 * \cK_\bes^{00} \ \simeq \ \scr{Y}_0 * \cK_\bes^{00} \ = \ \cO (h^{2L + 1}),
\end{equation}
which vanishes in the $h\to 0$ limit. The interested reader can find a detailed discussion in Appendix~\ref{app:K massless^2}.

\subsection{Equations for auxiliary particles}
\label{sec:smalltension-aux}

In the equations for auxiliary particles \eqref{eq:auxiliary-ym-TBA} and \eqref{eq:auxiliary-yp-TBA-naive}, the following convolutions with massive Y-functions are found,
\begin{equation}
-\log{(1+Y_Q)} \star K_\mp^{Qy} + \log{(1+\bar{Y}_Q)} \star K_\pm^{Qy}\,.
\end{equation}
We can safely neglect these terms because both kernels are regular and negligible at small $h$, as can be seen from \eqref{weak KpmQy}.

Using the $\gamma$-variable, we can rewrite the convolution with $Y_0$ by the Cauchy kernel as in Appendix \ref{app:K aux}.
We may neglect the term $\log (1+Y_0)^2 \cstar \delta (\gamma)$ because $Y_0 (\gamma)$ vanishes at~$\gamma=0$.

\subsection{Exact Bethe equations}
\label{sec:smalltension-exact-bethe}

The exact Bethe equations \eqref{eqBethe renorm} contain two more ingredients, the mirror-string BES kernel and the string-string BES factor. 
The mirror-string BES kernel $\cK_\bes^{00_*} (\gamma_1, \gamma_2)$ is singular around $\gamma_1 \to 0$, but the convolution $\log{(1+Y_0)^2} * \cK_\bes^{00_*}$ is regular and small in the $h \to 0$ limit.
The string-string BES phase $\theta_\bes^{0_*0_*} (\gamma_1, \gamma_2)$ is regular and small in the limit $\gamma_1 \to 0$ with $h$ fixed. Thus, it does not contribute at the leading order of small $h$.
The detailed discussion will be given in Appendix \ref{app:K exactBethe}.

\section{Small-tension TBA equations}
\label{sec:smalltensionTBA}
Having found above that, at the first nontrivial order in~$h$ --- which is at order $\mathcal{O}(h^1)$ for the energy, and at order  $\mathcal{O}(h^0)$ for the Y-functions themselves --- only massless and auxiliary Y-functions contribute to our equations, it is convenient to rewrite the resulting equations in terms of a the same, real $\gamma$-variable, appropriately shifting the kernels to account for the relation~\eqref{eq:string-mirror-gamma}.
These are the equations which we need to solve.

\subsection{TBA equations}
We assume that all excited Bethe roots come in pairs, $\{ \gamma_{*j} \,, - \gamma_{*j} \}$ for $j=1,2, \dots M$, so that the zero-momentum condition is trivially satisfied.\footnote{The Bethe roots are located $(i \pi/2)$ below the real axis of the mirror region. We choose $\gamma_{*j}$ as a real parameter, which satisfies $x_s(\gamma_{*j}) \, x_s(-\gamma_{*j}) = -1$.}
This restricts us to parity-even states. The form of the small-tension TBA equations for generic states is given in appendix~\ref{App:MasslessSourceTerms}.

Following our argument in Section \ref{sec:smalltension} and rewriting the equations in terms of $\gamma$ we have that, at leading order in~$h$,
\begin{equation}
\begin{aligned} \label{eq:Y0almostSimplyfied}
    -\log Y_0(\gamma) &= L \tilde{\mathcal{E}}_0 (\gamma) -  \(\log(1+Y_0)^{N_0}*s\) (\gamma)
   - \sum_{j=1}^{M} \log (S_{*}(-\gamma_j^{\dot{\alpha}_j},\gamma)\,S_{*}(\gamma_j^{\dot{\alpha}_j},\gamma))\\
    &-  \(\log{\left(1-Y_+\right)^2} *s\)(\gamma) -  \(\log{\left(1-\frac{1}{Y_-}\right)^2} *s\)(\gamma) \,,
\end{aligned}    
\end{equation}
where we have paired the roots of opposite sign.
For the auxiliary particles
\begin{equation}
\begin{aligned} \label{eq:YalmostSimplyfied}
    \log Y_-(\gamma) =& -\(\log{(1+Y_0)^{N_0}} *s\)(\gamma) 
   - \sum_{j=1}^{M} \log \left(S_{*}(-\gamma_j^{\dot{\alpha}_j}-\gamma)S_{*}(\gamma_j^{\dot{\alpha}_j}-\gamma)\right)\,,\\
    \log Y_+(\gamma) =& +\(\log{(1+Y_0)^{N_0}} *s\)(\gamma) 
   + \sum_{j=1}^{M} \log \left(S_{*}(-\gamma_j^{\dot{\alpha}_j}-\gamma)S_{*}(\gamma_j^{\dot{\alpha}_j}-\gamma)\right)\,.
\end{aligned}
\end{equation}
It is worth noting that the fact that the massless and auxiliary TBA equations pick up the \textit{same} source terms is due to our choice of parity-even states, \textit{i.e.} $(-\gamma_j^{\dot{\alpha}_j},\gamma_j^{\dot{\alpha}_j})$. We give the equations for an arbitrary number of roots in Appendix~\ref{App:MasslessSourceTerms}.
We see that at leading order
\begin{equation}
\label{eq:auxiliaryidentity}
    Y_+(\gamma)\,Y_-(\gamma)=1\,,
\end{equation}
so that we can set
\begin{equation}
\label{eq:auxiliaryidentification}
     Y(\gamma)= Y_+(\gamma)=\frac{1}{Y_{-}(\gamma)}\,,
\end{equation}
and finally write
\begin{equation}
\begin{aligned}
    -\log Y_0(\gamma) &= L \tilde{\mathcal{E}}_0(\gamma) -  \(\log(1+Y_0)^{N_0}*s\) (\gamma)
   - \sum_{j=1}^{M} \log (S_{*}(-\gamma_j^{\dot{\alpha}_j}-\gamma)\,
   S_{*}(\gamma_j^{\dot{\alpha}_j}-\gamma))\\
    &-  \(\log{\left(1-Y\right)^4} *s\)(\gamma) \,,\\
    \log Y(\gamma) =& \(\log{(1+Y_0)^{N_0}} *s\)(\gamma) 
   + \sum_{j=1}^{M} \log \left(S_{*}(-\gamma_j^{\dot{\alpha}_j}-\gamma)S_{*}(\gamma_j^{\dot{\alpha}_j}-\gamma)\right)\,.
\end{aligned}    
\label{weak TBA Y0Yy}
\end{equation}

These equations are to be supplemented by the exact Bethe equations%
\footnote{%
As before, for these equation it is important to pick the branch of the logarithm, and $\ln(z)$ denotes the principal branch. It is worth noting that different prescriptions for analytic continuation would have shifted the value of~$\nu$, which we fix by comparison with the asymptotic Bethe ansatz.
}
\begin{equation}
\begin{aligned}
    i\pi(2\nu_k^{\dot{\alpha}}+1)=&
    -i L p(\gamma^{\dot{\alpha}_k})-  \(\ln{(1+Y_0)^{N_0}} * s_*\)(\gamma_k^{\dot{\alpha}_k})
    + \sum_{j=1}^{2M} \ln S(\gamma_j^{\dot{\alpha}_j}-\gamma_k^{\dot{\alpha}_k})\\
    &- \(\ln{\left(1-Y\right)^4}  * s_*\)(\gamma_k^{\dot{\alpha}_k}) \,.
\end{aligned}
\label{weak exact Bethe eqs}
\end{equation}
Finally, the energy is
\begin{equation}
\mathcal{E}(L) =- \int\limits_{-\infty}^{+\infty} \frac{\de \gamma}{2 \pi} \frac{\de \tilde{p}}{\de \gamma} \log\left(1+Y_0(\gamma)\right)^{N_0}
+\sum_{j=1}^{2M}\mathcal{E}(\gamma_j^{\dot{\alpha}_j})\,.
\label{weak exact energy}
\end{equation}
For convenience, let us collect the definition of the relevant functions of~$\gamma$:
\begin{equation}
\begin{aligned}
&s(\gamma) = \frac{1}{2 \pi i} \, \frac{\de}{\de \gamma} \log S (\gamma)
=  \frac{1}{2 \pi \cosh{\gamma}} \,,& \qquad
&S (\gamma) =-i \tanh \( \frac{\gamma}{2} - \frac{i \pi}{4} \)\,,&\\
&s_* (\gamma) = s(\gamma+\tfrac{i\pi}{2})
=\frac{1}{2\pi i\,\sinh\gamma}\,,& \qquad
&S_* (\gamma) =S (\gamma+\tfrac{i\pi}{2}) =-i \tanh  \frac{\gamma}{2} \,.&
\end{aligned}
\label{def:ss-SS}
\end{equation}
and 
\begin{equation}
    \tilde{p}(\gamma)=-\frac{2h}{\sinh\gamma}\,,\qquad
    \frac{\de\tilde{p}}{\de\gamma}=\frac{2h\cosh\gamma}{\sinh^2\gamma}\,,\qquad
    \tilde{\mathcal{E}}(\gamma)=-\ln\left(\frac{1-e^\gamma}{1+e^\gamma}\right)^2\,,
\end{equation}
and in the string region
\begin{equation}
    \mathcal{E}(\gamma)=i\tilde{p}(\gamma-\tfrac{i\pi}{2})=\frac{2h}{\cosh\gamma}\,,\qquad
    {p}(\gamma)=i\tilde{\mathcal{E}}(\gamma-\tfrac{i\pi}{2})=-i\ln\left(\frac{e^\gamma-i}{e^\gamma+i}\right)^2\,.
\end{equation}

To begin with, we will solve these equations for $L$ a positive integer and $M=1$ (\textit{i.e.}, a state with two excitations of opposite momentum) as well as $M=2$. In the latter case, we will further assume for simplicity that the momenta come in pairs (\textit{i.e.}, that the state is even under parity).

\subsection{Y system}
It is worth noting that we can straightforwardly obtain a set of equations called Y system starting from the small-tension TBA of the previous subsection. To this end, let us introduce the (pseudo-)inverse of the Cauchy kernel, $s^{-1}$, and the notation 
\begin{equation}
    [F*s^{-1}](\gamma)=\lim_{\epsilon\to0^+}\left[F(\gamma+\frac{i\pi}{2}-i\epsilon)+F(\gamma-\frac{i\pi}{2}+i\epsilon) \right]\,.
\end{equation}
In the sense of distributions,
\begin{equation}
    [s*s^{-1}](\gamma) = \delta(\gamma)\,.
\end{equation}
However, $s^{-1}$ is not a left-inverse because it has non-trivial null space. In particular, note that
\begin{equation}
    [\tilde{\mathcal{E}}_0*s^{-1}](\gamma) = 0\,,
\end{equation}
exactly as it is the case in the relativistic case. The source terms involving~$\log S_*$ are also in the null space of~$s^{-1}$.
Hence, applying~$s^{-1}$ to the two TBA equations we get
\begin{equation}
\begin{aligned}
    Y_0(\gamma+\tfrac{i\pi}{2}-i0)\,Y_0(\gamma-\tfrac{i\pi}{2}+i0)&=\left[1+Y_0(\gamma)\right]^{N_0}\,\left[1-Y(\gamma)\right]^{4}\,,\\
    Y(\gamma+\tfrac{i\pi}{2}-i0)\,Y(\gamma-\tfrac{i\pi}{2}+i0)&=\left[1+Y_0(\gamma)\right]^{N_0}\,,
\end{aligned}
\end{equation}
which takes the form of a standard Y-system.

It is also possible to refine this analysis by twisting the $\su(2)_{\bullet}$ symmetry. To this end, it is sufficient to repeat our analysis starting with the twisted TBA equations~\cite{Frolov:2021bwp} by means of a chemical potential~$e^{+i\mu}$ for the $\alpha=1$ auxiliary Y~function and $e^{-i\mu}$ for the  $\alpha=2$ auxiliary Y~function.%
\footnote{In fact, with respect to the notation of~\cite{Frolov:2021bwp} it is convenient to redefine $Y_{\pm}^{(\alpha)}|_{\alpha=1}\to e^{-i\mu}Y_{\pm}^{(\alpha)}|_{\alpha=1}$ and $Y_{\pm}^{(\alpha)}|_{\alpha=2}\to e^{+i\mu}Y_{\pm}^{(\alpha)}|_{\alpha=2}$ so that the twist appears only in the right-hand side of the auxiliary TBA equations.}
In this case, the auxiliary TBA equations become
\begin{equation}
\begin{aligned}
    Y^{(\alpha)}_\pm\big|_{\alpha=1}&= +i\mu \mp \sum_{\dot{\alpha}=1}^{N_0}\log\left(1+Y_0^{\dot{\alpha}}\right)\check{\star}K^{0y}+\mathcal{O}(h)\,,\\
     Y^{(\alpha)}_\pm\big|_{\alpha=2}&= -i\mu \mp \sum_{\dot{\alpha}=1}^{N_0}\log\left(1+Y_0^{\dot{\alpha}}\right)\check{\star}K^{0y}+\mathcal{O}(h)\,.
\end{aligned}
\end{equation}
In this case we cannot proceed with the identification~\eqref{eq:auxiliaryidentification} but we must instead distinguish two types of auxiliary Y~functions,
\begin{equation}
    Y_1(\gamma)=Y^{(1)}_+(\gamma)=\frac{1}{Y_{-}^{(2)}(\gamma)},\qquad
    Y_2(\gamma)=Y^{(2)}_+(\gamma)=\frac{1}{Y_{-}^{(1)}(\gamma)}.
\end{equation}
With this in mind, the derivation of the TBA and Y~system follows closely the case above and it gives eventually
\begin{equation}
\begin{aligned}
    Y_0(\gamma+\tfrac{i\pi}{2}-i0)\,Y_0(\gamma-\tfrac{i\pi}{2}+i0)&=\left[1+Y_0(\gamma)\right]^{N_0}\,\left[1-Y_1(\gamma)\right]^{2}\,\left[1-Y_2(\gamma)\right]^{2}\,,\\
    Y_1(\gamma+\tfrac{i\pi}{2}-i0)\,Y_1(\gamma-\tfrac{i\pi}{2}+i0)&=\left[1+Y_0(\gamma)\right]^{N_0}\,e^{+2i\mu},\\
    Y_2(\gamma+\tfrac{i\pi}{2}-i0)\,Y_2(\gamma-\tfrac{i\pi}{2}+i0)&=\left[1+Y_0(\gamma)\right]^{N_0}\,e^{-2i\mu}.
\end{aligned}
\end{equation}

In the case of relativistic systems with ADE symmetries (as well as their supersymmetric generalisations), there is a well-known relation between the Y-system (or TBA) and the Cartan matrix of the model~\cite{Zamolodchikov:1991et}. Let us define the incidence matrix $I_{ij}$ by
\begin{equation}
    Y_i(\gamma+\tfrac{i\pi}{2}-i0)\,Y_i(\gamma-\tfrac{i\pi}{2}+i0)=\prod_{j=1}^r [1\pm Y_j(\gamma)]^{I_{ij}}\,,\qquad
    i=1,\dots, r\,,
\end{equation}
where the plus or minus sign depends on the statistics of the excitation and we drop the twists. Then the Cartan matrix is, at least for ADE models and their supersymmetrisations, $C_{ij}=2\delta_{ij}-I_{ij}$.
It is tempting to map our Y~systems to some generalised Cartan matrix. In fact, the result matrices~$C_{ij}$ are reminiscent of those of almost affine  Lie superalgebras, see~\cite{Frappat:1987ix,Chapovalov:2009ni}.
While it is not immediately clear to us how to do so, it would be interesting to better understand the symmetry properties underlying this weak-tension Y~system.

\section{Numerical evaluation of the tensionless spectrum}
\label{sec:tensionlessspectrum}

We numerically solved the TBA equations \eqref{weak TBA Y0Yy} and the exact Bethe equations \eqref{weak exact Bethe eqs}, and computed the energy \eqref{weak exact energy}.
The exact energy for excited states is a sum of asymptotic terms and integration over the massless Y-function. Both terms contribute at $\cO(h)$ for the states with massless particle excitations.

\subsection{Numerical algorithm}

The numerical solution of TBA equations follows a standard iterative route~\cite{Zamolodchikov:1990tba,Tongeren:2016tba}. However, the additional presence of the exact Bethe equations \eqref{weak exact Bethe eqs} gives rise to subtleties worth discussing.
The TBA equations \eqref{weak TBA Y0Yy} have the schematic form
\begin{equation}
\label{eq:tbaschematic}
\begin{split}
    &\log Y_0 = D_0 + \log\qty(1+Y_0)^{N_0}*s  + 
    \log\qty(1-Y)^{4}*s\,, \\
    &\log Y = D + \log\qty(1+Y_0)^{N_0}*s\,,
\end{split}
\end{equation}
where $D_0(\gamma)$ and $D(\gamma)$ are the driving terms that in our case include also the logarithm of the S~matrices evaluated at the positions of the excitations.
The bottom equation is just auxiliary, and in fact the expression for $Y(\gamma)$ could be directly substituted into the top equation. In contrast, in the first equation $Y_0$ appears on both sides.
To find $Y_0(\gamma)$ we will therefore proceed by iterations (starting from some reasonable guess) which generally leads to a rapid convergence.

\paragraph{Convolutions.}
To numerically evaluate the convolutions we need to set a cutoff on the rapidity, $|\gamma|\leq \Lambda$, discretise the resulting finite interval~$[-\Lambda,\Lambda]$ and use the fast Fourier transform algorithm.
In cutting off the space of rapidity, we are introducing an error when the convolution is computed close to~$\gamma=\pm\Lambda$~\cite{Tongeren:2016tba}. To address this, we can subtract the constant asymptotic value of the Y functions.
These are defined as
\begin{equation}
    y_0=\lim_{\gamma\to\pm\infty}Y_0(\gamma)\,,\qquad
    y=\lim_{\gamma\to\pm\infty}Y(\gamma)\,,
\end{equation}
where we used that the states which we consider are parity-even. 
We rewrite~\eqref{eq:tbaschematic} as
\begin{align}
\label{eq:regularisedtba}
\nonumber
    &\log Y_0 = D_0 + \log(\tfrac{1+Y_0}{1+y_0})^{N_0}*s  + \frac{1}{2}\log\qty(1+y_0)^{N_0} +
    \log(\tfrac{1-Y}{1-y})^{4}*s + \frac{1}{2}\log(1-y)^4,\\
    &\log Y = D + \log(\tfrac{1+Y_0}{1+y_0})^{N_0}*s  + \frac{1}{2}\log(1+y_0)^{N_0}   .
\end{align}
so that the argument of the logarithms in the convolutions goes to zero when $\gamma\to\pm\infty$, and is small at $\gamma=\pm\Lambda$ if $\Lambda$ is sufficiently large. In fact, $Y_0(\Lambda)$ approaches~$y_0$ exponentially as $\Lambda\to\infty$, so that it is easy to bound the error due to the cutoff. The same is true for~$Y(\Lambda).$

To find the values of~$y_0$ and $y$ we can therefore drop the convolutions and use the asymptotic behaviour of~$D(\gamma)$ and $D_0(\gamma)$.  The only non-vanishing contribution is due to the S~matrices, and results in
\begin{equation}
    \lim_{\gamma\to\pm\infty} D_0(\gamma)
    =
     \lim_{\gamma\to\pm\infty} D(\gamma)
     =\log(-1)^M\,.
\end{equation}
Therefore we find that it must be
\begin{equation}
\begin{split}
    &y_0 = (-1)^M(1+y_0)^{\frac{N_0}{2}}(1-y)^2\,,  \\
    &y = (-1)^M(1+y_0)^{\frac{N_0}{2}}\,.
\end{split}
\end{equation}
We collect some numerical solutions of these equations for different values of $N_0$ and $M$ in Table \ref{table:AsymYfunc}.
Finally, we discretise the interval to a lattice. 
We found that the cutoff~$\Lambda\approx40$ and a discretisation over~$N=2^{12}\approx 4000$ points is sufficient  to achieve a precision beyond the twelfth decimal place in the energies of the states. More precisely, if we double both the cutoff $\Lambda$ and $N$, thus keeping $\Lambda/N$ constant, the energies and the Bethe roots change after the twelfth decimal place. This is consistent with the fact that we found   at the cutoff $\Lambda=40$ both $Y_0(\pm\Lambda)$ and $Y(\pm\Lambda)$ differ from their asymptotic values $y_0$ and $y$ by less than~$10^{-15}$.
A similar check has been done increasing $N$ with~$\Lambda$ fixed; this also only affects the result beyond the twelfth decimal~place.

\begin{table}[t]
\centering
\begin{tabular}{l|| l| l |l |l} 

  & $N_0=1$ & $N_0=2$ & $N_0=3$& $N_0=4$ \\ [0.5ex] 
 \hline\hline
 $M$~odd & $
\begin{aligned}
    y_0&\approx -0.828427\\
    y&\approx -0.413214
\end{aligned}
 $ & $\begin{aligned}
    y_0&\approx -0.646790\\
    y&\approx -0.353210
\end{aligned}
 $ & $\begin{aligned}
    y_0&\approx -0.539171\\
    y&\approx -0.312830
\end{aligned}
 $ & $\begin{aligned}
    y_0&\approx -0.467940\\
    y&\approx -0.283658
\end{aligned}
 $\\ [3.5ex] 
 \hline
 $M$~even & 
 $
 \begin{aligned}
    y_0&=0\\
    y&=1
\end{aligned}
 $
 & $
 \begin{aligned}
    y_0&=0\\
    y&=1
\end{aligned}
 $
 & $
 \begin{aligned}
    y_0&=0\\
    y&=1
\end{aligned}
 $
 & $
 \begin{aligned}
    y_0&=0\\
    y&=1
\end{aligned}
 $\\[3.5ex] 
\end{tabular}
\caption{Asymptotic values of the Y functions for various choices of $M$ and $N_0$. The main case of interest for us is~$N_0=2$, but it is worth including more general cases for later convenience. For $M$~odd, all values are between $0$ and~$-1$. For $M$~even (\textit{i.e.}, when the total number of excitations is a multiple of~4), the asymptotic values are the same as for the vacuum. This makes the regularisation~\eqref{eq:regularisedtba} ill defined (see the paragraph ``the case of even~$M$'').}
\label{table:AsymYfunc}
\end{table}

\paragraph{Exact Bethe equations.}
To fully specify the right-hand side of the TBA equations, we also need to compute the driving terms, which in turn depend on the position of the exact Bethe roots~$\gamma_k^{\dot{\alpha}_k}$. These in turn are fixed by the exact Bethe equations~\eqref{weak exact Bethe eqs}, whose solution gives rise to some subtleties.
Let us discuss these in the case where we have two roots that have opposite values; to lighten the notation we may drop the $\su(2)_\circ$ indices, which are irrelevant here (they only matter in case of repeated roots) and denote the two roots by $(\gamma_{1}, -\gamma_{1})$.%
\footnote{We stress that these are \textit{real} roots, which morally are defined on  the real string line. The TBA equations are already written in such a way to account for the $i\pi/2$ shifts to and from the mirror line.}
First of all, the integral kernel diverges when the integration variable $\gamma$ approaches $\pm\gamma_1$, but since $Y_0(\pm\gamma_1)=Y(\pm\gamma_1)=0$, in both the convolutions the simple pole is canceled out by the zero from the logarithm. 
Thus, in the numerical evaluation of the convolutions, we approximate the integrand in a small neighborhood around~$\pm\gamma_1$ by
\begin{equation}
\lim_{\gamma\to\gamma_1}\frac{\log(1+Y_0(\gamma))^{N_0}}{2\pi i\sinh(\gamma_1-\gamma)}=
\frac{N_0}{2\pi i}\,Y_0{}'(\gamma_1),\qquad
\lim_{\gamma\to\gamma_1}\frac{\log(1-Y(\gamma))^4}{2\pi i\sinh(\gamma_1-\gamma)}=
-\frac{2}{i\pi}\,Y{}'(\gamma_1)\,.
\end{equation}
The derivatives of the Y-functions $Y_0{}'(\gamma_1)$ and $Y'(\gamma_1)$ can be computed by taking the derivative of both sides of~\eqref{weak TBA Y0Yy}. It is worth noting that the right-hand side does not depend on the derivative of $Y(\gamma)$ and $Y_0(\gamma)$; in fact, the only non-vanishing terms are those where the derivative acts on~$\log S_*$.
Another issue that requires some care is the identification of the mode numbers~$\nu_k^{\dot{\alpha}}$ --- or, in our lighter notation, $(\nu_1,-\nu_1)$. The left-hand side of the exact Bethe equation is of the form $i\pi(2\nu_1+1)$, hence it is important not to introduce spurious monodromies of the logarithm  $2\pi i$ in the right-hand side, as this would effectively shift the mode number.%
\footnote{This is not important in the TBA equations, as we are interested in $Y_0$ and $Y$, so that the monodromies drop when exponentiating~\eqref{eq:tbaschematic}.}
We define the driving term to be given by the sum of the principal branch of the logarithm of the S~matrices, 
\begin{equation}
    \sum_{j=1}^{2M}\ln S(\gamma_j-\gamma_k)=\ln S(0)+\ln S(-2\gamma_k)=i\pi+\ln S(-2\gamma_k)\,,
\end{equation}
where in the first equality we evaluated the expression for the case at hand~$M=1$, and $\ln(z)$ indicates the principal branch of the logarithm.

Finally, we have to understand the eligible values of the modes $\nu_k^{\dot{\alpha}}$.
To this end, we consider the asymptotic Bethe Ansatz equations.
It is possible to neglect the BES phase in the massless-massless S-matrix at small $h$, as explained in Appendix \ref{app:K exactBethe}. Thus the asymptotic Bethe Ansatz Equations for massless particles are
\begin{equation}
1=e^{i p_k L} \, \prod_{j \neq k}^{2M} S_{00} (p_j, p_k), \qquad 
S_{00} (p_j, p_k) \simeq S (\gamma_{j}-\gamma_k) \, e^{2 \phiSG (\gamma_{j}-\gamma_k)} + \cO(h).
\end{equation}
where the S-matrices are given in Appendix \ref{app:notation}. 
By taking the logarithm, we find%
\footnote{%
A similar expression, without the Sine-Gordon terms~$\varphi(\gamma)$, could also be obtained from dropping the convolutions in~\eqref{weak exact Bethe eqs}. In that case one would find similar numerical values for the roots, equally suitable for the purpose of using them as seed in the iterative procedure.
}
\begin{equation}
\label{eq:ABA}
2 \pi i \nu_k = -i p(\gamma_k) L + \sum_{j \neq k}^{2M} \Big( \ln S (\gamma_{j}-\gamma_k) + 2 \phiSG (\gamma_{j}-\gamma_k) \Big)\,.
\end{equation}
In the case discussed before with $M=1$, we have
\begin{equation}
2 \pi i \nu_1 = -i p(\gamma_1) L + \ln S (-2\gamma_1) + 2 \phiSG (-2\gamma_1) \,.
\end{equation}
Since we have two equations, one for $\nu_1$ and the other for $-\nu_1$, we can assume $\nu_1\geq0$. Moreover, the momentum is defined modulo $2\pi i$ and we choose as fundamental region $[-\pi,\pi]$, so that's why the possible modes are $\nu_1=0,\ldots,L/2$ for $L$ even. 
In particular, for $\nu_1=0$ we have the formal solution $\gamma\to\infty$ which correspond to zero momentum.

\paragraph{Iterative procedure.}
To start the iterative procedure we need to specify the initial values for $Y_0(\gamma)$ and $Y(\gamma)$ on the real mirror line, and of the Bethe roots on the real string line.
We initially assume that $Y_0$ is constant, and equal to its asymptotic value $y_0$; we do not need an initial value for $Y$ as it is computed from~$Y_0$.
For the Bethe roots, we take them the solutions of~\eqref{eq:ABA}. 
With this choice of the initial values, we can start the iterative procedure. If at the step $n$ we have $Y_0^{[n]}$,$Y^{[n]}$ and $\gamma_k^{[n]}$, we evolve to the next step by computing
\begin{equation}
\begin{split}
    Y^{[n+1]} &= \exp\left[D(\gamma_k^{[n]}) + \log\qty(1+Y_0^{[n]})^{N_0}*s \right], \\ 
    Y_{0,tmp}^{[n+1]} &= \exp[D_0(\gamma_k^{[n]}) + \log\qty(1+Y_0^{[n]})^{N_0}*s  + 
    \log\qty(1-Y^{[n+1]})^{4}*s],
\end{split}
\end{equation}
where $D(\gamma_k^{[n]})$ and $D_0(\gamma_k^{[n]})$ are the driving terms computed using the set of Bethe roots $\gamma_k^{[n]}$, and the reason for defining $Y_{0,tmp}^{[n+1]}$ will be clear in a moment.
We stress that the  Y functions at order $(n+1)$ are computed using $\gamma_k^{[n]}$, hence $Y_{0,tmp}^{[n+1]}(\gamma_k^{[n]}) = Y^{[n+1]}(\gamma_k^{[n]}) = 0$. This is important when solving the exact Bethe equations, as that zero is needed to cancel a pole in the integration kernel (see the previous paragraph). Hence, we solve
\begin{equation}
\label{eq:exactbetheiterative}
\begin{aligned}
    i\pi(2\nu_k^{\dot{\alpha}}+1)=&
    -i L p(\gamma_k^{[n+1]})-  \(\ln(1+Y_{0,tmp}^{[n+1]})^{N_0}* s_*\)(\gamma_k^{[n]})
    \\
    &
    - \(\ln{\left(1-Y^{[n+1]}\right)^4}  * s_*\)(\gamma_k^{[n]})+ \sum_{j=1}^{2M} \ln S(\gamma_j^{[n+1]}-\gamma_k^{[n+1]}) \,,
\end{aligned}
\end{equation}
where convolutions are computed with reference to~$\gamma_k^{[n]}$ rather than of~$\gamma_k^{[n+1]}$ precisely to make the convolution well-defined. This is then solved as an equation (or a system of equations, if $M>1$) for $\gamma_k^{[n+1]}$.
Finally, we define $Y_0^{[n+1]}$ as
\begin{equation}
    Y_0^{[n+1]} = (1-a)~Y_{0,tmp}^{[n+1]} + a~Y_0^{[n]} \\
\end{equation}
where $a$ is a damping parameter that helps with the convergence~\cite{Dorey:1996re} which we set $a=0.6$.
In our evaluation, we terminated the iteration if both $\|Y_0^{[n+1]}(\gamma)-Y_0^{[n]}(\gamma)\|<10^{-15}$ in the uniform norm, and $|\gamma_k^{[n+1]}-\gamma_k^{[n]}|<10^{-13}$ for all~$k$.

\paragraph{The case of even $M$.}
As it can be seen from table~\ref{table:AsymYfunc}, if $M$ is even (that is, if the total number of excitation is a multiple of four), the asymptotic value of the Y-function is particularly simple. 
In fact, the values of~$y$ and~$y_0$ are the same that we would find for a vacuum solution~\cite{Frolov:2023wji}.
However, this makes the prescription around~\eqref{eq:regularisedtba} ill-defined when it comes to adding and subtracting the logarithm of $(1-y_0)$. In this case, to solve numerically the TBA equations, we just take $Y_0(\gamma)$ to be zero outside of the interval~$[-\Lambda,\Lambda]$.%
\footnote{To ensure that $|Y_0(\pm\Lambda)|<10^{-12}$ at the cutoff and keep the numerical errors small, we take $\Lambda=80$ for~$M=2$.} 
If needs be, to obtain the desired accuracy, we can always make the cutoff~$\Lambda$ bigger. We then use the original form of the equations~\eqref{eq:tbaschematic}. The equation for $Y(\gamma)$ has no issue. The only possible issue comes from the equation for~$Y_0(\gamma)$ when~$|\gamma|$ is sufficiently large, as there is a logarithmic divergence from $\log(1-Y)^4$. In the exponential form of the equations, however, this just makes it clear that $Y_0(\gamma)\approx0$ when~$|\gamma|$ is sufficiently large, as it should be.
In the exact Bethe equations, which are necessarily written in logarithmic form, see~\eqref{eq:exactbetheiterative}, we may worry about the appearance of large negative numbers from $(\log(1-Y)*s_*)(\gamma_k)$. However, this is also not an issue: when $\gamma\approx\gamma_k$, where the integration kernel is large (in fact, divergent), the logarithm is small because $Y(\gamma_k)=0$. When the integration variable is $|\gamma|\gg \gamma_k$, the term~$\log(1-Y)$ does diverge logarithmically, but this is more than compensated by $s_*(\gamma-\gamma_k)$ which goes to zero exponentially. Hence, also in this case we do not encounter any issue in the numerical evaluation.

\subsection{Numerical results}
Here we present the result of the numerical evaluation of the mirror TBA equations.  After calculating the Y-functions and the exact Bethe roots as explained in the previous subsection, we obtained the energies as in~\eqref{weak exact energy}. In that formula, the integrand has a second order pole for $\gamma=0$ but $Y_0(\gamma)\approx \gamma^{2L}$ for $\gamma\ll1$ owing to the $L\tilde{\mathcal{E}}_0(\gamma)$ term in~\eqref{weak TBA Y0Yy}. Thus, the potential divergence is cured and the integrand can be approximated by zero in a small neighborhood around $\gamma=0.$
 We will discuss the case $N_0=2$ in detail and comment on the generalisation in the last paragraph of this section.

\begin{figure}
     \centering
\includegraphics[width=0.47\textwidth]{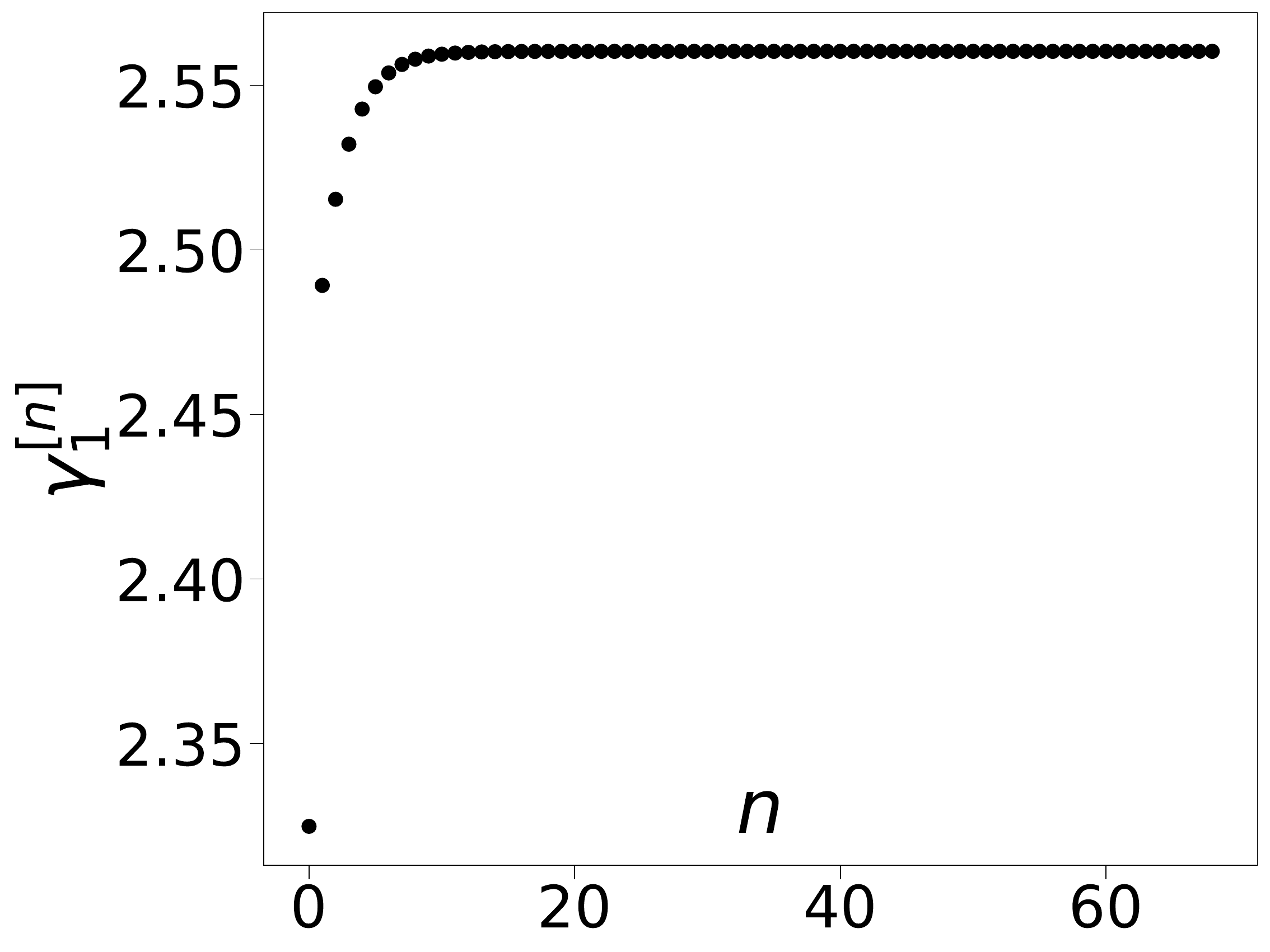}
\includegraphics[width=0.47\textwidth]{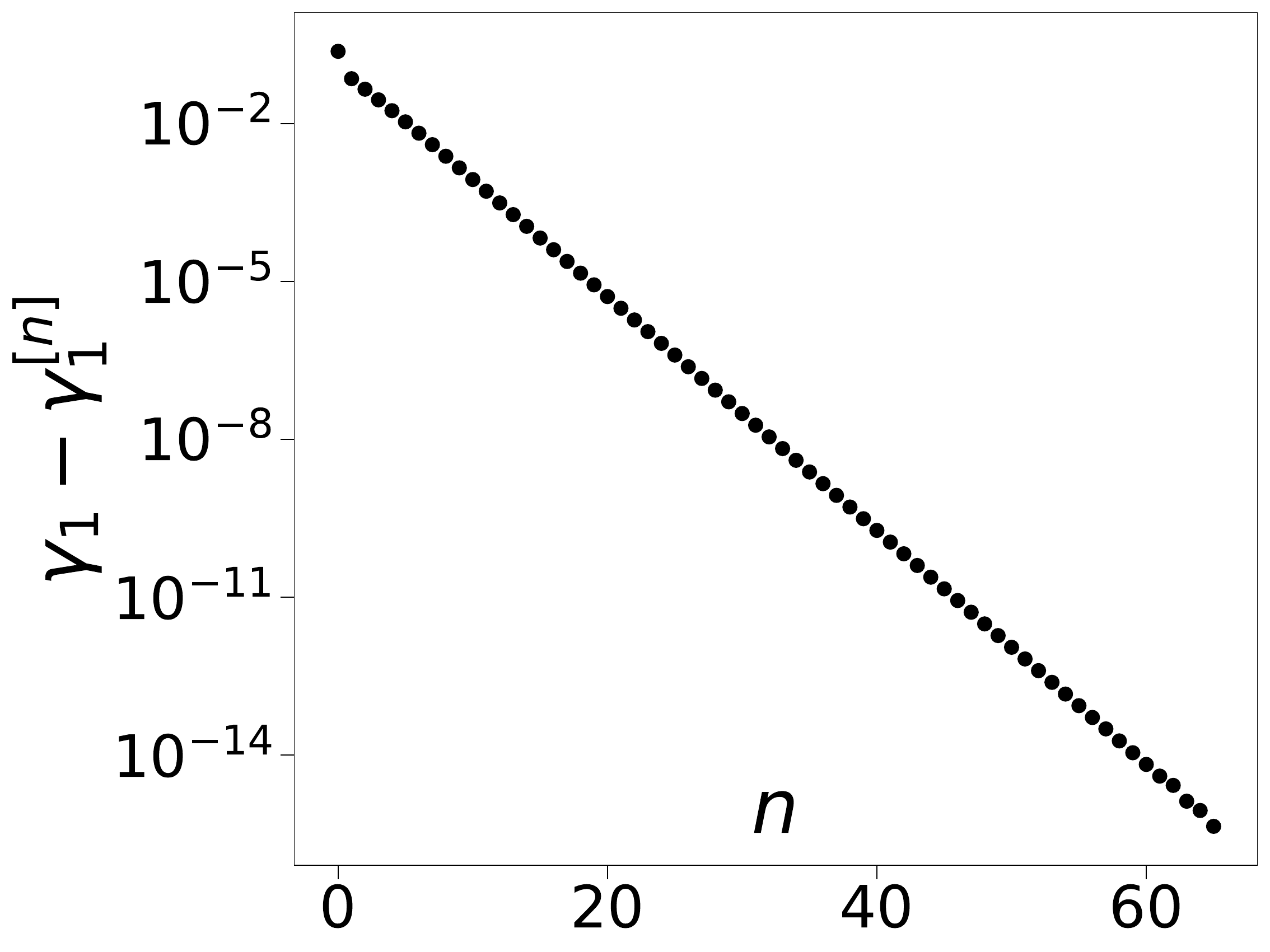}
\caption{Convergence of the exact Bethe root $\gamma_1^{[n]}$ at the~$n$-th iteration for world-sheet volume $L=32$ and mode number $\nu=1$. Each dot represents an iteration. On the left the plot with the values of $\gamma_1^{[n]}$, while on the right the deviations with respect to the final value. The plot on the right has y-axis in log scale to emphasise the exponential convergence. The starting value is the asymptotic Bethe root, as explained in the previous subsection.}
\label{fig:exactBethe}
\end{figure}

\begin{figure}
     \centering
        \includegraphics[width=\textwidth]{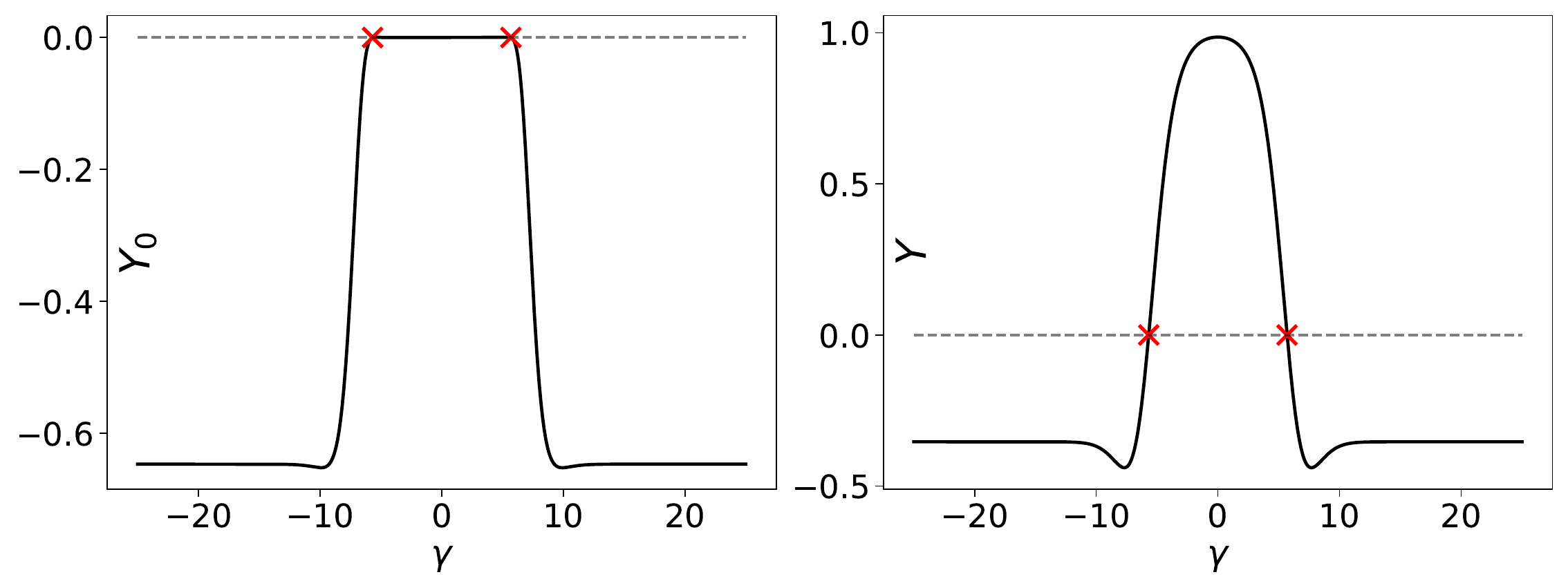}
        \caption{Y~functions for a state with $L=256$ and mode number~$\nu=1$. 
        On the left $Y_0(\gamma)$ and on the right $Y(\gamma)$. The red crosses indicate the positions of the Bethe roots, where both the Y~functions are equal to zero and both change sign. $Y_0(\gamma)$ takes very small positive values in the interval~$[-\gamma_1,\gamma_1]$, since it rapidly converges to zero as $\gamma\to 0$. For $|\gamma|\gg\gamma_1$ both Y functions quickly converge to their asymptotic values, see table~\ref{table:AsymYfunc}. Even though the plots show only the region  $|\gamma|<25$, the cutoff has been set to $\Lambda=40$.}
        \label{fig:Yfunctions}
\end{figure}

\begin{figure}
     \centering
         \includegraphics[width=0.49\textwidth]{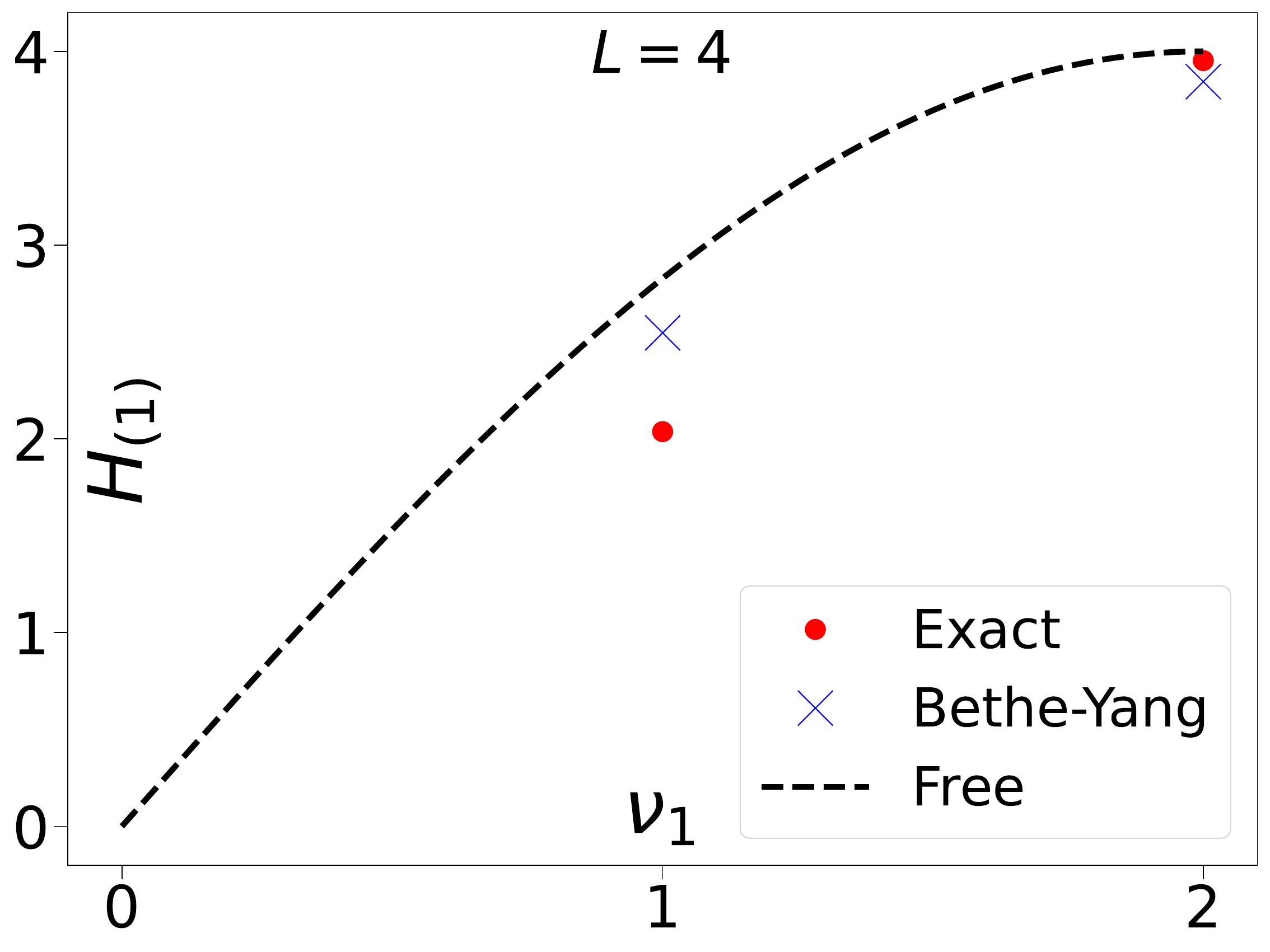}~%
         \includegraphics[width=0.49\textwidth]{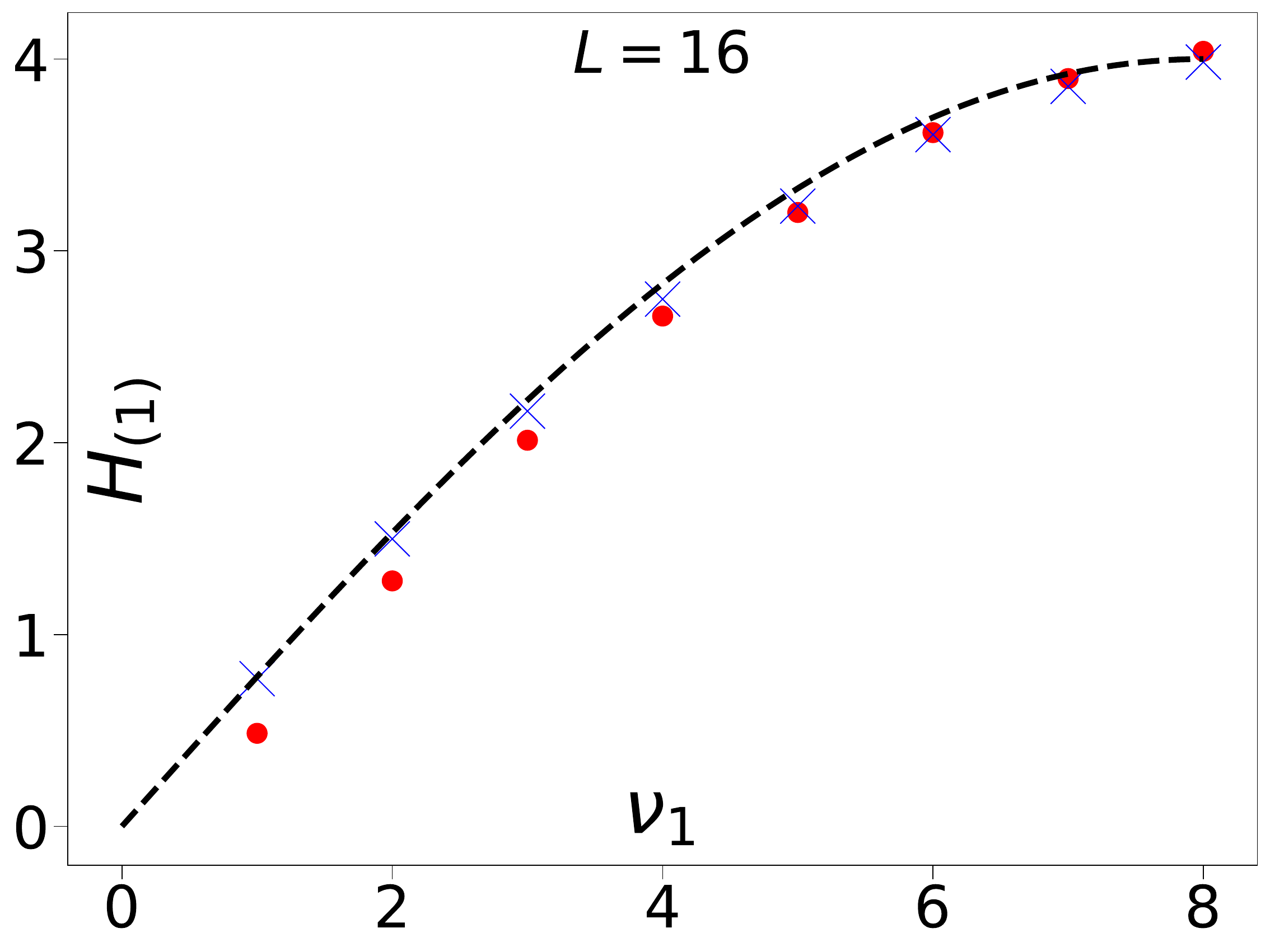}\\
         \includegraphics[width=0.49\textwidth]{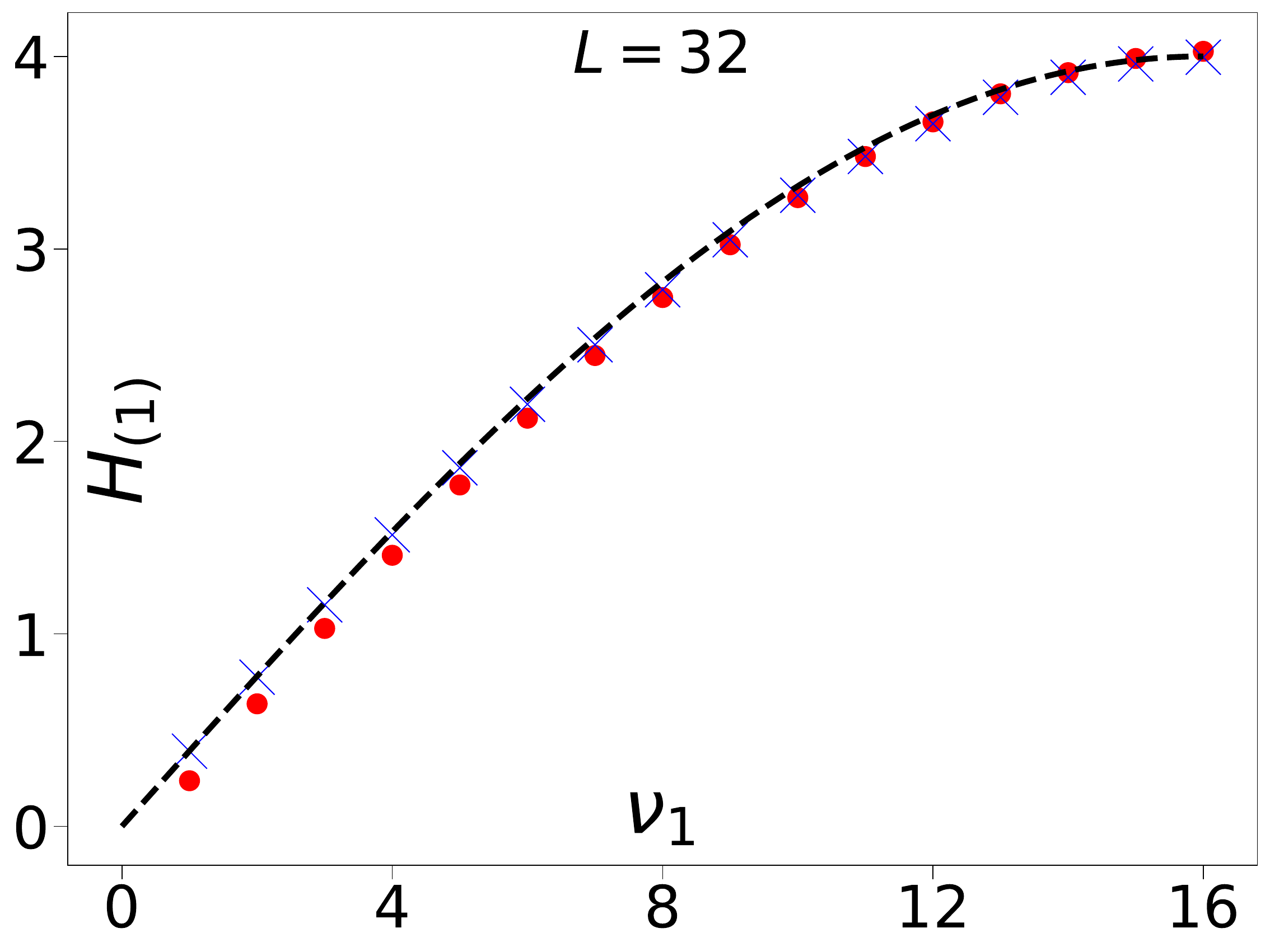}~%
         \includegraphics[width=0.49\textwidth]{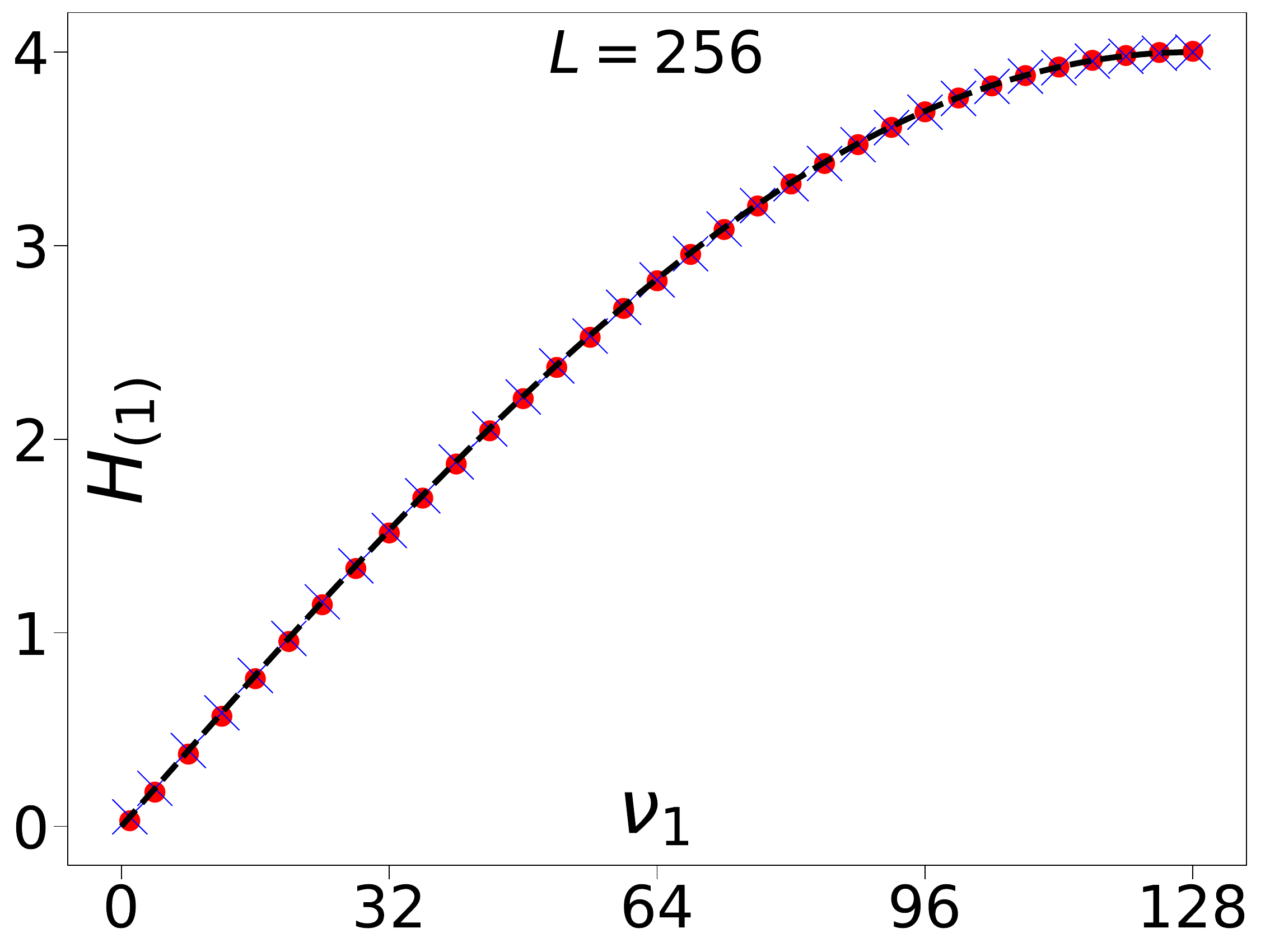}
        \caption{Anomalous dimensions $H_{(1)}$ at order $\order{h}$ in the string tension. The plots are for different values of L, namely $L=4,16,32,256$. In each plot are represented the exact energy from~\eqref{weak exact energy}, the asymptotic energy from the Bethe-Yang equations, and the energy for a free model, see~\eqref{eq:freeapprox}.}
        \label{fig:plotsMequals1}
\end{figure}

\begin{figure}
     \centering
\includegraphics[width=0.6\textwidth]{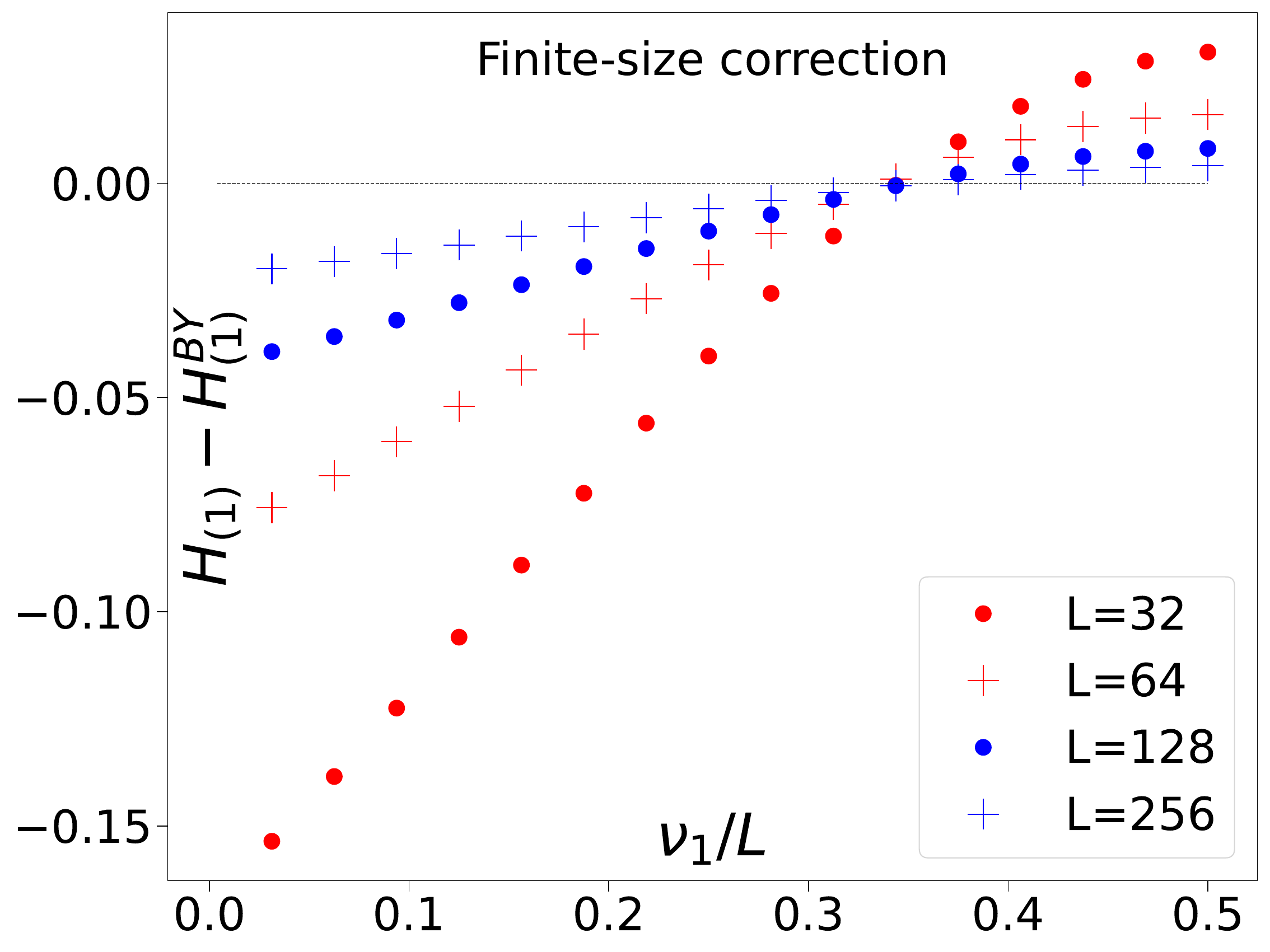}
         \caption{Deviation of the asymptotic energy $H_{(1)}^{BY}$ from the exact energy $H_{(1)}$ for different values of the world-sheet length $L$. The deviation goes like $1/L$.}
        \label{fig:scaling}
\end{figure}

\paragraph{Two excitations.}
In the case of two excitations (\textit{i.e.}~$M=1$) we have to fix the worldsheet volume~$L$, which is quantised, and then solve the excited-state TBA equations for excitations of mode number $(-\nu,\nu)$, with $\nu=0,1,2,\dots \lfloor L/2\rfloor$. The case of $\nu=0$ is special, because in that case the Bethe roots sit an infinity and the TBA equations are singular, corresponding to a BPS state~\cite{Baggio:2017kza}. All other excitation numbers give well-defined equations which can be readily solved numerically. We find that, after a small number of iterations, the results stabilise, see figure~\ref{fig:exactBethe}. The $Y$ functions also converge and take a form similar to those of figure~\ref{fig:Yfunctions}.
It is worth stressing that we always find that $Y_0(\gamma)>-1$, so that the energy formula and the main TBA equation are well-defined --- there are no imaginary terms coming from~$\log(1+Y)$; similarly, $Y(\gamma)<1$ so that the convolutions involving $\log(1-Y_0)$ are also well-defined. This also justifies a posteriori the fact the we interchangeably wrote $N_0 \log(1-Y)$ or $\log(1-Y)^{N_0}$.
 The final result of the evaluation is the leading order correction to the energy of the state, meaning that the energy of a state~$|\Psi^{(L,\nu)}\rangle$ identified by~$L$ and $\nu$, is
 \begin{equation}
     \left(\mathbf{L}_0+\bar{\mathbf{L}}_0\right)|\Psi^{(L,n)}\rangle = H^{(L,\nu)}\,|\Psi^{(L,n)}\rangle,\qquad
     H^{(L,\nu)}=L+ H_{(1)}^{(L,n)}\,h+\mathcal{O}(h^2)\,,
 \end{equation}
where $\mathbf{L}_0$ and $\bar{\mathbf{L}}_0$ are the chiral and antichiral $\mathfrak{sl}(2)$ Cartan elements in the dual CFT, and $h$ is the tension.
In other words, we are after~$H_{(1)}^{(L,n)}$ which is the leading part of the anomalous dimensions and appears at order~$h^1$.
In figure~\ref{fig:plotsMequals1} we plot the energies as a function of $n$ for various values of~$L$, namely $L=4,16, 32, 256$. As it can be expected, the exact value of the energies gets closer to the asymptotic value (\textit{i.e.}, the value predicted by the Bethe-Yang equations) as $L\to\infty$.
As it can be seen in figure~\ref{fig:scaling}, the deviation from the asymptotic prediction goes like~$1/L$. This is expected for a system with gapless excitations in the spectrum.
It is also interesting to note that the energy, already for relatively modest values of~$L$, such as~$L\gtrsim16$, is well approximated by the black dashed line, corresponding to a free theory with
\begin{equation}
\label{eq:freeapprox}
    p_1=-p_2=\frac{2\pi\,\nu}{L}\,,\qquad
    H_{(1)}^{(L,\nu)}=\sum_{j=1}^2  \big| 2 \sin(\tfrac{1}{2}p_j) \big|=4\sin\left(\frac{\pi \nu}{L}\right)\,,
\end{equation}
with $\nu=0,1,2,\dots \lfloor\frac{L}{2}\rfloor$. That is, with good approximation the system behaves similarly to weakly-interacting massless magnons.
In figure 4, we see that the difference between the exact solution and the asymptotic one depends on the mode number~$\nu_1$ and it has a change of sign around $\nu_1/L\approx 0.331$ (note however that this value is $L$-dependent). Technically, this can be understood by looking at the \textit{exact Bethe equations} \eqref{weak exact Bethe eqs} and the form of the Y-functions in Figure~2.
In \eqref{weak exact Bethe eqs}, the corrections to the Bethe-Yang equation are given by the convolutions involving the Y-functions. In particular, one can check from~\eqref{weak exact Bethe eqs} that the exact Bethe root~$p_1$ is always smaller than the momentum~$p_1^{\text{BY}}$ of the asymptotic Bethe root (equivalently, $\gamma_1>\gamma_1^{(\text{BY})}$).
First, note that in the region where $\nu/L$ is small, the contribution of the exact roots to the energy is $2|\sin(p/2)|\sim|p|$, and a discrepancy in $p_1-p_1^{\text{BY}}$ has a big (linear) effect on the energy, which is sufficient to make $H_{(1)}-H_{(1)}^{\text{BY}}<0$. As we increase mode number, both $p_1$ and $p_1^{\text{BY}}$ get larger with $\nu_1/L$, and eventually we get to a regime where $p_1\lesssim \pi$. In this region, $|\sin(p/2)|$ is flatter and small deviations in $p_1-p_1^{\text{BY}}$ affect the energy less and less.
Secondly,  adding to this effect is the fact that by increasing $\nu/L$, the deviation $|p_1-p_1^{BY}|$ gets smaller.
That deviation is given by the convolutions in \eqref{weak exact Bethe eqs}. As $\nu/L$ gets larger, the Bethe root $\gamma_1$ gets closer to zero; when this happens, the integrand is more and  more well approximated by an odd function (it is exactly odd for $\gamma_1=0$). Thus, the integral gets smaller and smaller.
Finally, as we increase the mode, the integral term in \eqref{weak exact energy} becomes more and more important. Looking at figure~\ref{fig:Yfunctions}, we see that $Y_0(\gamma)$ is approximately constant and negative in a region $|\gamma|\gtrsim \gamma_1$. As the mode number grows, $p_1$ increases, and $\gamma_1$ decreases; hence the contribution of $-\log(1+Y_0)$ to the energy convolution is larger and larger (and positive). Hence, for $\nu_1/L$ sufficiently close to~$1/2$, the convolution term dominates, and $H_{(1)}-H_{(1)}^{\text{BY}}>0$. The point where $H_{(1)}-H_{(1)}^{\text{BY}}$ changes sign due to the balancing of these effects does not seem to correspond to a physical mode number. It would be interesting to understand whether this behaviour is fundamentally tied to any underlying physics of the model, or it has no deeper meaning.

\paragraph{Four excitations.}
\begin{figure}
     \centering
        \includegraphics[width=\textwidth]{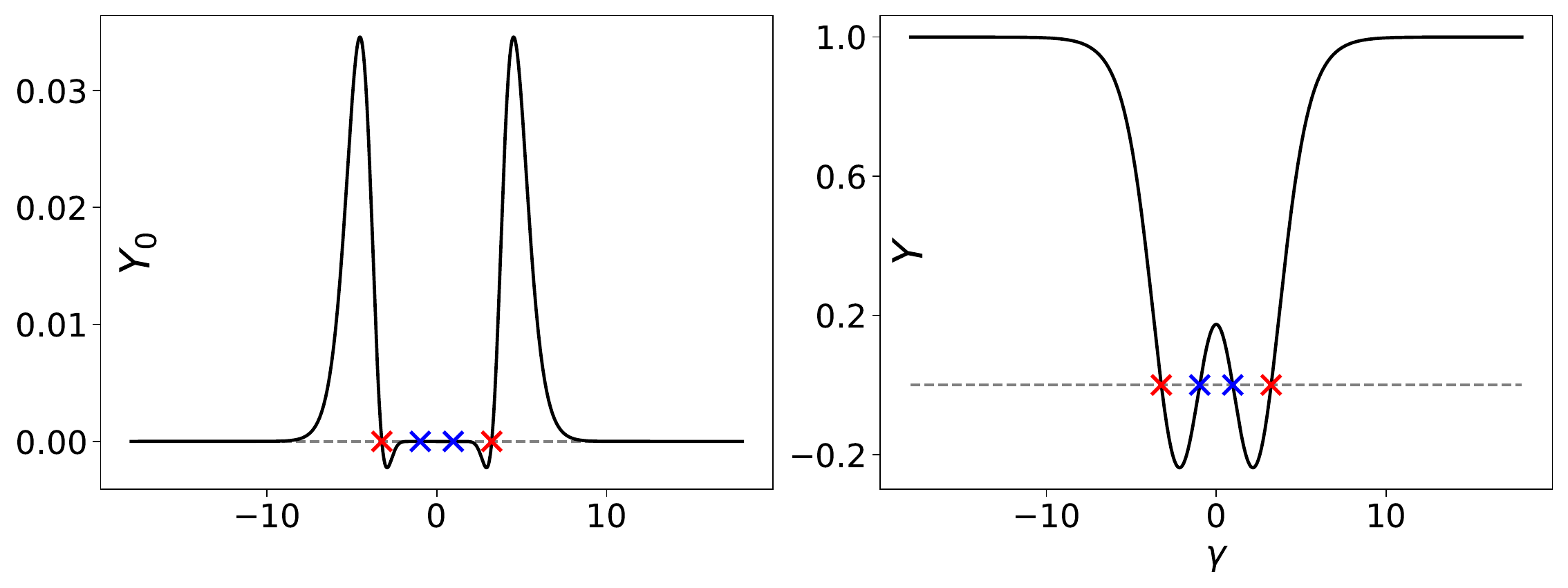}
        \caption{Y~functions for a state with $L=16$ and mode numbers~$\nu_1=1$ and~$\nu_3=4$. On the left $Y_0(\gamma)$ and on the right $Y(\gamma)$. The crosses indicate the positions of the 4 Bethe roots, where both the Y~functions are equal to zero and both change sign. The red ones are associated to the modes $(-\nu_1,\nu_1)$ and the blue ones to the modes $(-\nu_3,\nu_3)$.  In $Y_0$ the changes of sign due to the bigger Bethe roots (marked by the red crosses) are distinguishable, while the other two (marked by the blue crosses) are not, since again $Y_0(\gamma)$ rapidly converges to zero as $\gamma\to 0$. Again, the large $\gamma$ behaviour is the expected one and they quickly converge to their asymptotic values as stated in Table \ref{table:AsymYfunc}. The plots shows the region $\abs{\gamma}<18$, but again the cutoff has been set to $\Lambda=40$.}
        \label{fig:Mis2Yfunctions}
\end{figure}
Let us now turn to the case of $M=2$, \textit{i.e.}~of four excitations coming in pairs of opposite momentum. Let the mode numbers be~$(\nu_1,-\nu_1,\nu_3,-\nu_3)$ and let us take $\nu_1$ and $\nu_3$ to be non-negative integers. A special case is that of $\nu_1=\nu_3$; this is an allowed state, as long as the excitations carry distinct~$\su(2)_\circ$ quantum numbers.
Let us first discuss the case of generic (different) quantum numbers. In this case, the Y~functions go to their vacuum values as~$\gamma\to\pm\infty$, see figure~\ref{fig:Mis2Yfunctions}.  It still remains true that~$Y_0(\gamma)>-1$ and $Y(\gamma)<1$ for any finite~$\gamma$ on the real mirror line, so that the TBA equations are well-defined.
\begin{figure}
     \centering
         \includegraphics[width=0.49\textwidth]{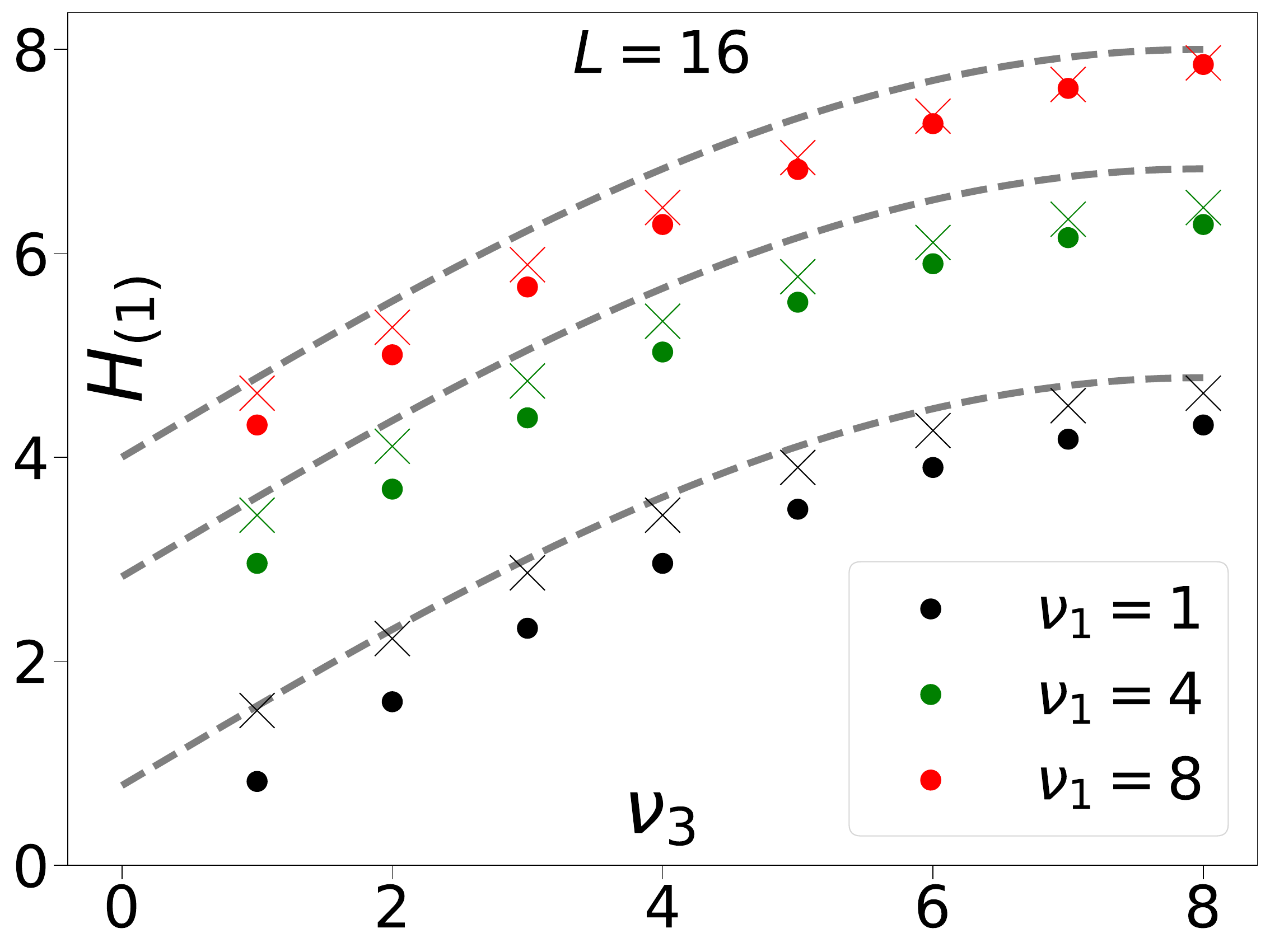}~%
         \includegraphics[width=0.49\textwidth]{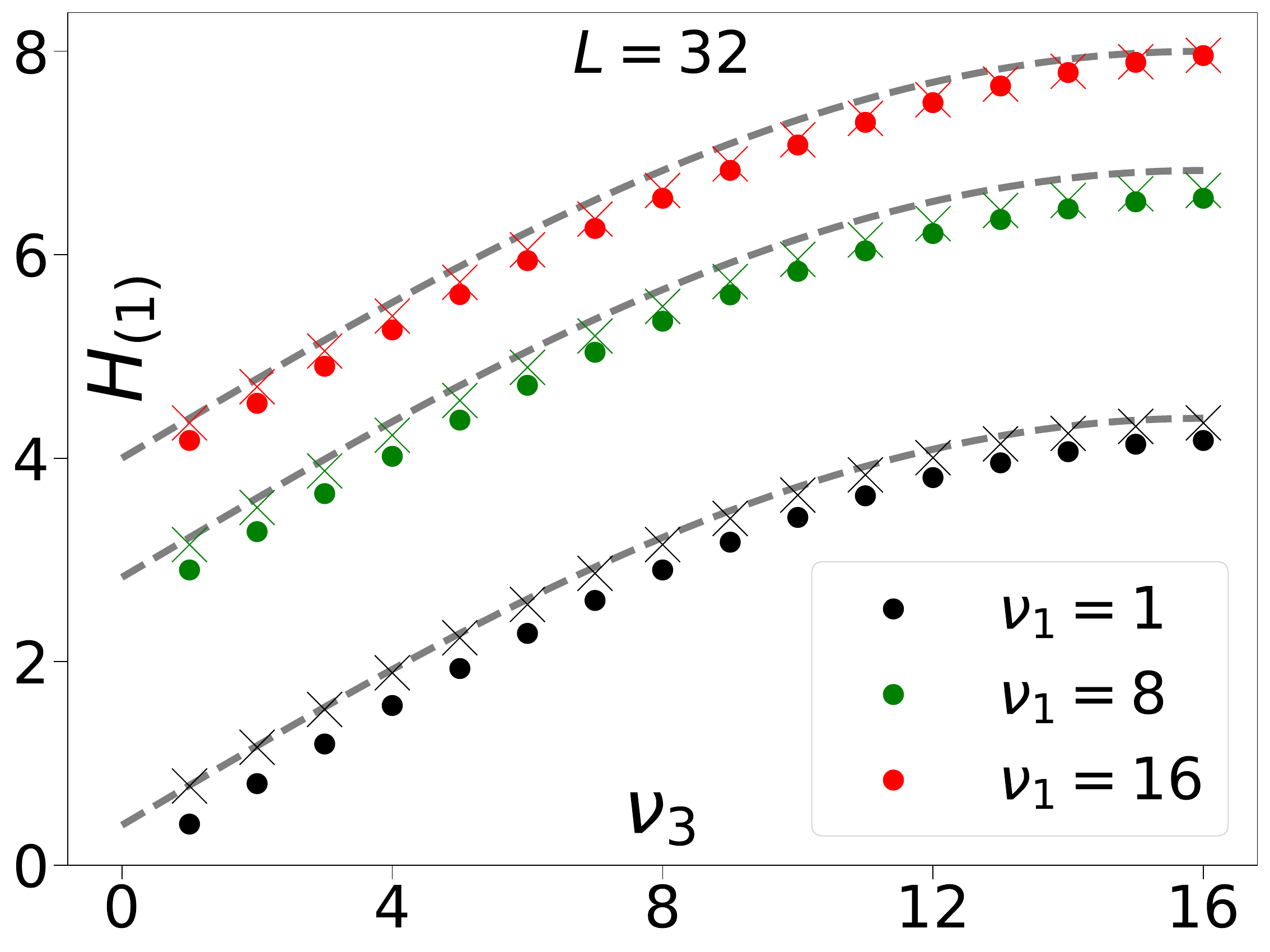}
        \caption{Anomalous dimensions $H_{(1)}$ at order $\order{h}$ in the string tension for $L=16$ and~$L=32$. In each plot are represented the energies as function of the mode $\nu_3$ for fixed values of $\nu_1$. As in the previous energy plots in figure~\ref{fig:plotsMequals1}, the dots mark the exact energy from~\eqref{weak exact energy}, the crosses are for the asymptotic energy from the Bethe-Yang equations, and the dashed lines for the energy for a free model, see~\eqref{eq:freeapprox}.}
        \label{fig:Mis2energies}
\end{figure}
For each given~$L$, the energy can be computed as a function of~$\nu_3$ for fixed~$\nu_1$. Some of the resulting curves are plotted in figure~\ref{fig:Mis2energies} for $L=16$ and~$L=32$. Once again, we find that the deviation is relatively small with respect to the asymptotic, or even free, result.
Let us now turn to the case of $\nu_1=\nu_3$. Here, already asymptotically, we see that there are two possible solutions: one with the Bethe roots $\gamma_{1}=\gamma_3$ and the other with~$\gamma_1\neq\gamma_3$. Interestingly, it is the solution with~$\gamma_1\neq\gamma_3$ which fits the trajectories of energies in figure~\ref{fig:Mis2energies}; the other solution would look like an outlier.%
\footnote{It should also be stressed that strictly speaking our TBA equations have been derived under the assumptions that all roots are distinct.} 
It is not clear what would be the interpretation of the configuration with identical roots.

\clearpage

\paragraph{Other values of $N_0$.}
\begin{figure}
     \centering
\includegraphics[width=0.49\textwidth]{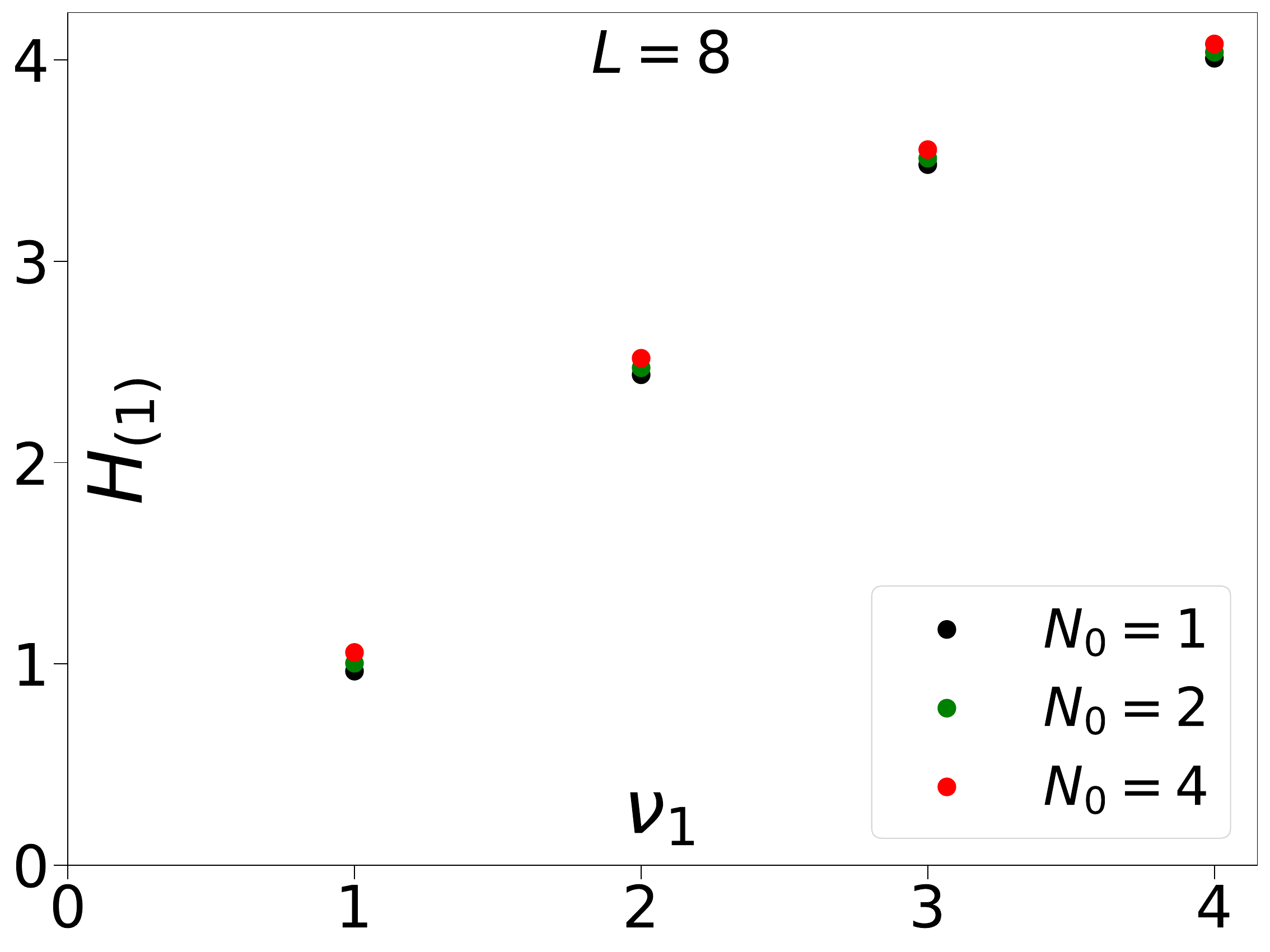}~
\includegraphics[width=0.49\textwidth]{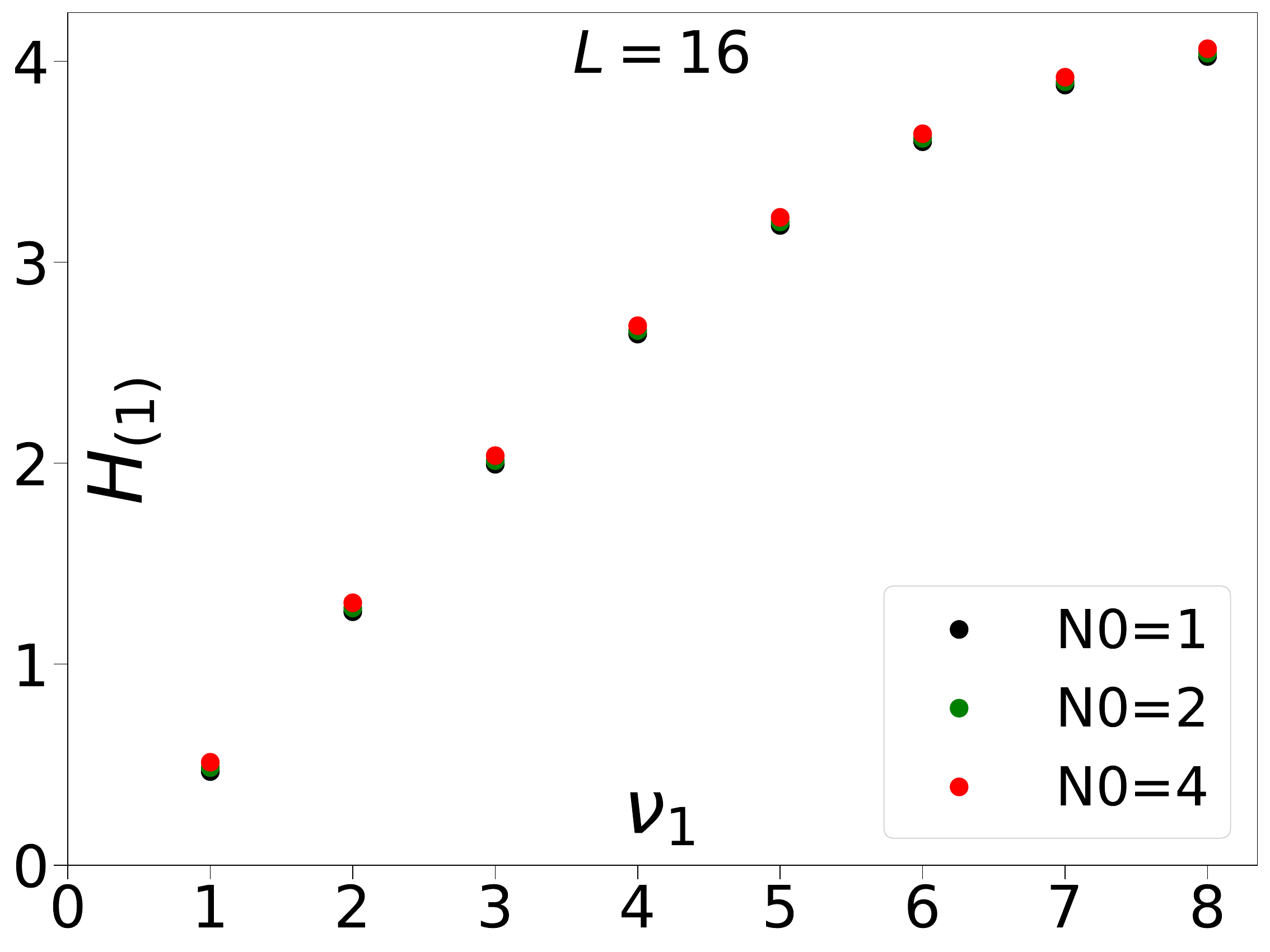}
         \caption{Comparison of the anomalous dimensions of two-particle states ($M=1$) for different values of $N_0=1,2,4$. For definiteness, we consider $L=8$ and~$L=16$. For states with larger $L$, the difference becomes even smaller.}
        \label{fig:N0dependence}
\end{figure}
In~\cite{Frolov:2023wji}, the mirror TBA for a twisted vacuum was discussed, and the ground state energies were computed. This is done by taking the volume~$L\to\infty$ while keeping the tension fixed. The computation can be then formally compared with a semiclassical one, where $L\to\infty$ and~$h\to\infty$ with~$\mathcal{J}=L/h$ fixed.
In both approaches, it is easy to keep track of the contribution of gapless and gapped (mirror) particles. Indeed, at large~$\mathcal{J}$ the gapless particles contribute at order~$1/\mathcal{J}$ while the gapped ones at order~$e^{-\mathcal{J}}/\sqrt{\mathcal{J}}$.
For the gapped particles, the precise form of the contribution seems to match with the semiclassical prediction (up to identifying the twist in a specific way). For gapless ones, it does not, unless one assumes~$N_0=1$ in the mirror TBA --- in which case it does.
While a discrepancy might be explained by the different ways in which the limit is taken in the mirror TBA and in the semiclassical computation, it is highly suggestive that the results do match as far as the massive contribution goes (and they do match in $\text{AdS}_5\times \text{S}^5$~\cite{Frolov:2009in,Frolov:2023wji}). While the physical interpretation of putting $N_0=1$ is unclear, it is worth exploring the effect of such a choice (or on  setting~$N_0\neq2$ in general) on the spectrum.
What we find is that nothing special seems to happen. Indeed, the numerical value of the energies changes in a mild way as we change~$N_0$, see figure~\ref{fig:N0dependence}.\footnote{Roughly speaking, the energy $H_{(1)}$ changes by the order of $\sim 1 \%$ by changing $N_0=2$ to $N_0=1,4$, as can be found by the numerical table in Appendix \ref{app:numerical}.} Unfortunately, this does not suggest how to resolve the puzzle of~\cite{Frolov:2023wji}, as no choice of $N_0$ appears pathological or particularly ``nice''.
The only way to obtain a sure answer would be a direct quantitative comparison with predictions either from string theory, which would necessitate extending this analysis to the semiclassical regime.

\section{Conclusions and outlook}
\label{sec:conclusions}

We derived the weak-tension expressions for the mirror TBA for pure-RR $\AdSST$ with massless particle excitations and solved the corresponding equations numerically. We find that the leading order correction to the energy comes from the massless sector at order~$\mathcal{O}(T)$ in the tension~$T\ll1$, while the massive sector is suppressed to~$\mathcal{O}(T^{2L})$, where~$L$ is the R-charge of a reference vacuum (\textit{i.e.}, a reference BPS state). This is strikingly different from $\text{AdS}_5\times \text{S}^5$, where there is no massless sector at all.
It was natural to expect that the dynamics at~$\mathcal{O}(T)$ should be substantially simpler than the full worldsheet dynamics at arbitrary tension. What was unclear was whether such the weak-tension physics could be understood as a nearest-neighbour spin chain (like in $\text{AdS}_5\times \text{S}^5$) or as a symmetric orbifold CFT of a free model (like in the case of the tensionless limit of pure-NSNS backgrounds).
By solving numerically the equations we determine that it is neither. The fact that a nearest-neighbour description would be too simple a dynamics was already strongly implied by the presence of gapless (\textit{i.e.}, long range) excitations.

At weak tension, the model is given by a system of TBA equations of difference-form, with a nonrelativistic energy and momentum. From the numerical solution of the spectrum, we find that it represents a system of weakly-interacting massless magnons,%
\footnote{%
By ``weakly-interacting'' we mean that, even for relatively small values of the volume, the energies are relatively well-approximated by those of a free model of magnons. However, the model is not free: for a free model we expect that the energy of multi-particle states is the sum of energies of single-particle ones \textit{with the same allowed momenta}, leading to huge degeneracies in the spectrum which we do not observe here.
}
with energy $\mathcal{E}(p)\sim |\sin(p/2)|$. The deviation of the exact energies from the asymptotic model is of order~$1/L$ --- as it is the of deviation of the asymptotic result from the free one.
It would be interesting to see if one may reverse-engineer a scattering phase~$\varphi_{\text{eff}}$ so that the exact energies are given by
\begin{equation}
    e^{i p_j L}\prod_{k\neq j}e^{i \varphi_{\text{eff}}(p_j,p_k)}=1,\qquad\mathcal{E}=\sum_j\mathcal{E}(p_j)\,.
\end{equation}
Of course this is generally impossible for a system of TBA equations, but perhaps at leading order the dynamics of the model is so simple to allow for such a simplification. A brief exploration of this idea however did not result in a perfect match of the energies when using a few known scattering phases. It is also interesting to explore whether this spectrum may correspond to any known quantum-mechanical integrable model with long-range interactions.

We have also explored the question of how many species of massless particles should be included in the TBA. From perturbation theory we expect to have $N_0=2$ types of excitations, but in recent work it was observed~\cite{Frolov:2023wji} that setting $N_0=1$ seem to better account for the energy of a twisted ground state. We find that, for excited states at small tension, any value of~$N_0$ yields an apparently reasonable solution, and in fact the deviations between $N_0=1$, $N_0=2$, and (for the sake of generality) $N_0=4$ are numerically small. Hence, this does not seem to resolve the confusion on this point.
One way to resolve this puzzle might be to derive the mirror TBA equations for the ${\rm AdS}_3 \times {\rm S}^3 \times {\rm S}^3 \times {\rm S}^1$ background \cite{Tong:2014yna,Borsato:2015mma}, and study their $\text{T}^4$ limit.

Aside from the TBA, there is another proposal for a system of equations describing the spectrum: the quantum spectral curve~\cite{Ekhammar:2021pys,Cavaglia:2021eqr,Cavaglia:2022xld}. It would be interesting to use those equations to extract the small-tension spectrum and compare with our results. Unfortunately, it is currently unclear if and how the QSC equations account for states including massless excitations, which are the ones with the largest anomalous dimension in this regime.
Hence, our results cannot be compared with that framework, at least until it is possible to adapt the QSC to describe states involving massless excitations.

\subsubsection*{Acknowledgements}

We thank Sergey Frolov, Alessandro Torrielli, Arkady Tseytlin, and Linus Wulff for useful related discussions.
We are particularly grateful to Stefano Scopa for discussions and help with the Python code for the numerical solution of the TBA equations.
DlP~acknowledges support from the Stiftung der Deutschen Wirtschaft.
The work of RS is supported by NSFC grant no.~12050410255.
AS thanks the participants of the workshop ``Integrability in Low-Supersymmetry Theories'' in Filicudi, Italy, for a stimulating environment where part of this work was carried out.
AS acknowledges support from the European Union – NextGenerationEU, and from the program STARS@UNIPD, under project ``Exact-Holography – A new exact approach to holography: harnessing the power of string theory, conformal field theory, and integrable models.''

\appendix
\section{Notation}\label{app:notation}

Here we summarise our notation, various S-matrix elements and kernels.

\subsection{Rapidity variables}

We will use three types of variables to express the S-matrices: $x$-variable, $u$-rapidity and $\gamma$-variable. When working with the $u$-rapidity, we introduce the notation$f^\pm (u) \equiv f ( u \pm i|m|/h )$, where the value of $|m|=Q=\bar{Q}$ is the mass of the particle.

\paragraph{Mirror momentum-carrying particles.}
The map from Zhukovsky variables to rapidity variables in the mirror region is
\begin{equation}
x^+ ( \gamma^+) = \frac{1 + e^{ \gamma^+}}{1 - e^{ \gamma^+}} \,, \qquad 
x^- ( \gamma^-) = \frac{1 - e^{ \gamma^-}}{1+ e^{ \gamma^-}} \,, \qquad
x ( \gamma) = -\tanh\frac{\gamma}{2} \,.
\label{def:tilde-gamma-mirror}
\end{equation}
Here $x^\pm$ and $\gamma^\pm$ refer to massive particles, and for real mirror particles obey the reality constraint
\begin{equation}
    (x^\pm)^*=\frac{1}{x^\mp}\,,\qquad
    (\gamma^\pm)^{*}=\gamma^\mp\,,
\end{equation}
while $\gamma$ is real for real-momentum mirror particles.
The same formulae can be found by using the $u$-paramatrisation
\begin{equation}
\label{eq:ugammarelation}
    u(\gamma)= -2\coth\gamma\,,\qquad
    \gamma(u)=\frac{1}{2}\log\frac{2-u}{2+u}+\frac{i\pi}{2}
\end{equation}
and recalling that
\begin{equation}
    x(u)=\frac{1}{2}\left(u-i\sqrt{4-u^2}\right)\,,
\end{equation}
which confirms that indeed for $\gamma$ just above the real line
\begin{equation}
    x(\gamma):=x(u(\gamma+i0))\,.
\end{equation}
Vice versa we have
\begin{equation}
\gamma(x)=-2\,\text{atanh}\, x\,,\qquad
\gamma(u):=\gamma(x(u))\,.
\end{equation}
We can use $\gamma(u)$ to define
\begin{equation}
\begin{aligned}
\gamma^+ (u) &= \gamma(u+\tfrac{i}{h}|m|)-i \pi ,
&\quad x^+ (u) &= x^+ (\gamma^+ (u)) ,
\\
\gamma^- (u) &= \gamma(u-\tfrac{i}{h}|m|) ,
&\quad x^- (u) &= x^- (\gamma^- (u)) ,
\end{aligned}
\label{def:gamma-u mirror}
\end{equation}
which is compatible with our mirror reality.
The relation between $\gamma^\pm$ and $x^\pm$ is then
\begin{equation}
    \gamma^\pm(x^\pm) = \log\left(i\frac{x^\pm-1}{x^\pm+1}\right)\mp i\frac{\pi}{2}\,.
\end{equation}
Finally we recall that both $\gamma^\pm$ and $x^\pm$ are not sets of independent variables, because
\begin{equation}
    \frac{h}{i}\left(\coth\gamma^- - \coth\gamma^+\right)=
    \frac{h}{2i}\left(x^++\frac{1}{x^+}-x^+-\frac{1}{x^+}\right)=|m|\,.
\end{equation}

\paragraph{Mirror branch cuts and massless physical region.}
It is important to note that massless mirror particles, are defined for
\begin{equation}
\label{eq:mirrorphysical}
    x\in(-1,+1),\quad
    \gamma\in(-\infty+i0,+\infty+i0)\,,\quad
    u\in(-\infty+i0,-2+i0)\cup(+2+i0,+\infty+i0)\,.
\end{equation}
In other words, the region in the $u$-plane is \textit{just above} the long cut. The mirror momentum and energy for massless particles are given by
\begin{equation}
    \tilde{p}_0(\gamma)=-\frac{2h}{\sinh\gamma}\,,\qquad\tilde{\mathcal{E}}_0(\gamma)=-\log\left(\frac{1-e^\gamma}{1+e^\gamma}\right)^2\,.
\end{equation}
Note that the second formula is not analytic along the imaginary axis of the $\gamma$-plane (much like $\log z^2$). It is sometimes useful to treat separately the positive and negative momentum regions. For this purpose note that $\tilde{p}_0>0$ corresponds to
\begin{equation}
    \tilde{p}_0>0 \, : \qquad
    x\in(0,+1),\qquad
    \gamma\in(-\infty+i0,0+i0)\,,\qquad
    u\in(+2+i0,+\infty+i0)\,.
\end{equation}

\paragraph{String momentum-carrying particles.}
The kinematics of the string region can be obtained by analytic continuation but for us it is most useful to use distinct $\gamma$-parametrisation. We indicate string-kinematics expressions by a subscript ``$s$''. We have
\begin{equation}
    x^+_s(\gamma^+_s) = \frac{i-e^{\gamma^+_s}}{i+e^{\gamma^+_s}}\,,\qquad
    x^-_s(\gamma^+_s) = \frac{i+e^{\gamma^-_s}}{i-e^{\gamma^-_s}}\,,\qquad
    x_s(\gamma_s) = \frac{i-e^{\gamma_s}}{i+e^{\gamma_s}}\,.
\label{def:xspm gamma_s}
\end{equation}  
For massive particles the reality condition is
\begin{equation}
    (x^+_s)^*=x^-_s\,,\qquad
    (\gamma^+_s)^*=\gamma^-_s\,,
\end{equation}
whereas $\gamma_s$ is once again real for real (string) particles.
In terms of the  $u$-rapidity we have
\begin{equation}
    u=-2\tanh\gamma_s\,,\qquad
    \gamma_s(u)=\frac{1}{2}\log\frac{u-2}{u+2}-i\frac{\pi}{2}\,.
\end{equation}
In contrast to~\eqref{eq:ugammarelation}, now the cuts on the $u$-plane are short (they run from $-2$ to $+2$), as it is the case in the formula for
\begin{equation}
    x_s(u)=\frac{u}{2}\left(1+\sqrt{1-\frac{4}{u^2}}\right)\,.
\end{equation}
Indeed it can be checked that
\begin{equation}
    x_s(\gamma_s):=x_s(u(\gamma_s-i0))\,,
\end{equation}
with $\gamma_s$ just below the real line so that $u$ is just above the short cut and $x_s$ is on the upper half-circle.
We can also define
\begin{equation}
    \gamma_s^-(u)=\gamma_s(u-\tfrac{i}{h}|m|)-i\pi\,,\qquad
    \gamma_s^+(u)=\gamma_s(u+\tfrac{i}{h}|m|).
\end{equation}
In terms of $x^\pm$ we have
\begin{equation}
    \gamma^\pm_s(u)=\log\left(\mp i\frac{x^\pm - 1}{x^\pm+1}\right)\,.
\end{equation}
Finally, we have the constraint
\begin{equation}
    \frac{h}{i}\left(\tanh\gamma^-_s - \tanh\gamma^+_s)\right)=
    \frac{h}{2i}\left(x^+_s+\frac{1}{x^+_s}-x^+_s-\frac{1}{x^+_s}\right)=|m|\,.
\end{equation}

\paragraph{String branch cuts and massless physical region.}
Also in this region, string massless particles live on the $u$-plane branch cut, which is now short. Real-momentum particles satisfy
\begin{equation}
\label{eq:stringphysical}
    x_s\in\mathbb{S}^1_+,\quad
    \gamma_s\in(-\infty-i0,+\infty-i0)\,,\quad
    u\in(-2+i0,+2+i0)\,,
\end{equation}
where $\mathbb{S}^1_+$ is upper-half circle. 
The string energy and momentum are
\begin{equation}
    \mathcal{E}_0(\gamma_s)=\frac{2h}{\cosh\gamma_s}\,,\qquad
    p_0(\gamma_s)=-i\log
    \left(\frac{i-e^{\gamma_s}}{i+e^{\gamma_s}}\right)^2,
\end{equation}
where the latter is not analytic across the imaginary axis.

\paragraph{Auxiliary (mirror) particles.}
Finally, we have the auxiliary $y^+$ and $y^-$ particles. In the mirror region, they lie on the upper and lower half-circle respectively. As such, we can parametrise them using the same variables as massless string particles.
\begin{itemize}
    \item \underline{$y^+$ particles}
    \begin{equation}
       y(u)= x_s(u+i0)=x_s(u(\gamma_s-i0)), \qquad -2\le u \le \,,\quad \gamma_s\in\mathbb{R}\,,
    \end{equation}
        \item \underline{$y^-$ particles}
    \begin{equation}
       y(u)= \frac{1}{x_s(u+i0)}=\frac{1}{x_s(u(\gamma_s-i0))}, \qquad -2<u<2\,,\quad \gamma_s\in\mathbb{R}\,.
    \end{equation}
\end{itemize}
Note that $x_s(u+i0) = 1/x(u)$ for $-2 \le u \le 2$.

\paragraph{Shift-identity.}
While we will generally think of the string and mirror variables $\gamma$ and $\gamma_s$ as independent, there is a simple relation that allows us to go from one to the other, namely
\begin{equation}
    \gamma_s=\gamma - \frac{i\pi}{2}\,.
\label{analytic continuation to string gamma}
\end{equation}
This can be useful in finding various kernels and S~matrices. In particular, note that
\begin{equation}
    i\,\mathcal{E}_0(\gamma - \tfrac{i\pi}{2})=\tilde{p}_0(\gamma)\,,\qquad
    i\,p_0(\gamma - \tfrac{i\pi}{2})=\tilde{\mathcal{E}}_0(\gamma)\,.
\end{equation}

\paragraph{S-matrices.}
In view of the above, it is natural to define the  following notation. Given the S-matrix $S (x,x')$ which depends on two massive variables, we rewrite it in terms of the $u$- and $\gamma$-variable as
\begin{equation}
S(u, u') = S^{AB}(x(u), x(u')), \qquad
S(\gamma, \gamma') = S^{AB}(x(\gamma), x(\gamma')).
\label{def:S-matrix u-gamma}
\end{equation}
Here the $u$ and $\gamma$ variables are in the mirror physical region~\eqref{eq:mirrorphysical}.
For massive particles we always use the standard $u$-parametrisation to define the S matrix and kernels, that is \textit{e.g.}
\begin{equation}
S^{AB}(u, u') = S^{AB}(x^\pm(u), x^\pm(u')).
\end{equation}
For the scattering of auxiliary particles in the mirror region we use
\begin{equation}
S^{Ay} (u, u') = S^{AB}(x(u), x_s(u'+i0)), \qquad
S^{Ay}(\gamma, \gamma') = S^{AB}(x(\gamma), x_s(\gamma'-i0))
\end{equation}
for {\it both} $y^+$ and $y^-$ particles, as exemplified in \eqref{def:S0y Sy0 uu'}. The mirror auxiliary particle lives is parametrised in terms of the physical region as the massless string particle, that is~\eqref{eq:stringphysical}.

\subsection{Kernels in \texorpdfstring{$\gamma$}{gamma}-rapidity parametrisation}

We define the kernels as
\begin{equation}
\label{eq:calligraphickernel}
    K^{AB} (u, u') = \frac{1}{2\pi i} \frac{\de}{\de u} \log S^{AB} (u, u')\,,\qquad
    \mathcal{K}^{AB} (\gamma, \gamma') = \frac{1}{2\pi i} \frac{\de}{\de \gamma} \log S^{AB} (\gamma, \gamma')\,,
\end{equation}
so that 
\begin{equation}
    K^{AB}(u,u')=\frac{\de\gamma}{\de u}\,\mathcal{K}^{AB} (\gamma(u), \gamma(u'))\,.
\end{equation}
With these kernels, define the left-convolution
\begin{equation}
    (f*\mathcal{K})(\gamma')= \int\limits_{-\infty}^{+\infty}\de\gamma \, f(\gamma)\,\mathcal{K}(\gamma,\gamma'),,
\end{equation}
which we will use to rewrite the TBA equations. When a kernel is of difference type, this means
\begin{equation}
    (f*\mathcal{K})(\gamma')= \int\limits_{-\infty}^{+\infty}\de\gamma f(\gamma)\,\mathcal{K}(\gamma-\gamma')\,.
\end{equation}
We also note that the sign may change,
\begin{equation}
\label{eq:convolutionchangevar}
\begin{aligned}
\( f \, \hat{\star} \, K^{yA} \) (v)
&= \int \limits_{-2}^{+2} \de u\, f (u)\,K^{yA}(u,v)
\\
&= - \int \limits_{-\infty}^{+\infty} \de \gamma \, f ( u(\gamma) ) \,\frac{\de u}{\de\gamma} \, K^{yA}(u (\gamma),v(\gamma'))
= - \( f * \mathcal{K}^{yA} \) (\gamma')\,.
\end{aligned}
\end{equation}
This is because $x_s (\gamma)$ parameterises the upper half plane counter-clockwise, while $x_s(u+i0)$ parameterises it clockwise.
This is not the case for the integration over $(-\infty,-2) \cup (2,\infty)$, where $(\de \gamma/\de u)$ remains positive.
We have schematically
\begin{equation}
    f \, \hat{\star} \, g \to -f*g,\qquad
    f \, \check{\star} \, g \to +f*g.
\end{equation}

\subsection{List of S~matrices}\label{app:S-mat list}

We basically use the same notation as Appendix B in \cite{Frolov:2021bwp}. 
Whenever the $u$-rapidity lies on the branch cut, we always take the $u+i0$ prescription.

The standard bound state S~matrix is 
\begin{equation}
    S^{QQ^\prime} (u-u^{\prime}) = 
    S^{Q+Q^\prime} (u-u^{\prime}) S^{Q - Q^\prime} (u-u^{\prime})
    \prod_{j=1}^{Q^\prime -1} S^{Q - Q^\prime + 2j} (u-u^{\prime}),
\end{equation}
where $S^Q$ is the rational S-matrix
\begin{equation}
    S^Q (u, u') = \frac{u - u' - \frac{iQ}{h}}{u - u' + \frac{iQ}{h}}
\label{def:rational S-matrix}
\end{equation}

\paragraph{Left-anything scattering.}

\begin{align}
S_{\textit{sl}}^{Q_a Q_b} (u_a,u_b) &= S^{Q_a Q_b}(u_a-u_b)^{-1} (\Sigma_{ab}^{Q_a Q_b})^{-2} \,,
\\
\tilde{S}_{\textit{sl}}^{Q_a \bar{Q}_b}(u_a,u_b) &= e^{i p_a} \frac{1-\frac{1}{x_a^+ x_b^+}}{1-\frac{1}{x_a^- x_b^-}} \frac{1-\frac{1}{x_a^+ x_b^-}}{1-\frac{1}{x_a^- x_b^+}} (\Tilde{\Sigma}_{ab}^{Q_a \bar{Q}_b})^{-2} \,,
\\
S^{Q_a 0}(u_a,x_j) &= i e^{-\frac{i}{2}p_a} \frac{x_a^+ x_j-1}{x_a^- - x_j} \frac{(\Sigma_{\bes}^{Q_a 0}(x_a^\pm,x_j))^{-2}}{\PhiSG (\gamma_{aj}^{+\circ}) \PhiSG (\gamma_{aj}^{-\circ})} \,,
\label{def:SQa0} \\
S_+^{Q_a y}(u, v) &= 
e^{\tfrac{i}{2}p_a} \, \frac{x^- (u_a) - x (v) }{x^+ (u_a) - x (v) } \,,
\label{def:SQay+} \\
S_-^{Q_a y}(u, v) &= e^{\tfrac{i}{2}p_a} \,
\frac{x^- (u_a) - \frac{1}{ x (v) } }{x^+ (u_a) - \frac{1}{ x (v) } } \,,
\label{def:SQay-} 
\end{align}
where $\PhiSG = e^\phiSG$ is the sine-Gordon factor \eqref{eq:varphiexplicit}, and $\gamma^{\pm \circ}_{aj} =  \gamma^\pm_a - \gamma_j$\,.
The improved BES factors such as $\Sigma^{Q_a Q_b}$ and $\Sigma_{\bes}^{Q_a 0}$ will be discussed in Appendix \ref{app:BES phase definitions}.

\paragraph{Right-anything scattering.}\footnote{The S-matrix $\tilde S^{Q0}$ corresponds to the kernel $\tilde{K}^{Q0}$, which was denoted by $\bar{S}^{Q0}$ in \cite{Frolov:2021bwp}.} 
\begin{align}
S_{\textit{su}}^{\bar{Q}_a \bar{Q}_b} (u_a,u_b) &= e^{i p_a} e^{-i p_b} \left( \frac{x_a^+ - x_b^-}{x_a^- - x_b^+} \right)^{-2} S^{\bar{Q}_a \bar{Q}_b}(u_a - u_b)^{-1} (\Sigma_{ab}^{\bar{Q}_a \bar{Q}_b})^{-2} \,,
\\
\tilde{S}_{\textit{su}}^{\bar{Q}_a Q_b}(u_a,u_b) &= e^{-i p_b} \frac{1-\frac{1}{x_a^- x_b^-}}{1-\frac{1}{x_a^+ x_b^+}} \frac{1-\frac{1}{x_a^+ x_b^-}}{1-\frac{1}{x_a^- x_b^+}} (\Tilde{\Sigma}_{ab}^{\bar{Q}_a Q_b}(u_a,u_b))^{-2} \,,
\\
\tilde{S}^{\bar{Q}_a 0}(u_a,x_j) &= i e^{+\frac{i}{2}p_a} \frac{x_a^- - x_j}{x_a^+ x_j -1} \frac{(\Sigma_{\bes}^{\bar{Q}_a 0}(x_a^\pm,x_j))^{-2}}{\PhiSG (\gamma_{aj}^{+\circ}) \PhiSG (\gamma_{aj}^{-\circ})} \,.
\label{def:barS bQa0} 
\end{align}

\paragraph{Massless-anything scattering.}

\begin{align}
S^{00}(u_j,u_k) &= a(\gamma_{jk}) \PhiSG (\gamma_{jk})^2 (\Sigma_{\bes}^{0 0}(x_j,x_k))^{-2} \,,
\label{def:S00} \\
S^{0 Q_b}(x_j, u_b)  &= \frac{1}{ S^{Q_b0}(u_b,x_j) } \,, 
\\
\tilde{S}^{0 \bar{Q}_b}(x_j,u_b) &= \frac{1}{ \tilde{S}^{\bar{Q}_b0}(u_b,x_j) } \,,
\\
S^{0y}(u, v) &= \frac{1}{\sqrt{x(u+i0)^2}}\frac{x(u+i0)-x_s(v+i0 )}{\frac{1}{x(u+i0)}-x_s(v+i0 )} \,.
\label{def:S0y uu}
\end{align}

\paragraph{Auxiliary-anything scattering.} 

\begin{align}
S^{yQ}_- (v,u_j)
&= e^{+\tfrac{i}{2}p_j} \frac{\frac{1}{x(v)} - x^- (u_j) }{\frac{1}{x(v)} - x^+ (u_j) }
= S^{Q y}_- ( u_j, v ) ,
\label{def:SyQm vu} \\[1mm]
S^{y Q}_+(v ,u_j)
&= e^{-\tfrac{i}{2}p_j} \frac{x(v) - x^+ (u_j) }{x(v) - x^- (u_j) }
= \frac{1}{S^{Q y}_+ ( u_j, v )} \,,
\label{def:SyQp vu} \\[1mm]
S^{y0}(v,u_j)
&= \frac{1}{\sqrt{ x( u_j+i0)^2 }} \, \frac{ x(v) - x (u_j+i0) }{ x(v) - \frac{1}{x(u_j+i0)} } 
= \frac{1}{S^{0y}(u_j, v)} \,.
\label{def:Sy0 vu}
\end{align}
These $S^{yQ}_\mp (v,u)$ are identical to $S^{yQ}_\pm (v,u)$ in \cite{Arutyunov:2009ax}, and the corresponding kernels are positive in the mirror-mirror region.

\paragraph{Analytic continuation.}

The symbol $S^{AB} (u_1, u_2)$ usually denotes an S-matrix in the mirror-mirror region.
When one of the rapidities is analytically continued to the string region, say $u_1$, they are denoted interchangeably by 
\begin{equation}
S^{AB} (u_{*1} , u_2), \quad
S^{A_* B} (u_1, u_2) \quad {\rm or} \quad
S^{A_* B} (u_{*1}, u_2).
\end{equation}

When the S-matrix is written in the $\gamma$-parametrisation, $S^{AB} (\gamma_1, \gamma_2)$, then we perform the analytic continuation as explained in \cite{Frolov:2021fmj}. For massless particles, the string region is $i \pi/2$ above the mirror region. To describe the rapidity in the string region, we use the notation $\gamma - \frac{i \pi}{2} \equiv \gamma_* \in \bb{R}$ as in \eqref{analytic continuation to string gamma}. For massive particles, the string region is $i \pi/2$ below the mirror region.

\paragraph{Cauchy kernel.}

\begin{equation}
s(\gamma) = \frac{1}{2 \pi i} \, \frac{\de}{\de \gamma} \log S (\gamma)
=  \frac{1}{2 \pi \cosh{\gamma}} \,, \qquad
S (\gamma) =-i \tanh \( \frac{\gamma}{2} - \frac{i \pi}{4} \).
\label{eq:CauchyKernel}
\end{equation}
The multiplicative normalisaton of $S(\gamma)$ has been chosen so that
\begin{equation}
    S(0) =-1\,.
\end{equation}
The ``analytically continued'' (\textit{i.e.} shifted) kernel is
\begin{equation}
s_* (\gamma) = s(\gamma+\tfrac{i\pi}{2})
=\frac{1}{2\pi i\,\sinh\gamma}\,, \qquad
S_* (\gamma) =S (\gamma+\tfrac{i\pi}{2}) =-i \tanh  \frac{\gamma}{2} \,.
\label{eq:CauchyKernel-shifted}
\end{equation}
The (right-)inverse Cauchy kernel is the following difference operator,
\begin{equation}
(f * s^{-1})(\gamma) = 
f(\gamma + \frac{i \pi}{2} - i0) 
+ f(\gamma - \frac{i \pi}{2} + i0).
\label{def:Cauchy inverse}
\end{equation}

Cauchy kernel satisfies
\begin{equation}
(1 * s)(\gamma) = \frac{1}{2}, \qquad
(s * s) (\gamma) 
= \frac{\gamma}{2 \pi^2 \, \sinh (\gamma)}
= \mathcal{K}_{\text{SG}}(\gamma)\,,
\label{eq:CauchySGidentity}
\end{equation}
where $\mathcal{K}_{\text{SG}}(\gamma)$ is the sine-Gordon kernel,
\begin{equation}
\mathcal{K}_{\text{SG}}(\gamma)
=\frac{1}{2\pi i} \, \frac{\de}{\de \gamma} \log \PhiSG (\gamma)
=\frac{1}{2\pi i} \, \frac{\de}{\de \gamma} \phiSG (\gamma)\,,
\end{equation}
as defined in \eqref{eq:varphiexplicit}. Furthermore we have that
\begin{equation}
    \int\limits_{-\infty}^{+\infty}\de\gamma'\,\log \left(-S(\gamma-\gamma')\right)s(\gamma'-\gamma'')=
    \phiSG(\gamma-\gamma'')\,.
\end{equation}

\subsection{Massive dressing factors}\label{app:massive dressing}
The massive dressing factors in the mirror-mirror kinematics, with $Q,Q^\prime =1,2,\dots$ are given in appendix C of~\cite{Frolov:2021bwp} and read
\begin{equation}
\begin{aligned}
    (\Sigma_{12}^{QQ^\prime})^{-2} &= - \frac{\sinh{\frac{\gamma_{12}^{-+}}{2}}}{\sinh{\frac{\gamma_{12}^{+-}}{2}}} e^{\varphi^{\bullet \bullet}(\gamma_1^{\pm},\gamma_2^{\pm})}  (\Sigma_{\bes}^{QQ^\prime}(x_1^{\pm},x_2^{\pm}))^{-2} \,,\\
    (\tilde{\Sigma}_{12}^{QQ^\prime})^{-2} &= + \frac{\cosh{\frac{\gamma_{12}^{+-}}{2}}}{\cosh{\frac{\gamma_{12}^{-+}}{2}}} e^{\tilde{\varphi}^{\bullet \bullet}(\gamma_1^{\pm},\gamma_2^{\pm})}  (\Sigma_{\bes}^{QQ^\prime}(x_1^{\pm},x_2^{\pm}))^{-2} \,.
\end{aligned}
\end{equation}
We define the corresponding kernels by
\begin{equation}
\begin{aligned}
K_\Sigma^{Q_1 Q_2} (u_1, u_2) &= \frac{1}{2\pi i} \frac{\partial}{\partial u_1} 
\log \Sigma_{12}^{Q_1 Q_2} (x_1^\pm, x_2^\pm),
\\
\tilde{K}_\Sigma^{Q_1 Q_2} (u_1, u_2) &= \frac{1}{2\pi i} \frac{\partial}{\partial u_1} 
\log \tilde{\Sigma}_{12}^{Q_1 Q_2} (x_1^\pm, x_2^\pm).
\end{aligned}
\label{def:massive dressing kernels}
\end{equation}

The phases $\varphi^{\bullet \bullet}$ and $\tilde{\varphi}^{\bullet \bullet}$ are given by eq. (5.6) in \cite{Frolov:2021fmj} and can be expressed as
\begin{equation}
\begin{aligned}
e^{\varphi^{\bullet \bullet}(\gamma_1^{\pm},\gamma_2^{\pm})} &= 
\exp \Bigl(
\varphi_+ (\gamma_{12}^{--}) + \varphi_+ (\gamma_{12}^{++}) 
+ \varphi_- (\gamma_{12}^{-+}) + \varphi_- (\gamma_{12}^{+-}) 
\Bigr) ,
\\[1mm]
e^{\tilde{\varphi}^{\bullet \bullet}(\gamma_1^{\pm},\gamma_2^{\pm})} &= 
\exp \Bigl(
\varphi_- (\gamma_{12}^{--}) + \varphi_- (\gamma_{12}^{++}) 
+ \varphi_+ (\gamma_{12}^{-+}) + \varphi_+ (\gamma_{12}^{+-})
\Bigr) ,
\end{aligned}
\label{def:massive sG factors}
\end{equation}
where the functions
\begin{equation}
\label{eq:varphiexplicit}
\begin{aligned}
\varphi_-(\gamma) &=+ \frac{ i}{\pi} \text{Li}_2\left(+e^{\gamma}\right)- \frac{i}{4\pi} \gamma^2+\frac{i}{\pi} \gamma\,  \log
   \left(1-e^{\gamma }\right)-\frac{i \pi }{6}\,,
   \\
\varphi_+(\gamma) &=-\frac{ i}{\pi} \text{Li}_2\left(-e^{\gamma}\right)+ \frac{i}{4\pi}  \gamma^2-\frac{i}{\pi}\gamma \,  \log
   \left(1+e^{\gamma}\right)-\frac{i \pi }{12} \,,
\end{aligned}
\end{equation}
were introduced. 
These formulae are valid when $\gamma$ is in the strip between zero and~$i\pi$.
We will also use\footnote{This formula is given for $\gamma$ in the vicinity of the real line. More precisely, the phase are regular in the strip ${\rm Im} \, \gamma \in (-\pi, \pi)$. More general values of $\gamma$ can be reached by analytic continuation through the cuts of the logarithm and dilogarithm~\cite{Frolov:2021fmj}.}
\begin{equation}
\begin{aligned}
\phiSG(\gamma) &\equiv \varphi_+(\gamma) + \varphi_-(\gamma)
\\
&=
\frac{i}{\pi}\text{Li}_2(-e^{-\gamma})-\frac{i}{\pi}\text{Li}_2(e^{-\gamma})+\frac{i\gamma}{\pi}\log(1-e^{-\gamma})-\frac{i\gamma}{\pi}\log(1+e^{-\gamma})+\frac{i\pi}{4}\,,
\\[1mm]
\phiMON(\gamma) &\equiv \varphi_+(\gamma) - \varphi_-(\gamma) 
\\
&= \frac{i}{2 \pi }\text{Li}_2\left(e^{-2\gamma }\right)-\frac{i}{2 \pi}\gamma^2 -\frac{i}{\pi}  \log
\left(1-e^{-2\gamma }\right)-\frac{i\pi }{12} \,,
\end{aligned}
\label{eq:varphiexplicit2}
\end{equation}
noting that the $\phiSG(\gamma)$ is the Sine-Gordon dressing factor.
We find crossing-like relations
\begin{equation}
\label{eq:crossingregularisation}
\begin{aligned}
e^{ \varphi_+ (\gamma) + \varphi_- (\gamma+\pi i)}
&=\frac{1}{2 \cosh\tfrac{\gamma}{2}}\,,
&\qquad
e^{ \varphi_+ (\gamma) + \varphi_- (\gamma-\pi i)}
&= 2 \cosh\frac{\gamma}{2} \,,
\\[1mm]%
e^{ \varphi_- (\gamma) + \varphi_+ (\gamma+\pi i)}
&= 2 i \sinh\frac{\gamma}{2}\,,\qquad
&\qquad
e^{ \varphi_- (\gamma) + \varphi_+ (\gamma-\pi i)}
&= \frac{i}{2 \sinh\tfrac{\gamma}{2}}\,,
\\[1mm]%
e^{ \varphi_+ (\gamma) - \varphi_+ (\gamma+2 \pi i)}
&= \frac{-1}{4 \cosh^2 \tfrac{\gamma}{2}}\,,
&\qquad
e^{ \varphi_- (\gamma) - \varphi_- (\gamma-2 \pi i)}
&= - 4 \sinh^2 \frac{\gamma}{2} \,.
\end{aligned}
\end{equation}

The improved BES factor for $QQ'$ particles is the same as in~\cite{Arutyunov:2009kf},
\begin{equation}
\label{mass^2 mir^2 BES}
\begin{aligned}
&\frac{1}{i}\log\Sigma^{QQ'}_\bes (x_1^\pm,x_2^\pm)
 =
\Phi(x_1^+,x_2^+)-\Phi(x_1^+,x_2^-)-\Phi(x_1^-,x_2^+)+\Phi(x_1^-,x_2^-)\\
&\qquad\qquad-\frac{1}{2}\left(\Psi(x_1^+,x_2^+)+\Psi(x_1^-,x_2^+)-\Psi(x_1^+,x_2^-)-\Psi(x_1^-,x_2^-)\right) \\
&\qquad\qquad+\frac{1}{
2}\left(\Psi(x_{2}^+,x_1^+)+\Psi(x_{2}^-,x_1^+)-\Psi(x_{2}^+,x_1^-)
-\Psi(x_{2}^-,x_1^-) \right)
 \\
&\qquad\qquad+\frac{1}{i}\log
\frac{ i^{Q}\,\Gamma\big[Q'-\frac{i}{2}h\big(x_1^++\frac{1}{x_1^+}-x_2^+-\frac{1}{x_2^+}\big)\big]}
{i^{Q'}\Gamma\big[Q+\frac{i}{2}h\big(x_1^++\frac{1}{x_1^+}-x_2^+-\frac{1}{x_2^+}\big)\big]}
\frac{1-\frac{1}{x_1^+x_2^-}}{1-\frac{1}{x_1^-x_2^+}}\sqrt{\frac{x_1^+x_2^-}{x_1^-x_2^+}} \,,
\end{aligned}
\end{equation}
where $\Phi(x_1,x_2)$ and $\Psi(x_1,x_2)$ will be given in Appendix \ref{app:BES phase}.

\subsection{Mixed-mass dressing factors}

To define mixed-mass scattering elements in the mirror TBA it is sufficient to use the Sine-Gordon dressing factor~$\PhiSG$,
\begin{equation}
\PhiSG(\gamma)=e^{\phiSG(\gamma)} \,,
\label{def:Phi-phi SG}
\end{equation}
given in the previous subsection, as well as an appropriate generalisation of the BES phase~\cite{Frolov:2021fmj}.
We can obtain the (improved) mixed-mass BES dressing factor from~\eqref{mass^2 mir^2 BES} by setting~$Q=0$ or~$Q'=0$ as needed.

We will also need to consider the string-mirror and mirror-string kinematics, when the massless excitation is on the string region. 
By setting $Q=0$, we find the Zukovsky variable $x$ of massless particles in the string region sitting on the upper-half circle.
There is an apparent problem, however, with the BES phase: the massless particle sits on the integration contour, leading to a potential single pole. Consider for definiteness $\Sigma_{\BES}^{0Q}(x,y^\pm)$. Potential issues arise from the integrals $\Phi(x,y^\pm)$, $\Phi(\tfrac{1}{x},y^\pm)$, $\Psi(y^\pm, x)$, and $\Psi(y^\pm, \tfrac{1}{x})$. In any of these integrals we can shrink or enlarge a bit the radius of the integration contour without hitting any singularity as long as $h$ is finite.

In practice, we should use a principal-value prescription and add the relevant residues at the poles, when we want to compute the explicit values of the BES dressing factor. 
Following \cite{Frolov:2021fmj}, we should take the limit $Q \to 0$ of the Zukovsky variable keeping the relation $x^+ x^- = 1$. 
A more detailed expression will be given in Appendix \ref{app:BES phase}.

\section{Simplifying the renormalised kernels}\label{app:Kren}

The renormalised kernels are defined in \eqref{def:kernels ren} as
\begin{equation*}
K_{\rm ren}^{00} = K^{00} + 2 K^{0y} \hstar K^{y0},\quad
K_{\rm ren}^{Q 0} = K^{Q 0} + 2 K^{Qy}_+ \hstar K^{y0}, \quad
\tilde{K}_{\rm ren}^{Q 0} = \tilde{K}^{Q 0} - 2 K^{Qy}_- \hstar K^{y0} \,.
\end{equation*}
Below we will show that these kernels take a simple form,\footnote{Note that $K_{\rm ren}^{Q 0} (u_Q, u_0 ) = \tilde{K}_{\rm ren}^{Q 0} (u_Q, u_0 )$.}
\begin{align}
\cK_{\rm ren}^{00} (\gamma_j, \gamma_k) 
&= s (\gamma_{jk}) 
- 2 \cK_\bes^{00} \Big(x (\gamma_j) ,x (\gamma_k) \Big) 
\label{CK00 ren-ev} \\
K_{\rm ren}^{Q 0} (u_Q, u_0  ) &= 
K_{\rm aux} ( u_Q, u_0 )
- 2 K_\bes^{Q 0} \Big( x^\pm (u_Q) , x(u_0) \Big)
\label{KQ0 ren-ev} 
\\
\tilde{K}_{\rm ren}^{\bQ 0} (u_Q, u_0 ) &= 
K_{\rm aux} ( u_Q, u_0 )
- 2 K_\bes^{\bQ 0} \Big( x^\pm (u_Q) , x(u_0) \Big) , 
\label{tKQ0 ren-ev}
\end{align}
where 
\begin{equation}
\begin{aligned}
\cK_\bes^{00} \Big(x (\gamma_j) ,x (\gamma_k) \Big) &= \frac{1}{2\pi i} \, \frac{\partial}{\partial \gamma_j} 
\log \Sigma_{\bes}^{0 0}(x (\gamma_j) , x (\gamma_k))
\\
K_\bes^{Q 0} (u_Q, u_0  ) &= 
\frac{1}{2\pi i} \, \frac{\partial}{\partial u_Q} 
\log \Sigma_{\bes}^{Q 0} \Big( x^\pm (u_Q) , x(u_0) \Big) ,
\\
K_{\rm aux} ( u_Q, u_0 ) &=
\frac{1}{2\pi i} \, \frac{\partial}{\partial u_Q} 
\log \frac{S ( \gamma^- (u_Q) - \gamma (u_0) + \frac{\pi i}{2} )}
{S ( \gamma^+ (u_Q) - \gamma (u_0)+ \frac{\pi i}{2} )} \,.
\end{aligned}
\label{def:K_bes}
\end{equation}

\subsection{Massless-massless renormalised kernels}\label{app:massless^2 Kren}

The S-matrix $S^{00}$ is given by \eqref{def:S00}, where the ``auxiliary factor'' introduced in~\cite{Frolov:2021fmj} is precisely
\begin{equation}
    a(\gamma)=S(\gamma)\,.
\end{equation}
The calligraphic kernel $\cK^{00}$ is given by
\begin{multline}
\mathcal{K}^{00}_{\text{ren}}(\gamma,\gamma') =
s(\gamma-\gamma')+2\mathcal{K}_{\text{SG}}(\gamma-\gamma')
- 2 \cK_\bes^{00} \Big(x (\gamma) ,x (\gamma' ) \Big)
\\
+ 2 \( \frac{d u}{d \gamma} \)  (K^{0y} \hstar K^{y0}) (\gamma, \gamma') .
\label{cK00 ren comp}
\end{multline}

We observe that
\begin{equation}
S^{0y}(x(\gamma),x_s(\gamma')) = \frac{-i\,\text{sgn}(\gamma)\,}{S(\gamma-\gamma')}\,,\qquad
S^{y0}(x_s(\gamma),x(\gamma')) = \frac{+i\,\text{sgn}(\gamma')}{S(\gamma-\gamma')}\,,
\label{comp: S0y Sy0}
\end{equation}
so that
\begin{equation}
    \mathcal{K}^{0y}(\gamma,\gamma')=-s(\gamma-\gamma')+\frac{1}{2}\delta(\gamma)\,,\qquad
    \mathcal{K}^{y0}(\gamma,\gamma')=-s(\gamma-\gamma')\,.
\label{def:cK0y cKy0}
\end{equation}
From the identity \eqref{eq:CauchySGidentity}, we see that the convolution $\mathcal{K}^{0y}*\mathcal{K}^{y0}$ cancels $\mathcal{K}_{\text{SG}}$. 
More precisely, repeating the argument of \eqref{eq:convolutionchangevar},
\begin{equation}
\begin{aligned}
K^{0y} \hstar K^{y0} (u, u') &= \int_{-2}^2 dv \, K^{0y} (u, v) K^{y0} (v, u')
\\
&= - \int_{-\infty}^\infty d \gamma_y \,
K^{0y} \Big(u (\gamma) , u_s (\gamma_y) \Big) \,
\frac{dv}{d \gamma_y} \,
K^{y0} \Big(u_s (\gamma_y) , u(\gamma') \Big) 
\\
&= \( \frac{du}{d \gamma} \)^{-1} \Bigl\{  - \cK_{\rm SG} (\gamma, \gamma') + \frac14 \delta (\gamma) \Bigr\} ,
\end{aligned}
\label{conv K0y-Ky0}
\end{equation}
where $v=u_s (\gamma_y)$.
Substituting this into \eqref{cK00 ren comp} and neglecting the $\delta$ functions (which we can do because the $Y$-functions of massless particles must vanish at~$\gamma=0$), we obtain \eqref{CK00 ren-ev}.

\subsection{Massive-massless renormalised kernels}\label{app:massive-massless Kren}

Our goal is to prove the identity \eqref{convolution KQys-2} and remove some factors from the kernels of $S^{Q_a 0}, \bar{S}^{\bar{Q}_a 0}$\,, which are given in \eqref{def:SQa0}, \eqref{def:barS bQa0}, respectively.

To begin with, we find that the S-matrices $S_\pm^{Q_a y}(u, v)$ in \eqref{def:SQay+}, \eqref{def:SQay-} can be summarised as
\begin{equation}
\begin{gathered}
S_+^{Q y} ( u, v ) = S^{Q y} ( x_a^\pm (u), x_s (v+i0) ) , \qquad
S_-^{Q y} ( u, v ) = S^{Q y} ( x_a^\pm (u), \frac{1}{x_s (v+i0)} ) ,
\\[1mm]
S^{Q y}( x_a^\pm , y) 
= \sqrt{ \frac{x_a^+}{x_a^-} } \,\frac{x_a^- - 1/y }{x_a^+ - 1/y } \,.
\end{gathered}
\label{def:SQy pm}
\end{equation}
The function $S^{Q y}( x_a^\pm , y)$ satisfies the identity
\begin{equation}
\begin{aligned}
\log {S}^{Q y}( x_a^\pm , y) &= \frac12 \Bigl\{
\log S^{0y}(x_a^-, y) - \log S^{0y}(x_a^+, y) + \log S^Q (u_a, v) \Bigr\},
\end{aligned}
\end{equation}
where $x_a + 1/x_a = u_a, \ y+1/y=v$, and $S^Q$ is the rational S-matrix \eqref{def:rational S-matrix} and $S^{0y}$ is given in \eqref{def:S0y uu}.
The kernels $K^{Qy}_\pm$ are 
\begin{equation}
\begin{aligned}
K^{Qy}_+ (u, v) &= \frac12 \Bigl\{
K^Q (u-v) - K^{0y} (u^- , v) + K^{0y}(u^+, v) \Bigr\} , 
\\[1mm]
K^{Qy}_- (u, v) &= \frac12 \Bigl\{
K^Q (u-v) + K^{0y} (u^- , v) - K^{0y}(u^+, v) \Bigr\} ,
\end{aligned}
\label{KQypm identity}
\end{equation}
and both of them are positive in the mirror-mirror region.

Na\"ively we cannot uniquely define the calligraphic kernel $\cK^{Qy}_\pm$ because $S^{Qy}_\pm$ depends on both $\gamma^\pm$ while $\gamma^+$ and $\gamma^-$ are not independent. 
Thus we work with the $u$ rapidities by inverting the relation $u = u(\gamma)$ if necessary.

\bigskip
Consider the convolutions
\begin{equation}
K^{Qy}_\pm \hstar K^{y0} = 
\frac12 \Bigl\{
 K_Q (u_Q , v) \mp K^{0y} (x^-, y) \pm K^{0y} (x^+, y) 
\Bigr\} \hstar K^{y0} (y, x_0) ,
\label{convolution KQys}
\end{equation}
where the $u$-rapidities should stay a little bit above the real axis whenever necessary.
The first term of \eqref{convolution KQys} can be written as
\begin{equation}
\begin{aligned}
K_Q \hstar K^{y0} (u_Q, u_0) &=
\int_{-2}^2 dv_y \, K^{Q} (u_Q - v_y) K^{y0} (v_y, u_0)
\\
&=
\int_{-\infty}^\infty d \gamma_y \, \( - \frac{d v_y}{d \gamma_y} \) 
K^{Q} (u_Q - u_s (\gamma_y)  ) K^{y0} \Big( u_s (\gamma_y) , u(\gamma_0) \Big)
\end{aligned}
\end{equation}
where we introduced $v_y = u_s (\gamma_y)$ and $u_0 = u(\gamma_0)$ defined in \eqref{def:u to gamma's}.
Using \eqref{def:cK0y cKy0} we obtain
\begin{equation}
\begin{aligned}
K_Q \hstar K^{y0} (u_Q, u_0) &=
\int_{-\infty}^\infty d \gamma_y \, K^{Q} (u_Q - u_s (\gamma_y)  ) s \Big( \gamma_y - \gamma (u_0) \Big)
\\[1mm]
&= \frac{1}{2 \pi i} \frac{\partial}{\partial u_Q} 
\log \(  \sqrt{ \frac{x^+ (u_Q)}{x^- (u_Q)} } \frac{x^- (u_Q) - x (u_0)}{x^+ (u_Q) - 1/x (u_0)} \),
\end{aligned}
\end{equation}
which cancel some factors in $S^{Q 0}$ and $\tilde S^{Q 0}$.
The remaining terms in \eqref{convolution KQys} are
\begin{equation}
K^{0y} (x^\mp (u) , y) \hstar K^{y0} (y, x (u_0) )
= \pm \bar K^\mp_{\rm SG} ( u, u_0 ) ,
\end{equation}
where
\begin{equation}
\bar K^\mp_{\rm SG} ( u, u_0 ) =
\frac{1}{2 \pi i} \, \frac{\partial}{\partial u} \, 
\log \PhiSG \Big( \gamma^\mp (u) - \gamma (u_0) \mp \pi i \Big) ,
\label{def:bar KSG}
\end{equation}
The $\gamma$-variables are defined in \eqref{def:gamma-u mirror}. In particular, $\gamma (u_0)$ is real when $u_0 \in (-\infty,-2) \cup (2,\infty)$. 
We rewrite this result using
\begin{equation}
\PhiSG(\gamma)\PhiSG(\gamma+i\pi)=i\tanh\frac{\gamma}{2}\,,
\label{PhiSG quasi-periodic}
\end{equation}
to find
\begin{equation}
\bar K^\pm_{\rm SG} ( u_j , u_0 ) =
- \frac{1}{2 \pi i} \, \frac{\partial}{\partial u_j} \log \PhiSG ( \gamma^{\pm \circ}_{j0} ) 
+ \frac{1}{2 \pi i} \, \frac{\partial}{\partial u_j} 
\log S \Big( \gamma^{\pm \circ}_{j0} \pm \frac{\pi i}{2} \Big).
\end{equation}

In summary, the equation \eqref{convolution KQys} becomes
\begin{multline}
K^{Qy}_\pm \hstar K^{y0} (u_Q, u_0) = 
\frac{1}{4 \pi i} \frac{\partial}{\partial u_Q} 
\log \Biggl[  \sqrt{ \frac{x^+ (u_Q)}{x^- (u_Q)} } \frac{x^- (u_Q) - x (u_0)}{x^+ (u_Q) - 1/x (u_0)} 
\Biggr] 
\\
\pm
\frac{1}{4 \pi i} \frac{\partial}{\partial u_Q} 
\log \Biggl[  \frac{S ( \gamma^{- \circ}_{j0} + \frac{\pi i}{2} )}
{S ( \gamma^{+ \circ}_{j0} + \frac{\pi i}{2} )} \,
\PhiSG ( \gamma^{- \circ}_{j0} ) \, \PhiSG ( \gamma^{+ \circ}_{j0} )  
\Biggr] .
\label{convolution KQys-2}
\end{multline}

\subsection{Massless source terms and TBA equations}
\label{App:MasslessSourceTerms}

Considering massless excitations, we pick up various source terms in the TBA equations, \textit{cf.} eqs. \eqref{ex-TBA for YQ} - \eqref{eq:auxiliary-yp-TBA-naive}. Using the renormalised kernels the equation for the massless modes is given in eq. \eqref{eqY0 renorm}. At leading order in the small tension limit, the BES phase does not contribute and the source term is given by
\begin{equation}
S^{0_*0}_{\text{ren}}(\gamma_j^{\dot{\alpha}_j},\gamma')=-\frac{1}{S_*(\gamma_j^{\dot{\alpha}_j}-\gamma')}+\mathcal{O}(h)\,.
\end{equation}
Hence, for an arbitrary number of massless excitations $\mathscr{M}$ (without necessarily assuming $\mathscr{M}=2M$), the massless TBA equation is given by
\begin{equation}
\begin{aligned}
    -\log Y_0(\gamma) = &L \tilde{\mathcal{E}}_0 (\gamma) -  \(\log(1+Y_0)^{N_0}*s\) (\gamma)
   - \sum_{j=1}^{\mathscr{M}} \log (-S_{*}(\gamma_j^{\dot{\alpha}_j}-\gamma))\\
    &-  \(\log{\left(1-Y_+\right)^2} *s\)(\gamma) -  \(\log{\left(1-\frac{1}{Y_-}\right)^2} *s\)(\gamma) \,,
\end{aligned}    
\end{equation}

For the auxiliary modes we pick up the source terms
\begin{equation} 
S^{0_*y}(\gamma_j^{\dot{\alpha}_j},\gamma') = i \, \text{sgn}(\gamma_j^{\dot{\alpha}_j}) \, S(\gamma_j^{\dot{\alpha}_j}-\gamma')\,.
\end{equation}
Considering an arbitrary number of massless excitations $\mathscr{M}$, the auxiliary TBA equations can be written as
\begin{equation} 
    \log Y(\gamma) = \left( \log{(1+Y_0)^{N_0}} *s \right)(\gamma) 
   + \sum_{j=1}^{\mathscr{M}} \log \left(i\, \text{sgn}(\gamma_j^{\dot{\alpha}_j}) \, S_{*}(\gamma_j^{\dot{\alpha}_j}-\gamma)\right)\,.
\end{equation}
For an even number of massless excitations $\mathscr{M}=2M$ \textit{and by picking the rapidities to come in pairs} of the form $(-\gamma_j^{\dot{\alpha}_j},\gamma_j^{\dot{\alpha}_j})$, we see that the source terms picks up a sign due to the signum functions $\text{sgn}(-\gamma_j^{\dot{\alpha}_j}) \, \text{sgn}(\gamma_j^{\dot{\alpha}_j})  = -1$ inside the logarithm. However, there is also the factor $i^2=-1$ compensating the previous sign. Hence, the source terms in the auxiliary TBA equations from \eqref{eq:YalmostSimplyfied} are identical to the ones in the massless TBA given in eq.~\eqref{eq:Y0almostSimplyfied}.

Finally, we also have the exact Bethe equations. Here the source terms are given by 
\begin{equation}
    S_{\text{ren}}^{0_*0_*}(\gamma_j^{\dot{\alpha}_j},\, \gamma_k^{\dot{\alpha}_k}) = S(\gamma_j^{\dot{\alpha}_j}-\gamma_k^{\dot{\alpha}_k}) + \mathcal{O}(h)\,,
\end{equation}
in the small tension limit.
Therefore, for $\mathscr{M}$ massless excitations the exact Bethe equations read
\begin{equation}
\begin{aligned}
    i\pi(2\nu_k^{\dot{\alpha}}+1)=&
    -i L p(\gamma^{\dot{\alpha}_k})-  \(\log{(1+Y_0)^{N_0}} * s_*\)(\gamma_k^{\dot{\alpha}_k})
    + \sum_{j=1}^{\mathscr{M}} \log S(\gamma_j^{\dot{\alpha}_j}-\gamma_k^{\dot{\alpha}_k})\\
    &- \(\log{\left(1-Y\right)^4}  * s_*\)(\gamma_k^{\dot{\alpha}_k}) \,.
\end{aligned}
\end{equation}

\section{Weak-coupling expansions}\label{app:weak-coupling exp}

We start by rescaling the rapidity as $\tilde u = u/h$. Note that, for $Q>0$ and $u$ real,
\begin{equation}
x \Big( \frac{\tilde u + iQ}{h} \Big) = \frac{h}{\tilde u + iQ} + {\cal O} (h^3), 
\qquad
x \Big( \frac{\tilde u - iQ}{h} \Big) = \frac{\tilde u - iQ}{h} +\mathcal{O}(h),
\label{rescaled xu small-h}
\end{equation}
which shows that $\abs{x^+} \le h/Q \,, \ \abs{x^-} \ge Q/h$ for real $\tilde u$ at the leading order of small $h$.\footnote{It is convenient to redefine $x(v) = x_s(v)$ for ${\rm Im} \, v <0$ and $1/x_s(v)$ for ${\rm Im} \, v >0$ for deriving these expansions.}
In terms of the $\gamma$-rapidities we have
\begin{equation}
\label{eq:gammapmexpansion}
\begin{aligned}
\gamma_s^\pm  \Big( \frac{\tilde u}{h} \Big)
&=\mp\frac{i\pi}{2} - \frac{2h}{\tilde u \pm iQ} +\mathcal{O}(h^3)\,,
&\qquad
\gamma_s \Big( \frac{\tilde u}{h} + i 0 \Big)
&= - \frac{i \pi}{2} - \frac{2h}{\tilde u} +\mathcal{O}(h^2)\,,
\\
\gamma^\pm \Big( \frac{\tilde u}{h} \Big)
&=\mp i \pi - \frac{2h}{\tilde u \pm iQ} +\mathcal{O}(h^3)\,,
&\qquad
\gamma \Big( \frac{\tilde u}{h} + i 0 \Big)
&= - \frac{2h}{\tilde u} +\mathcal{O}(h^2)\,,
\end{aligned}
\end{equation}
and thus
\begin{equation}
\label{eq:gamma-pm-expansion2}
\begin{aligned}
\gamma^{\pm\pm}&= - 2h \( 
\frac{1}{\tilde{u}_1 \pm i Q_1} - \frac{1}{\tilde{u}_2 \pm i Q_2} \) +\mathcal{O}(h^3),
\\
\gamma^{\pm\mp}&= \mp 2 \pi i 
+ 2 h \( \frac{1}{\tilde{u}_1 \pm i Q_1}-\frac{1}{\tilde{u}_2 \mp i Q_2} \)
+\mathcal{O}(h^3),
\\
\gamma^{\circ\pm}&= \pm i \pi
- 2 h \(  \frac{1}{\tilde u_1} - \frac{1}{\tilde u_2 \pm iQ} \)
+\mathcal{O}(h^2) .
\end{aligned}
\end{equation}

\subsection{List of kernels}

The rational kernels are written as
\begin{equation}
K^Q \Big( \frac{\tilde u_1}{h}, \frac{ \tilde u_2 }{h} \Big) = 
\frac{h Q}{\pi} \, \frac{1}{(\tilde u_1 - \tilde u_2)^2+Q^2} \,,
\end{equation}
and
\begin{multline}
K^{Q_1 Q_2} \Big( \frac{\tilde u_1}{h}, \frac{ \tilde u_2 }{h} \Big) = 
-\frac{h}{4 \pi } \Bigg\{
\psi\left(\frac{Q_1-Q_2 - i \, (\tilde{u}_1-\tilde{u}_2)}{2}\right)
-\psi\left(\frac{Q_1+Q_2 - i \, (\tilde{u}_1-\tilde{u}_2)}{2}\right)
\\
-\psi\left(\frac{Q_1+Q_2 - i \, (\tilde{u}_1-\tilde{u}_2)}{2}+1\right)
+ \psi\left(\frac{Q_1-Q_2 - i \, (\tilde{u}_1-\tilde{u}_2)}{2}+2\right)
+ ({\rm c.c.})
\Bigg\} ,
\end{multline}
where (c.c.) is the complex conjugate.
Both expressions are $O(h)$ and regular for $Q \in \bb{N}_0$ and real rapidities.

\paragraph{Equations for $Q$-particles \eqref{ex-TBA for YQ}.}

The kernels and S-matrices are
\begin{alignat}{9}
K_{\textit{sl}}^{Q_1 Q_2} &= 
- K ^{Q_1 Q_2}(u_1-u_2) 
- 2 K_{\Sigma}^{Q_1 Q_2} (u_1,u_2)
\label{weak Ksl12} \\[1mm]
\tilde{K}_{\textit{su}}^{\bar{Q}_1 Q_2} &\simeq 
\frac{h}{\pi} \Bigl\{ \frac{Q_1}{\tilde{u}_1^2+Q_1^2}
-\frac{Q_1+Q_2}{(\tilde{u}_2-\tilde{u}_1)^2+(Q_1+Q_2)^2}
\Bigr\}
- 2 \tilde K_{\Sigma}^{\bar{Q}_1 Q_2} (u_1,u_2)
\label{weak tKsu12} \\[1mm]
K^{0 Q_2} &\simeq
\frac{2 h Q_2}{\pi  u_1^2 \sqrt{1-\frac{4}{u_1^2}} \( 1 + \sqrt{1-\frac{4}{u_1^2}} \)  \left(\tilde{u}_2^2+Q_2^2\right)}
- \frac{1}{2 \pi i} \frac{\partial}{\partial u_1} 
\log \( \frac{(\Sigma_{\bes}^{Q_2 0}(x_2^\pm,x_1))^{-2}}{\PhiSG (\gamma_{21}^{+\circ}) \PhiSG (\gamma_{21}^{-\circ})} \)
\label{weak K0Q2} \\
\log S^{0_* Q_2} &\simeq
\log \left( i \, \frac{\left(\tilde{u}_1-\tilde{u}_2+i Q_2\right) \sqrt{\tilde{u}_2+i Q_2}}{\left(\tilde{u}_1-\tilde{u}_2-i Q_2\right) \sqrt{\tilde{u}_2-i Q_2}}\right)
- \log \( \frac{\Sigma_{\bes}^{Q_2 0}(x_2^\pm, x_{1*})^{-2}}{\PhiSG (\gamma_{2 1_*}^{+\circ}) \PhiSG (\gamma_{2 1_*}^{-\circ})} \),
\label{weak lS0Q2} \\
K^{yQ_2}_+ &\simeq \frac{1}{2 \pi  \sqrt{4-u_1^2}}
+\frac{h \left(Q_2 \, (u_1^2-4)- \tilde{u}_2 \, u_1 \sqrt{4-u_1^2} \right)}
{2 \pi (u_1^2-4) (Q_2^2+\tilde{u}_2^2)}
\label{weak KyQ2p} \\
K^{yQ_2}_- &\simeq
\frac{1}{2 \pi \sqrt{4-u_1^2}}
-\frac{h \left(Q_2 \, (u_1^2-4)+ \tilde{u}_2 \, u_1 \sqrt{4-u_1^2} \right) }
{2 \pi (u_1^2-4) (Q_2^2+\tilde{u}_2^2)}
\label{weak KyQ2m}
\end{alignat}
where $\simeq$ means that we take the leading terms in the small $h$ expansion.

We find that the kernels \eqref{weak Ksl12} and \eqref{weak tKsu12} are finite, using the properties of the massive dressing kernels which will be discussed in Appendix \ref{app:massive dressing kernels exp}.
Some kernels are potentially dangerous at $u_1 = \pm 2$, but this singularity is integrable as long as $Y_0 (u_1), Y_\pm (u_1)$ remain finite as $u_1 \to \pm 2$.\footnote{Note that we did not rescale $u_1$ in $K^{0 Q_2} (u_1, u_2)$ and $K^{yQ_2}_\pm (u_1, u_2)$.}

\paragraph{Equations for $\bQ$-particles \eqref{ex-TBA for YbQ}.}

The kernels and S-matrices are
\begin{alignat}{9}
K_{\textit{su}}^{\bQ_1 \bQ_2} &\simeq 
\frac{h Q_1}{\pi  \left(\tilde{u}_1^2+Q_1^2\right)}
- K ^{Q_1 Q_2}(u_1-u_2) 
- 2 K_\Sigma^{\bar{Q}_1 \bar{Q}_2} (u_1,u_2)
\label{weak Ksu12} \\[1mm]
\tilde{K}_{\textit{sl}}^{\bar{Q}_1 Q_2} &\simeq 
-\frac{h \left(Q_1+Q_2\right)}{\pi  \left(\left(\tilde{u}_1-\tilde{u}_2\right){}^2+\left(Q_1+Q_2\right){}^2\right)}
- 2 \tilde K_\Sigma^{\bar{Q}_1 \bar{Q}_2} (u_1,u_2)
\label{weak tKsl12} \\[1mm]
\tilde K^{0 \bar{Q}_2} &\simeq 
\frac{-2 h \bar{Q}_2}{\pi  u_1^2 \sqrt{1-\frac{4}{u_1^2}} \( 1 + \sqrt{1-\frac{4}{u_1^2}} \)  \left(\tilde{u}_2^2+\bar{Q}_2^2\right)}
- \frac{1}{2 \pi i} \frac{\partial}{\partial u_1} 
\log \( \frac{(\Sigma_{\bes}^{\bar{Q}_2 0}(x_2^\pm,x_1))^{-2}}{\PhiSG (\gamma_{21}^{+\circ}) \PhiSG (\gamma_{21}^{-\circ})} \) ,
\label{weak tK0Q2} \\
\log \bar{S}^{0_* \bar{Q}_2} &\simeq
\log \left( i \, \frac{\left(\tilde{u}_1-\tilde{u}_2-i \bar{Q}_2\right) \sqrt{\tilde{u}_2-i \bar{Q}_2}}{\left(\tilde{u}_1-\tilde{u}_2+i \bar{Q}_2\right) \sqrt{\tilde{u}_2+i \bar{Q}_2}}\right)
- \log \( \frac{(\Sigma_{\bes}^{\bar{Q}_2 0}(x_2^\pm,x_{1*}))^{-2}}{\PhiSG (\gamma_{21_*}^{+\circ}) \PhiSG (\gamma_{21_*}^{-\circ})} \)
\label{weak ltS0Q2}
\end{alignat}
together with the kernels $K^{yQ}_\pm$ given above.
These kernels have similar properties as those in the equations for $Q$-particles.

\paragraph{Equations for massless particles \eqref{eq:TBA0-excited}.}

The mixed-mass kernels are
\begin{align}
K^{Q_1 0} &\simeq -\frac{h Q_1}{2 \pi  \left(\tilde{u}_1^2+Q_1^2\right)}
+ \frac{1}{2 \pi i} \frac{\partial}{\partial u_1} 
\log \( \frac{(\Sigma_{\bes}^{Q_1 0}(x_1^\pm,x_2))^{-2}}{\PhiSG (\gamma_{12}^{+\circ}) \PhiSG (\gamma_{12}^{-\circ})} \)
\\ 
\tilde{K}^{Q_1 0} &\simeq \frac{h Q_1}{2 \pi  \left(\tilde{u}_1^2+Q_1^2\right)}
+ \frac{1}{2 \pi i} \frac{\partial}{\partial u_1} 
\log \( \frac{(\Sigma_{\bes}^{\bar{Q}_1 0}(x_1^\pm,x_2))^{-2}}{\PhiSG (\gamma_{12}^{+\circ}) \PhiSG (\gamma_{12}^{-\circ})} 
\).
\end{align}

\paragraph{Equations for auxiliary particles \eqref{eq:auxiliary-ym-TBA} and \eqref{eq:auxiliary-yp-TBA-naive}.}

The massive-auxiliary kernels are
\begin{equation}
\begin{aligned}
K_-^{Qy} &\simeq
\frac{h Q_1}{2 \pi  \left(\tilde{u}_1^2+Q_1^2\right)}
+h^2 \left(\frac{u_2 Q_1 \tilde{u}_1}{\pi  \left(\tilde{u}_1^2+Q_1^2\right){}^2}-\frac{i u_2 \sqrt{1-\frac{4}{u_2^2}} \left(Q_1^2-\tilde{u}_1^2\right)}{2 \pi  \left(\tilde{u}_1^2+Q_1^2\right)^2}\right) ,
\\
K_+^{Qy} &\simeq
\frac{h Q_1}{2 \pi  \left(\tilde{u}_1^2+Q_1^2\right)}
+h^2 \left(\frac{u_2 Q_1 \tilde{u}_1}{\pi  \left(\tilde{u}_1^2+Q_1^2\right){}^2}+\frac{i u_2 \sqrt{1-\frac{4}{u_2^2}} \left(Q_1^2-\tilde{u}_1^2\right)}{2 \pi  \left(\tilde{u}_1^2+Q_1^2\right)^2}\right) .
\end{aligned}
\label{weak KpmQy}
\end{equation}
Both kernels remain small at small $h$ for any $\tilde u_1 \in \bb{R}$ and $u_2 \in [-2,2]$.

\subsection{Massive dressing kernels}\label{app:massive dressing kernels exp}

The massive dressing kernels \eqref{def:massive dressing kernels} will affect the self-coupling of the massive modes. 
We only need to show that the kernels have a finite limit as~$h\to0$ in the mirror-mirror region, because we consider only massless excitations in the TBA.
Below we examine the following quantities term by term,
\begin{equation}
\begin{aligned}
\log (\Sigma_{12}^{QQ^\prime})^{-2} &= 
\log \Bigg( - \frac{\sinh{\frac{\gamma_{12}^{-+}}{2}}}{\sinh{\frac{\gamma_{12}^{+-}}{2}}} \Bigg)
+ \varphi^{\bullet \bullet}(\gamma_1^{\pm},\gamma_2^{\pm})
- 2 \log \Sigma_{\bes}^{QQ^\prime}(x_1^{\pm},x_2^{\pm}) \,,
\\
\log (\tilde{\Sigma}_{12}^{QQ^\prime})^{-2} &= 
\log \Bigg( + \frac{\cosh{\frac{\gamma_{12}^{+-}}{2}}}{\cosh{\frac{\gamma_{12}^{-+}}{2}}} \Bigg)
+ \tilde{\varphi}^{\bullet \bullet}(\gamma_1^{\pm},\gamma_2^{\pm})
-2 \log \Sigma_{\bes}^{QQ^\prime}(x_1^{\pm},x_2^{\pm}) \,.
\end{aligned}
\label{log of massive dressing}
\end{equation}

Let us consider the BES dressing factor~\eqref{mass^2 mir^2 BES}, in the mirror-mirror kinematics. 
From \eqref{rescaled xu small-h} it is clear that in the mirror-mirror region $x_1^\pm$ and $x_2^\pm$ never lie close to the unit circle. As a result, the integrand for $\Phi(x_1^\pm,x_2^\pm)$ in~\eqref{def:Phi orig} is regular and in fact goes to zero as $h\to0$. As for $\Psi(x_1^\pm,x_2^\pm)$, the integrand is regular too. However in this case the $\log\Gamma$ terms do not give zero as $h\to0$, but they go to a constant. Regardless, the integral vanishes at $h\to0$. The remaining term goes to a constant, namely to~$-i\log (i^{Q-Q'})$.
Hence, the BES kernel in the mirror-mirror kinematics is zero at leading order.

It remains to estimate the contribution of the $\varphi_{\pm}(\gamma)$ functions~\eqref{eq:varphiexplicit}.
First, let us use the crossing equations~\eqref{eq:crossingregularisation} to note that
\begin{equation}
e^{\phiSG_+(\gamma\pm2\pi i)-\phiSG_+(\gamma)}=
-\left( 2 \cosh\frac{\gamma}{2} \right)^{\pm2},\qquad
e^{\phiSG_-(\gamma\pm2\pi i)-\phiSG_-(\gamma)}=
-\left( 2 \sinh\frac{\gamma}{2} \right)^{\mp2}.
\label{phiSG monodromy weak}
\end{equation}
This double-crossing equation allows us to account for~\eqref{eq:gammapmexpansion} while working on the real line. 
In fact, we see immediately that the resulting functions $\varphi_\pm$ vanish at small~$h$,
\begin{equation}
\varphi_\pm (\varepsilon)=\mathcal{O}(\varepsilon^3)\,,
\qquad
\varphi_-(\varepsilon)=\mathcal{O}(\varepsilon)\,.
\end{equation}
The remain factors in \eqref{log of massive dressing} including the monodromy \eqref{phiSG monodromy weak} behave as
\begin{equation}
\begin{aligned}
\log \Bigg( - 
\frac{\sinh\frac{\gamma_{12}^{+-}}{2} }{\sinh\frac{\gamma_{12}^{-+}}{2}}
\Bigg) &=
\log \Bigg( \!
-\frac{(\tilde{u}_1-i Q_1) (\tilde{u}_1-\tilde{u}_2+i   (Q_1+Q_2)) (\tilde{u}_2+i Q_2)}{(\tilde{u}_1+i   Q_1) (\tilde{u}_1-\tilde{u}_2-i (Q_1+Q_2)) (\tilde{u}_2-i Q_2)}
\Bigg) + \cO(h^2) ,
\\
\log \Bigg( + \frac{\cosh\frac{\gamma_{12}^{-+}}{2} }{\cosh\frac{\gamma_{12}^{+-}}{2}}
\Bigg) &= \cO(h^2) ,
\end{aligned}
\end{equation}
and both of them are regular for $\tilde u_1 \,, \tilde u_2 \in \bb{R}$.

\subsection{Mixed-mass dressing factors and kernels}\label{app:weak mixed mass dressing}
These terms will affect the coupling of massive and massless modes. We will need to consider both the kernels in the mirror-mirror and mirror-string kinematics, as well as (some) S-matrix elements, namely
\begin{equation}
- \frac{1}{2 \pi i} \frac{\partial}{\partial u_1} 
\log \( \frac{(\Sigma_{\bes}^{Q_2 0}(x_2^\pm,x_1))^{-2}}{\PhiSG (\gamma_{21}^{+\circ}) \PhiSG (\gamma_{21}^{-\circ})} \)
\quad {\rm and} \quad
- \log \( \frac{\Sigma_{\bes}^{Q_2 0}(x_2^\pm, x_{1*})^{-2}}{\PhiSG (\gamma_{2 1_*}^{+\circ}) \PhiSG (\gamma_{2 1_*}^{-\circ})} \),
\end{equation}
as found e.g.\ in \eqref{weak K0Q2} and \eqref{weak lS0Q2}.

Let us start by considering the improved BES factor \eqref{mass^2 mir^2 BES} when either $Q=0$ or $Q'=0$  in the mirror-mirror kinematics. By the same token as above, the integrand of the $\Phi$-functions are regular and go to zero as $h\to0$.
Similarly, the related pieces of mirror-mirror kernels are regular and vanish at weak tension.
Things are a little more subtle for the $\Psi$-functions and for the $\log\Gamma$ functions. Now we have that the massless variable $x$ runs from $-1$ to~$+1$, which leads to a possible divergence at $x=0$. This is a distinguished point in the kinematics, and strictly speaking we distinguish
\begin{equation}
    \lim_{u\to+\infty}x(u+i0)=0^+\,,\qquad
    \lim_{u\to-\infty}x(u+i0)=0^-\,.
\end{equation}
This is the branch point for the massless dispersion. Moreover, we already know that $\tilde{\mathcal{E}}^0$ diverges at these points. Hence a singularity at this point is not completely unexpected. In the regime where $|x_1|\ll h\ll1$, the integrand of the $\Psi$-function behaves as\footnote{There is a bug in the asymptotic expansion of $\log \Gamma$ in {\tt Mathematica} 13.2.1.0, which is relevant to this particular expansion; please evaluate numerically the output of {\tt Table[Series[I LogGamma[I/z - I w + I z\^{}n q], {z, 0, 0}], {n, 5}]} at small {\tt z}. To obtain the correct series expansion consistent with numerics, we should expand $\log \Gamma (i/Z)$ at small $Z$ and substitute the solution of $i/Z = Q \pm \frac{i}{2}h\big( u_2^\pm -x_1-\frac{1}{x_1}\big)$.}
\begin{equation}
i\log\frac{\Gamma[1+\frac{ih}{2}(x_1+\tfrac{1}{x_1}-w-\tfrac{1}{w})]}
{\Gamma[1-\frac{ih}{2}(x_1+\tfrac{1}{x_1}-w-\tfrac{1}{w})]}
=
\frac{h}{2 x_1} \left[ 2 + \log \Big( \frac{4x_1^2}{h^2} \Big) \right]
+ \text{regular} ,
\label{psi-integrand-strongly-weak}
\end{equation}
so that the kernel $K^{0 Q_2}$ can diverge as $1/(x_1)^2$ in the vicinity of $x_1=0$ for $|x_1|\ll h\ll1$.\footnote{The $\Psi$-integrand is regular at small $h$ outside this region.}
Strictly speaking, here we should expand the integral rather than the integrand, which will be discussed in Appendix \ref{app:BES phase}.
The $\log \Gamma$ function behaves in the regime $\abs{x_1} \gg 1/h \gg 1$ as
\begin{multline}
i \log \frac{ \Gamma\big[-\frac{i}{2}h\big(x_2^\pm +\frac{1}{x_2^\pm}-x_1-\frac{1}{x_1}\big)\big]}
{\Gamma\big[ Q+\frac{i}{2}h\big(x_2^\pm +\frac{1}{x_2^\pm}-x_1-\frac{1}{x_1}\big) \big]}
=
\frac{h }{2 x_1} \left[ 2 + \log \left(\frac{4 x_1^2}{h^2 }\right) \right]
\\
- \frac{1}{2} \left\{ 
\left(\tilde{u}_2 - i Q \pm i Q \right) \log \left(\frac{4 x_1^2}{h^2 }\right) 
+ \pi  (Q-1) \, \text{sgn}\left(x_1\right)
\right\}
+ \frac{x_1}{3h}
+ \cO(h^2).
\label{log-gamma-strongly-weak}
\end{multline}
where $x_2^\pm = x ( (\tilde u_2 \pm i Q)/h )$.
Although the kernel diverges, the convolution integrals
\begin{equation}
\log{(1+Y_0)^2} \, \check{\star} \, {K}^{0Q} 
\qquad {\rm and} \qquad
\log{(1+Y_0)^2} \, \check{\star} \,  \tilde{K}^{0Q} \,.
\end{equation}
remain finite because $Y_0 (u) \to 0$ as $u \to 0$. See the discussion in Section \ref{sec:smalltension-massive}.

In the string-mirror kinematics, we first need to regularise the integrals as the massless particle lies on the unit circle. We can do this by shrinking a bit the contour as discussed, which is tantamount to introducing a principal value prescription and adding suitable $\Psi$-functions and $\log\Gamma$ functions to~\eqref{mass^2 mir^2 BES}.
 Regardless of the detail, it is clear that the principal value integrals in $\Sigma_{\bes}^{0Q}(x,y^\pm)$ are regular, as are the $\log\Gamma$ functions.\footnote{Strictly speaking, some terms in $\log \Sigma_{\bes}^{0_* Q_2}(x_1, x_2^\pm)$ acquire a large imaginary part in the regime $u_2 \gg 1/h \gg 1$. Such a behaviour is not important, because the $Y_Q(u)$ functions are suppressed by the driving term $e^{- L\tilde{\mathcal{E}}^Q} \sim u^{-2L}$ at large $u$.}
 A more detailed discussion of the singularities of the BES phase in various regions can be found in appendix~\ref{app:BES phase}.

The Sine-Gordon factors appear in the S-matrix elements as the product
\begin{equation}
\label{eq:mixedmassSG}
    \PhiSG(\gamma_1-\gamma^+_2)\,\PhiSG(\gamma_1-\gamma^-_2)
    =e^{\phiSG(\gamma_1-\gamma^+_2)}e^{\phiSG(\gamma_1-\gamma^-_2)}\,,
\end{equation}
and the kernels are defined as usual. Recalling that
\begin{equation}
    \PhiSG(\gamma)\,\PhiSG(\gamma+i\pi)=i\tanh\frac{\gamma}{2}\,,\qquad
    \PhiSG(\gamma)\,\PhiSG(\gamma-i\pi)=i\coth\frac{\gamma}{2}\,,
\label{phiSG-monodromy}
\end{equation}
we can recast~\eqref{eq:mixedmassSG} so that the argument takes value in the physical strip $(0,i\pi)$. 
\begin{align}
&\phiSG(\gamma_1-\gamma^+_2) + \phiSG(\gamma_1-\gamma^-_2)
\label{phiSG-monodromy-2} \\
&=
- \phiSG(\gamma_1-\gamma^+_2 - i \pi ) - \phiSG(\gamma_1-\gamma^-_2 + i \pi)
+ \log \( - \tanh \frac{\gamma_1-\gamma^+_2 - i \pi}{2} \, \coth \frac{\gamma_1-\gamma^-_2 + i \pi}{2} \) .
\notag
\end{align}
By explicitly evaluating it using \eqref{eq:gamma-pm-expansion2}, we find that the kernels are finite as $h\to0$.

The driving terms in the massive equations are $S^{0_* Q} = 1/S^{Q 0_*}$ and $\tilde{S}^{0_* Q} = 1/\tilde{S}^{Q 0_*}$\,, where the latter is given by the analytic continuation of \eqref{def:SQa0} and \eqref{def:barS bQa0},
\begin{equation}
\begin{aligned}
S^{Q 0_*}(u_2, u_1 ) &= i e^{-\frac{i}{2}p_2} \frac{x^+ (u_2) x_s (u_1) -1}{x^- (u_2) - x_s (u_1) } 
\frac{(\Sigma_{\bes}^{Q 0_*}(x_2^\pm, x_1 ))^{-2}}{\PhiSG (\gamma_{21_*}^{+\circ}) \PhiSG (\gamma_{21_*}^{-\circ})} \,,
\\
\tilde{S}^{\bar{Q} 0_*} (u_2, u_1) &= i e^{+\frac{i}{2}p_2} \frac{x^- (u_2) - x_s (u_1)}{x^+ (u_2) x_s (u_1) -1} 
\frac{(\Sigma_{\bes}^{\bar{Q} 0_*}(x_2^\pm, x_1))^{-2}}{\PhiSG (\gamma_{21_*}^{+\circ}) \PhiSG (\gamma_{21_*}^{-\circ})} \,.
\end{aligned}
\end{equation}
The rational functions of Zhukovsky variables are regular at small $h$. As for $S^{Q 0_*}(u_2, u_1 )$ we find
\begin{equation}
i e^{-\frac{i}{2}p_2} \frac{x^+ (u_2) x_s (u_1) -1}{x^- (u_2) - x_s (u_1) } = 
\frac{(\tilde{u}_1-\tilde{u}_2-i Q) \sqrt{Q+i \tilde{u}_2}}{(\tilde{u}_1-\tilde{u}_2+i Q) \sqrt{Q-i \tilde{u}_2}}
+ \cO(h^2)
\end{equation}
and similarly for $\tilde{S}^{\bar{Q} 0_*}$.
The sine-Gordon factors are given by \eqref{eq:mixedmassSG} analytically continued with $\gamma_1$ in the string region.
Recall that the explicit expression of $\varphi (\gamma)$ in \eqref{eq:varphiexplicit} can be used when ${\rm Im} \, \gamma \in [0, \pi]$.
If we introduce $\gamma_{*1} = \gamma_1 - i \pi/2$, the quantity \eqref{phiSG-monodromy-2} becomes
\begin{align}
&\phiSG(\gamma_{*1}-\gamma^+_2) + \phiSG(\gamma_{*1}-\gamma^-_2)
\label{phiSG-monodromy-str} \\
&=
\phiSG(\gamma_{*1}-\gamma^+_2 ) + \phiSG(\gamma_{*1}-\gamma^-_2 + 2 \pi i )
+ \log \tanh^2 \( \frac{\gamma_{*1}-\gamma^-_2 + 2 \pi i}{2} \) .
\notag
\end{align}
An explicit evaluation then shows that the $h\to0$ limit is 
\begin{equation}
\phiSG(\gamma_{*1}-\gamma^+_2) + \phiSG(\gamma_{*1}-\gamma^-_2)
= -i \pi -\frac{4 {\bf G}}{\pi }
+ 2 i h \, \frac{Q^2 -2 i Q \tilde{u}_1+\tilde{u}_2 (\tilde{u}_2-\tilde{u}_1)}{\tilde{u}_1 \, (\tilde{u}_2^2+Q^2)} 
+ \cO(h^2),
\end{equation}
where ${\bf G}$ is Catalan's constant. This quantity is regular at $\cO(h^0)$.\footnote{An apparent singularity at $\tilde u_1=0$ disappears if we evaluate $\gamma_{1*}$ without rescaling by $h$. Note that $Y_Q$ remains small even if the phase $S^{0_*Q}$ is non-zero at small $h$.}

In Appendix \ref{app:mixed mass BES}, we will find that the improved dressing phase $\Sigma_{\bes}^{Q 0_*}(x_2^\pm, x_1 )$ diverges at most logarithmically at small $h$. Such a contribution is small compared to the driving term $L \tilde{\mathcal{E}}^Q$ in the TBA.

\paragraph{Renormalised kernels.}

For completeness, consider the renormalised kernels \eqref{KQ0 ren-ev} and \eqref{tKQ0 ren-ev}.
The expansion of the BES kernel has already been given above. The auxiliary kernel $K_{\rm aux}$ behaves as
\begin{equation}
K_{\rm aux} ( \tilde u_1/h , u_2 ) 
= -\frac{2 h^2 Q_1 \tilde{u}_1 \sqrt{u_2^2-4} }{\pi  \left(\tilde{u}_1^2+Q_1^2\right){}^2}
+ \cO(h^3),
\label{weak Kaux}
\end{equation}
which is regular and small for real $\tilde u_1$\,.

\subsection{Massless-massless kernels and S matrices}\label{app:K massless^2}

We now come to the massless modes. Let us begin by considering the BES dressing factor. The idea is similar to what we discussed in the preceding subsection: the integrand that defines the dressing factors is regular as $h\to0$, expect possibly in the vicinity of~$x=0$ on the real mirror line. This does not result in any issue when taking the weak-coupling limit of the TBA equations because the massless Y-functions actually vanish quite fast at~$x=0$.

\subsubsection*{Refined asymptotics around $x=0$.}

We discuss the behaviour of $Y_0(x)$ around $x=0$ by refining our na\"ive estimate \eqref{eq:masslessYatxis0}.
If we denote the $\gamma$-parametrisation of $x$ as $x(\gamma)$, then $x=0$ corresponds to $\gamma=0$.

Let us first discuss the behaviour of $\scr{Y}_0 (\gamma)$ around $\gamma=0$ , where $\scr{Y}_0 (\gamma)$ is the ``asymptotic part'' of the massless TBA equation defined in \eqref{def:scr Y0}.
As shown in \eqref{def:improved massless BES factor}, the massless-massless dressing factor $\Sigma_\bes^{0_*0}(\gamma_{*} , \gamma )$ is given by the massless-massless BES phase. The massless-massless BES phase in the string-mirror region is given in \eqref{massless^2 str-mir BES}, and its asymptotic behaviour near $\gamma_2 = 0$ is given in \eqref{massless^2 str-mir BES asymp}. It follows that
\begin{equation}
\Sigma_\bes^{0_* 0} (\gamma_{*1} , \gamma_2 )^{-2} 
= \exp \( -2 i \theta^{0_* 0} \Big( x_s (\gamma_{*1} ) , x_m (\gamma_2) \Big) \)
= \( \frac{\gamma_2}{h} \)^{2  \cE_0 (x_1)} + O\left(x_2\right) ,
\end{equation}
where we used
\begin{equation}
\gamma_2 = - 2 x_2 + O(x_2^3).
\end{equation}
Then, by using
\begin{equation}
e^{- L\, \tE_{0}(\gamma) } = \( \tanh \abs{\frac{\gamma}{2}} \)^{2L} , \qquad
S (\gamma_{*} , \gamma ) = \frac{1}{i} \, \coth \( \frac{\gamma_* - \gamma}{2} \).
\label{app: exponentially small at gamma=0}
\end{equation}
we find
\begin{equation}
\begin{aligned}
\lim \limits_{\gamma \to 0} \, \scr{Y}_0 (\gamma) &\simeq
\( \tanh \abs{\frac{\gamma}{2}} \)^{2L}
\( \frac{\gamma}{h} \)^{2  \sum_{j=1}^{2M} \cE_0 (\gamma_{*j})} 
\prod_{j=1}^{2M} \frac{1}{i} \, \coth \( \frac{\gamma_{*j} - \gamma}{2} \)
\\
&\simeq \gamma^{2 E_L^{\rm (ex)} } \, \frac{(-1)^{M}}{2^{2L} \, h^{2 E_L^{\rm (ex)} - 2L  } } \, 
\prod_{j=1}^{2M} \coth \( \frac{\gamma_{*j} - \gamma}{2} \) ,
\end{aligned}
\label{scr Y0 asymp}
\end{equation}
where we assumed $\gamma_{*j} \neq 0$ and 
\begin{equation}
E_L^{\rm (ex)} \equiv L + \sum_{j=1}^{2M} \cE_0 (\gamma_{*j}) \ge 0.
\label{def:EL ex}
\end{equation}
This quantity is roughly equal to the asymptotic energy.
Since the magnon energy $\cE_0 (\gamma_{*j})$ is generally positive, the function $\scr{Y}_0 (x)$ at $x=0$ is more strongly suppressed than our na\"ive estimate \eqref{eq:masslessYatxis0}.

\subsubsection*{Convolution with the massless-massless dressing kernel.}

We argue that the convolution with the BES kernel
\begin{equation}
\log (1+Y_0)^2 * \cK_\bes^{00} \,, \qquad
\cK_\bes^{00} (\gamma_1, \gamma_2) = \frac{1}{2 \pi i} \frac{\partial}{\partial \gamma_1} \, 
\log \Sigma^{00}_\bes ( x (\gamma_1), x (\gamma_2)).
\label{convolution with massless BES}
\end{equation}
does not contribute to the equation for massless particles at the leading order of $h \to 0$.

The massless-massless improved dressing factor in the mirror-mirror region is given by \eqref{massless^2 mir^2 improved BES}.
If we take the limit $h \to 0$ with $\gamma_1$ fixed, we get
\begin{align}
\cK_\bes^{00} (\gamma_1, \gamma_2) =
\frac{h^3 \zeta_3 }{2 \pi} \, 
\frac{\cosh \left(2 \gamma _1\right)-\sinh \left(2   \gamma _1\right) \coth \left(\gamma _2\right)+3}{\sinh^3 \gamma_1 \, \sinh \gamma_2}
+ O(h^5),
\end{align}
which is small unless $\gamma_1$ or $\gamma_2 = O(h)$.
If we take the limit $\gamma_1 \to 0$ with $h$ fixed, we get
\begin{align}
\cK_\bes^{00} (\gamma_1, \gamma_2) =
-\frac{h \left(x_2^2-1\right)}{4 \pi  x_1 x_2} + O(1) 
= -\frac{h}{\pi \gamma_1 \, \sinh \gamma_2} + O(1) ,
\label{KBES00 around origin}
\end{align}
which is potentially singular.

Now consider the convolution integral \eqref{convolution with massless BES} over a small interval $[0, +\delta]$ with $\delta \sim \cO(h) \ll 1$. We approximate $\log (1+Y_0)^2 * \cK_\bes^{00}$ by $\scr{Y}_0 \star \cK_\bes^{00}$\,, and evaluate the integral $\scr{Y}_0 \star \cK_\bes^{00}$ over the interval $[0, \delta]$ as
\begin{equation}
\begin{aligned}
\int_{0}^\delta \de \gamma \, \scr{Y}_0 (\gamma) \, \cK_\bes^{00} (\gamma, \gamma_2) 
&\simeq C (\gamma_2) \int_{0}^\delta \de \gamma \ 
\gamma^{2 E_L^{\rm (ex)} } \, h^{2L - 2 E_L^{\rm (ex)} } \, \frac{h}{\gamma}
+ \dots
\\
&= C (\gamma_2)  \, \frac{\delta^{2 E_L^{\rm (ex)}} \, h^{2L - 2 E_L^{\rm (ex)} + 1} }{2  E_L^{\rm (ex)}}
+ \dots
\\
&= C (\gamma_2)  \, \frac{\tilde \delta^{2 E_L^{\rm (ex)}} \, h^{2L + 1} }{2  E_L^{\rm (ex)}}
+ \dots
\end{aligned}
\label{dressing convolution delta}
\end{equation}
for some function $C(\gamma_2)$.
Here we used $\tilde \delta \equiv \delta/h$ in the second line because $\delta$ is small.
When $L \ge 0$, this quantity should vanish when $h \to 0$ for a fixed $\tilde \delta$, which can also be checked numerically.\footnote{The function $C(\gamma_2)$ has a logarithmic divergence around $\gamma_2=0$. However, this does not significantly alter the behaviour of $Y_0 (\gamma)$ around $\gamma = 0$ in TBA.}

\subsubsection*{Renormalised kernel and S-matrix.}

When $h$ is small, the BES term drops off from the convolution with $\cK_{\rm ren}^{00}$.
From Appendix \ref{app:massless^2 Kren} we find
\begin{equation}
\mathcal{K}^{00}_{\text{ren}}(\gamma,\gamma')
=s(\gamma-\gamma')+\frac{1}{2}\delta(\gamma)+\mathcal{O}(h)\,,
\end{equation}
Furthermore, up to choosing an appropriate way of taking the branches of the logarithm, we have
\begin{equation}
S^{00}_{\text{ren}}(\gamma,\gamma')=S(\gamma-\gamma')+\mathcal{O}(h)\,.
\end{equation}
The analytic continuation of the renormalised S-matrix into the string region is
\begin{equation}
S^{0_*0}_{\text{ren}}(\gamma,\gamma')=-\frac{1}{S_*(\gamma-\gamma')}+\mathcal{O}(h)\,,
\end{equation}
where \eqref{eq:CauchyKernel-shifted} is used.

\subsection{Massless-auxiliary kernels and S matrices}\label{app:K aux}

In \eqref{comp: S0y Sy0} we expressed $S^{0y}$ and $S^{y0}$ as
\begin{equation}
S^{0y}(x(\gamma),x_s(\gamma')) = \frac{-i\,\text{sgn}(\gamma)\,}{S(\gamma-\gamma')}\,,\qquad
S^{y0}(x_s(\gamma),x(\gamma')) = \frac{+i\,\text{sgn}(\gamma')}{S(\gamma-\gamma')}\,,
\end{equation}
Their analytic continuation can be expressed by $S_*$ in \eqref{eq:CauchyKernel-shifted} as
\begin{equation}
S^{0_*y}(x_s(\gamma),x_s(\gamma')) =i\,\text{sgn}(\gamma)\,S_*(\gamma-\gamma')\,,\qquad 
S^{y0_*}(x_s(\gamma),x_s(\gamma')) =\frac{i\,\text{sgn}(\gamma')\,}{S_*(\gamma-\gamma')}\,, 
\end{equation}
so that
\begin{equation}
    \mathcal{K}^{0_*y}(\gamma,\gamma')=+s_*(\gamma-\gamma')+\frac{1}{2}\delta(\gamma)\,,\qquad
    \mathcal{K}^{y0_*}(\gamma,\gamma')=-s_*(\gamma-\gamma')\,.
\end{equation}
The delta function can be altogether avoided if we defined the kernel as coming from $\tfrac{\de}{\de\gamma}\log S(\gamma)^2$. We will see that in any case it will not play any role in the TBA equations.

\subsection{Kernel and S-matrix for the exact Bethe equation}\label{app:K exactBethe}

The mirror-string BES kernel $\cK_\bes^{00_*}$ is given by the derivative of \eqref{massless^2 mir-str BES-2}.
If we take the limit $h \to 0$ with $\gamma_1$ fixed, we get
\begin{align}
\cK_\bes^{00_*} (\gamma_1, \gamma_2) =
\frac{i h^3 \zeta_3 }{2 \pi } \,
\frac{\cosh \left(2 \gamma _1-\gamma _2\right) +3 \cosh \gamma_2}{\sinh^3 \gamma_1 \, \cosh^3 \gamma_2}
+ O(h^5),
\end{align}
which is small unless $\gamma_1 = O(h)$.
If we take the limit $\gamma_1 \to 0$ with $h$ fixed, we get
\begin{align}
\cK_\bes^{00_*} (\gamma_1, \gamma_2) = 
\frac{i h}{2 \pi  x_1} \, \frac{x_2^2-1}{x_2^2+1} + O(1)
= \frac{i h }{\pi  \gamma _1 \, \cosh(\gamma_2) } + O(1) .
\label{KBES00* around origin}
\end{align}
This kernel has the same degree of divergence as \eqref{KBES00 around origin}. Thus, we can safely neglect the convolution $\log{(1+Y_0)^2} \star \cK_\bes^{00_*}$ in the $h \to 0$ limit.

The string-string BES phase $\theta_\bes^{0_*0_*}$ is given in \eqref{massless^2 str^2 BES-2}.
If we take the limit $h \to 0$ with $\gamma_1$ fixed, we get
\begin{align}
&\theta_\bes^{0_*0_*} (\gamma_1, \gamma_2) =
\frac{-i h^3 \zeta (3)}{4 \, \cosh^2 ( \frac{\gamma _1}{2} ) \, \cosh^2 \gamma_1 \, \cosh^2 \gamma_2 } \,
\Biggl( \cosh \gamma_1 \Big[
-4 \sinh \gamma _1 \sinh \gamma _2
\notag \\
&\quad
+\left(1+i \sinh \gamma _1 + \cosh \gamma_1 \right) \left(\cosh (2 \gamma _2)-3\right) \Big]
\\
&\quad
+4 \cosh \gamma _2 \Big[ 
\sinh ^2 \gamma _1 -\cosh \gamma _1 -
\sinh \gamma _2 \left(\cosh \gamma _1 +i \sinh \gamma _1 \right)
\left(\sinh (\gamma _1) -i\right) 
\Big]
\Biggr)
+ O(h^5)
\notag
\end{align}
which is small for any $\gamma_1 \in \bb{R}$. Furthermore, the string-string BES phase is regular and small in the limit $\gamma_1 \to 0$ with $h$ fixed. Thus, it does not contribute at the leading order of small $h$.

Again we find that the renormalised kernel \eqref{CK00 ren-ev} is equal to the Cauchy kernel in the $h \to 0$ limit.
By analytic continuation, we have
\begin{equation}
S^{00_*}_{\text{ren}}(\gamma,\gamma')=+S_*(\gamma-\gamma')+\mathcal{O}(h)\,,\qquad
S^{0_*0_*}_{\text{ren}}(\gamma,\gamma')=S(\gamma-\gamma')+\mathcal{O}(h)\,,
\end{equation}
and
\begin{equation}
\mathcal{K}^{00_*}(\gamma,\gamma')=s_*(\gamma-\gamma')+\mathcal{O}(h)\,.
\end{equation}

\section{BES phase}\label{app:BES phase}

\subsection{Definitions}\label{app:BES phase definitions}

We introduce
\begin{equation}
\Sigma_\bes^{QQ'} = \sigma_\bes^{QQ'}\ 
\prod_{j=1}^Q\prod_{k=1}^{Q'} \frac{1-\frac{1}{ x_j^+ z_k^-}}{1-\frac{1}{x^-_j z^+_k}} \,.
\end{equation}
When $Q=0$, we use the notation \cite{Frolov:2021zyc,Frolov:2021bwp}
\begin{equation}
\Sigma_\bes^{0Q'}(u,u') =
\sigma_\bes^{0Q'}(u,u')\prod_{j=1}^{Q'}
\frac{\frac{1}{x}-x^-_j}{x-x^-_j}\,, \qquad
\Sigma_\bes^{00}(u,u') =
\sigma_\bes^{00}(u,u')\,.
\label{def:improved massless BES factor}
\end{equation}

The BES factor for the massive case is defined by
\begin{equation}
\BES(x^\pm_1,x^\pm_2) = e^{i\theta(x^\pm_1,x^\pm_2)}\,,
\end{equation}
where
\begin{equation}
\label{eq:thetaBES}
\theta(x_1^+,x_1^-,x_2^+,x_2^-) =\chi(x_1^+,x_2^+)
-\chi(x_1^+,x_2^-)-\chi(x_1^-,x_2^+)+\chi(x_1^-,x_2^-)\,.
\end{equation}
For $|x_1|>1$ and $|x_2|>1$ the function $\chi(x_1,x_2)$ is given by
\begin{equation}
\chi(x_1,x_2)=\Phi(x_1,x_2)\,, \qquad
|x_1|>1\,,\quad |x_2|>1 \,.
\label{eq:chiBES}
\end{equation}
We define the $\Phi$-, $\Psi$- and $\Omega$-functions by
\begin{align}
\Phi(x_1,x_2)&=\oint\frac{{\rm d}w_1}{2\pi i}\oint \frac{{\rm
d}w_2}{2\pi i}\frac{\Omega(w_1,w_2)}{(w_1-x_1)(w_2-x_2)} \,,
\label{def:Phi orig} \\
\Psi(x_1,x_2)&=\oint\frac{{\rm d
}w}{2\pi i} \frac{\Omega(x_1,w)}{w-x_2}\,,
\label{def:Psi orig} \\
\Omega (x_1, x_2) &= 
i \, \log \frac{\Gamma\big[1+\frac{i}{2}h \big(x_1 + \frac{1}{x_1} - x_2 - \frac{1}{x_2} \big)\big]}
{\Gamma\big[1-\frac{i}{2}h \big(x_1 + \frac{1}{x_1} - x_2 - \frac{1}{x_2} \big)\big]} \,,
\label{def:Omega x}
\end{align}
where the integration is over the unit circle.

\subsection{Basic properties}\label{app:BES basic}

\paragraph{$\Phi$-function.}

We have the identities
\begin{align}
\lim_{\epsilon\to 0^+} \Phi(e^{\epsilon} x_1,x_2) - \Phi(e^{-\epsilon} x_1,x_2)
&= - \Psi (x_1,x_2 ), \qquad ( \abs{x_1} = 1 ),
\label{Phi to Psi} \\
\lim_{\epsilon\to 0^+} \Phi(x_1 , e^{\epsilon} x_2 ) - \Phi(x_1, e^{-\epsilon} x_2)
&= + \Psi ( x_2 , x_1 ), \qquad ( \abs{x_2} = 1 ).
\label{Phi to Psi2} 
\end{align}
If $\abs{x_1} \,, \abs{x_2} \neq 1$, we find
\begin{equation}
\Phi (x_1, x_2) + \Phi \Big( \frac{1}{x_1} , x_2 \Big) = \Phi (0, x_2) , \qquad 
\Phi (x_1, x_2) + \Phi \Big( x_1, \frac{1}{x_2} \Big) = \Phi (x_1, 0) .
\label{identities Phi0}
\end{equation}
As a corollary,
\begin{equation}
\begin{aligned}
&\Phi (x_1, x_2) - \Phi \Big( x_1, \frac{1}{x_2} \Big) 
- \Phi \Big( \frac{1}{x_1} \,, x_2 \Big) + \Phi \Big( \frac{1}{x_1}  , \frac{1}{x_2} \Big)
\\[1mm]
&\quad = 2 \Phi (x_1, x_2) - \Phi (x_1,0) 
- 2 \Phi \Big( \frac{1}{x_1} \,, x_2 \Big) + \Phi \Big( \frac{1}{x_1} \,, 0 \Big)
\\[1mm]
&\quad = 4 \Phi (x_1, x_2) - 2 \Phi ( 0, x_2) - 2 \Phi (x_1,0) ,
\end{aligned}
\label{identities Phi-comb}
\end{equation}
where we used
\begin{equation}
\Phi(x_1,x_2)= -\Phi(x_2,x_1), \qquad
\Phi(0,0)=0.
\label{identity Phi00}
\end{equation}

\paragraph{$\Psi$-function.}

We have the identity
\begin{equation}
\lim_{\epsilon\to 0^+} \Psi( x_1 , e^{\epsilon} x_2 ) - \Psi(x_1, e^{-\epsilon} x_2 )
= - \Omega (x_1, x_2) \qquad (\abs{x_2} = 1 ) .
\label{Psi to logG}
\end{equation}
If $\abs{x_2} \neq 1$, we find
\begin{equation}
\Psi (x_1, x_2) + \Psi \Big( x_1, \frac{1}{x_2} \Big) = \Psi (x_1, 0) , \qquad
\Psi (x_1, x_2) = \Psi \Big( \frac{1}{x_1} , x_2 \Big).
\label{identities Psi}
\end{equation}

\paragraph{$\Omega$-function.}

We find
\begin{equation}
\Omega (x_1, x_2) = - \Omega (x_2, x_1), \qquad
\Omega (x_1, x_2) = \Big( \frac{1}{x_1} , x_2 \Big) 
= \Omega \Big( x_1, \frac{1}{x_2} \Big) = \Big( \frac{1}{x_1} , \frac{1}{x_2} \Big) .
\label{identities Omega}
\end{equation}
When $x_1^\pm$ is the variable of a fundamental particle, we get
\begin{equation}
\Omega (x_1^+, x_2) - \Omega (x_1^-, x_2) = i \log \( \frac{4}{h^2} \, 
\frac{1}{x_1 + \frac{1}{x_1} - x_2 - \frac{1}{x_2} + \frac{i}{h}} \,
\frac{1}{x_1 + \frac{1}{x_1} - x_2 - \frac{1}{x_2} - \frac{i}{h}} \)
\end{equation}

\paragraph{Branch cuts.}
If we analytically continue the $\Phi$-function inside the unit circle, we encounter branch points due to the singularities of the $\Psi$ function. As discussed in detail in~\cite{Arutyunov:2009kf} they arise when $w$ is such that
\begin{equation}
x_1 + \frac{1}{x_1} - w- \frac{1}{w} = \frac{2ik}{h} \,, \qquad (k \neq 0, k \in \bb{Z}).
\label{infinite genus eq}
\end{equation}
suggesting that the BES phase is defined over an infinite-genus Riemann surface \cite{Beisert:2006ib}.

\subsection{Analytic continuation}\label{app:analytic continuation}

The function $\Phi (x_1, x_2)$ is discontinuous when $x_1$ crosses the unit circle, as in \eqref{Phi to Psi}.
When $h$ is finite, we can analytically continue the $\Phi$ function around $\abs{x_1} = 1$ by shrinking the integration contour a little bit inside the unit circle.\footnote{We cannot apply this argument when $h=\infty$, because the contour is pinched by the branch points.}
In other words, since the $\Phi$ function is not analytic around the unit circle, we {\it define} a new function $\tilde \Phi$ which is analytic around the unit circle,\footnote{Our discussion is essentially the same as Section 3.3 of \cite{Arutyunov:2009kf}, where they used $\chi$ instead of $\tilde \Phi$. In order to check the sign, use $\Phi_\text{(large contour)} - \Phi_\text{(small contour)} = + \Psi$.}
\begin{equation}
\tilde \Phi (x_1, x_2) \equiv
\begin{cases}
\Phi(x_1, x_2) &\qquad (\abs{x_1} > 1, \abs{x_2} > 1) \\
\Phi_\pv (x_1, x_2) - \frac12 \, \Psi (x_1, x_2) &\qquad (\abs{x_1} = 1, \abs{x_2} > 1) \\
\Phi(x_1, x_2) - \Psi (x_1, x_2) &\qquad (\abs{x_1} < 1, \abs{x_2} > 1) .
\end{cases}
\label{def:tilde Phi-0}
\end{equation}
Next, we move $x_2$ inside the unit circle using \eqref{Phi to Psi2}.
For this purpose, we define another function $\tilde \Psi$ by\footnote{When $\abs{x_2}=1$, the principal value prescription is applied.} 
\begin{equation}
\tilde \Psi (x_1, x_2) \equiv
\begin{cases}
\Psi(x_1, x_2) &\qquad (\abs{x_2} > 1) \\
\Psi(x_1, x_2) - \Omega (x_1, x_2) &\qquad (\abs{x_2} < 1) 
\end{cases}
\label{def:tilde Psi}
\end{equation}
The function $\tilde \Phi$ can be analytically continued as
\begin{equation}
\tilde \Phi (x_1, x_2) \equiv
\begin{cases}
\Phi(x_1, x_2) &\qquad (\abs{x_1} > 1, \abs{x_2} > 1) \\
\Phi(x_1, x_2) - \Psi (x_1, x_2) &\qquad (\abs{x_1} < 1, \abs{x_2} > 1) \\
\Phi(x_1, x_2) + \Psi (x_2, x_1) &\qquad (\abs{x_1} > 1, \abs{x_2} < 1) \\
\Phi(x_1, x_2) - \Psi (x_1, x_2) + \Psi (x_2, x_1) + \Omega (x_1, x_2) &\qquad (\abs{x_1} < 1, \abs{x_2} < 1) .
\end{cases}
\label{def:tilde Phi}
\end{equation}
This result can be derived in two different ways; performing the analytic continuation of $x_1$ first and $x_2$ second, or $x_2$ first and $x_1$ second. The two procedures give the same result thanks to $\Omega (x_1, x_2) = - \Omega (x_2, x_1)$.

\bigskip
Let us rephrase the difference between $\Phi (x_1, x_2)$ and $\tilde \Phi (x_1, x_2)$.
The two functions agree if $\abs{x_1}>1$ and $\abs{x_2}>1$.
The function $\Phi (x_1, x_2)$ is defined by the integral expression \eqref{eq:chiBES} for any $x_1\,, x_2$\,, and is discontinuous. The function $\tilde \Phi (x_1, x_2)$ is defined by the analytic continuation of $\Phi (x_1, x_2)$ through a suitable path which avoids all the branch-cuts of $\Psi(x_1,x_2)$, and has the following integral representation
\begin{equation}
\tilde \Phi (x_1, x_2) = \oint\frac{{\rm d }w_1}{2\pi i} \oint\frac{{\rm d }w_2}{2\pi i} \, 
\frac{\Omega (w_1, w_2) - \Omega(x_1, w_2) - \Omega(w_1,x_2) + \Omega (x_1, x_2)}{(w_1-x_1)(w_2-x_2)} \,,
\label{def:tilde Phi integral}
\end{equation}
which is analytic across $\abs{x_1}=1$ and $\abs{x_2}=1$.
 In fact, the function is analytic in an annulus inside the unit circle, until the point where one encounters the branch points described in~\eqref{infinite genus eq}.

\subsection{Partial regularisation}\label{app:partial reg}

\subsubsection*{$\Psi$-function}

The function $\Psi(x_1, x_2)$ is discontinuous when $x_2$ crosses the unit circle, which can be seen from
\begin{equation}
\Psi (x_1, x_2) = \oint\frac{{\rm d }w}{2\pi i} \frac{\Omega(x_1, w) - \Omega(x_1, x_1)}{w-x_2} 
 + \oint \frac{{\rm d }w}{2\pi i} \frac{\Omega(x_1, x_2)}{w-x_2} \,,
\end{equation}
where the second term causes the discontinuity. Let us define a new function which depends on the regularisation as
\begin{equation}
\begin{aligned}
\Psi_{\epsilon_2} (x_1, x_2) &= \Psi_{\rm reg} (x_1, x_2) + \epsilon_2 \, \Omega(x_1, x_2) ,
\\[1mm]
\Psi_{\rm reg} (x_1, x_2) &= \oint\frac{{\rm d }w}{2\pi i} \frac{\Omega(x_1, w) - \Omega(x_1, x_2)}{w-x_2}  \,,
\end{aligned}
\label{def:Psi-epsilon} 
\end{equation}
where $\epsilon_2=1$ is $x_2$ is inside the unit circle, and $\epsilon_2=0$ if outside.

\subsubsection*{$\Phi$-function}

Consider the case where only $x_2$ lies around the unit circle. We write
\begin{equation}
\Phi (x_1, x_2) = \oint\frac{{\rm d }w_1}{2\pi i} \oint\frac{{\rm d }w_2}{2\pi i}  \frac{\Omega (w_1, w_2) - \Omega (w_1, x_2)}{(w_1-x_1)(w_2-x_2)} 
+ \oint\frac{{\rm d }w_1}{2\pi i} \oint\frac{{\rm d }w_2}{2\pi i}  \frac{\Omega (w_1, x_2)}{(w_1-x_1)(w_2-x_2)} \,.
\end{equation}
where the first term is analytic around $\abs{x_2}=1$, and the second term is proportional to $\Psi (x_2, x_1)$. Note that
$\Psi (x_2, x_1)$ is analytic around $\abs{x_2}=1$.
Let us define a new function which depends on the regularisation as
\begin{equation}
\begin{aligned}
\Phi_{\epsilon_2} (x_1, x_2) &= \Phi_{\rm reg} (x_1, x_2) - \epsilon_2 \, \Psi (x_2, x_1),
\\[1mm]
\Phi_{\rm reg} (x_1, x_2) &= \oint\frac{{\rm d }w_1}{2\pi i} \oint\frac{{\rm d }w_2}{2\pi i}  \frac{\Omega (w_1, w_2) - \Omega (w_1, x_2)}{(w_1-x_1)(w_2-x_2)}   \,,
\end{aligned}
\label{def:Phi-epsilon} 
\end{equation}
where $\epsilon_2=1$ is $x_2$ is inside the unit circle, and $\epsilon_2=0$ if outside.

If only $x_1$ lies around the unit circle, we use $\Phi(x_1, x_2) = - \Phi(x_2, x_1)$ to obtain
\begin{equation}
\Phi_{\epsilon_1} (x_1, x_2) = - \Phi_{\rm reg} (x_2, x_1) + \epsilon_1 \, \Psi (x_1, x_2),
\label{def:Phi-epsilon1}
\end{equation}
where $\epsilon_1=1$ is $x_1$ is inside the unit circle, and $\epsilon_1=0$ if outside.
Note that the function $\Phi_{\rm reg} (x_1, x_2)$ is not anti-symmetric.

\bigskip
Consider the case where both $x_1$ and $x_2$ lie on the unit circle. We write
\begin{multline}
\Phi (x_1, x_2) = \oint\frac{{\rm d }w_1}{2\pi i} \oint\frac{{\rm d }w_2}{2\pi i} \, \Biggl[
\frac{\Omega (w_1, w_2) - \Omega (w_1, x_2) - \Omega (x_1, w_2) + \Omega (x_1, x_2)}{(w_1-x_1)(w_2-x_2)} 
\\
+ \frac{\Omega (w_1, x_2) - \Omega (x_1, x_2)}{(w_1-x_1)(w_2-x_2)} 
+ \frac{\Omega (x_1, w_2) - \Omega (x_1, x_2)}{(w_1-x_1)(w_2-x_2)} 
+  \frac{\Omega (x_1, x_2)}{(w_1-x_1)(w_2-x_2)} \Biggr] .
\end{multline}
where the second line causes the discontinuity. Let us define a new regularised function
\begin{equation}
\begin{aligned}
\Phi_{\epsilon_1 , \epsilon_2} (x_1, x_2) &= \Phi_{\rm reg, reg} (x_1, x_2) 
- \epsilon_2 \, \Psi_{\rm reg} (x_2, x_1) 
+ \epsilon_1 \, \Psi_{\rm reg} (x_2, x_1)
+ \epsilon_1 \, \epsilon_2 \, \Omega (x_2, x_1),
\\[2mm]
\Phi_{\rm reg,reg} (x_1, x_2) &= \oint\frac{{\rm d }w_1}{2\pi i} \oint\frac{{\rm d }w_2}{2\pi i} \, 
\frac{\Omega (w_1, w_2) - \Omega (w_1, x_2) - \Omega (x_1, w_2) + \Omega (x_1, x_2)}{(w_1-x_1)(w_2-x_2)} 
\end{aligned}
\label{def:Phi-epsilon12}
\end{equation}
where $\epsilon_k =1$ is $x_k$ is inside the unit circle, and $\epsilon_k=0$ if outside.

\subsection{Integration over the \texorpdfstring{$\gamma$}{gamma}-rapidity}\label{app:BES gamma rap}

The functions appearing in the BES phase is defined as an integral over the unit circle. We rewrite this integral by introducing $w = x_s (\theta)$ as 
\begin{equation}
\oint dw f(w) 
= i \int_{-\infty}^\infty \frac{d \theta}{\cosh \theta} \Biggl\{ x_s (\theta) f ( x_s (\theta) ) 
+ \frac{f ( 1/x_s (\theta) )}{x_s (\theta)}   \Biggr\} 
\equiv \int_{-\infty}^\infty d \theta \, {\bf f} (\theta),
\end{equation}
where $x_s(\theta)$ is defined in \eqref{def:xspm gamma_s}.
In terms of $\theta$, the $\Omega$-function \eqref{def:Omega x} becomes
\begin{equation}
{\bf \Omega} (\theta_1, \theta_2) = 
i \log \Gamma \Big( 1-i h \frac{\sinh (\theta_1 - \theta_2)}{\cosh \theta_1 \cosh \theta_2} \Big)
- i \log \Gamma \Big( 1 + i h \frac{\sinh (\theta_1 - \theta_2)}{\cosh \theta_1 \cosh \theta_2} \Big) \,.
\end{equation}

When both particles are massless, the $\Phi$-function in the mirror-mirror and string-mirror regions become
\begin{align}
{\bf \Phi}_{mm}^{\circ\circ} (\gamma_1, \gamma_2) &= 
\int_{-\infty}^\infty d \theta_1 \, \int_{-\infty}^\infty d \theta_2 \, \frac{1}{4 \pi^2} \ \times
\\
&\qquad
\frac{\left(\cosh (\gamma _1-\theta _1)+\cosh \theta _1 \right) 
\left(\cosh (\gamma _2-\theta _2)+\cosh \theta _2 \right)}
{\cosh \theta_1 \, \cosh (\gamma _1-\theta _1) \, \cosh \theta_2 \, \cosh ( \gamma _2-\theta _2 )} \,
{\bf \Omega} (\theta_1, \theta_2) ,
\notag \\[1mm]
{\bf \Phi}_{sm}^{\circ\circ} (\gamma_1, \gamma_2) &= 
\int_{-\infty}^\infty d \theta_1 \, \int_{-\infty}^\infty d \theta_2 \, \frac{1}{4 \pi^2} \ \times
\\
&\qquad
\frac{\left(\sinh (\gamma _1-\theta _1)+ i \cosh \theta _1 \right) 
\left(\cosh (\gamma _2-\theta _2)+\cosh \theta _2 \right)}
{\cosh \theta_1 \, \sinh (\gamma _1-\theta _1) \, \cosh \theta_2 \, \cosh ( \gamma _2-\theta _2 )} \,
{\bf \Omega} (\theta_1, \theta_2) .
\notag
\end{align}
The $\Psi$-function in the mirror-mirror, string-mirror and mirror-string regions become
\begin{align}
{\bf \Psi}_{mm}^{\circ\circ} (\gamma_1, \gamma_2) &= 
\int_{-\infty}^\infty d \theta \, \frac{1}{2 \pi} \,
\frac{\left(\cosh (\gamma _2-\theta )+\cosh \theta  \right)}{\cosh \theta \, \cosh ( \gamma _2-\theta )} \,
{\bf \Omega} ( \gamma_1 + i \pi/2 , \theta) ,
\notag \\[1mm]
{\bf \Psi}_{sm}^{\circ\circ} (\gamma_1, \gamma_2) &= 
\int_{-\infty}^\infty d \theta \, \frac{1}{2 \pi} \,
\frac{\left(\cosh (\gamma _2-\theta )+\cosh \theta  \right)}{\cosh \theta \, \cosh ( \gamma _2-\theta )} \,
{\bf \Omega} ( \gamma_1, \theta) ,
\\[1mm]
{\bf \Psi}_{ms}^{\circ\circ} (\gamma_1, \gamma_2) &= 
\int_{-\infty}^\infty d \theta \, \frac{1}{2 \pi} \,
\frac{\left(\sinh (\gamma _2-\theta)+ i \cosh \theta \right) }{\cosh \theta \, \sinh ( \gamma _2-\theta )} \,
{\bf \Omega} ( \gamma_1 + i \pi/2, \theta) .
\notag 
\end{align}

For massless-massive kinematics, the $\Phi$-function in the mirror and string regions become\footnote{We use the $\gamma$-rapidity only for the massless particles. The symbol $x$ denotes the Zhukovsky variable of massive particles.}
\begin{align}
{\bf \Phi}_{mx}^{\circ\bullet} (\gamma_1, x_2) &= 
\int_{-\infty}^\infty d \theta_1 \, \int_{-\infty}^\infty d \theta_2 \, \frac{1}{2 \pi^2} \ \times
\\
&\qquad
\frac{\left( \cosh (\theta_1) + \cosh (\gamma _1-\theta _1)\right) 
\left(x_2 \tanh \left(\theta _2\right)+1\right)}{
\cosh (\theta _1) \, \cosh (\theta _2) \, \cosh (\gamma _1-\theta _1) \,
 \left(2 x_2 \tanh \left(\theta _2\right)+x_2^2+1\right)} \,
{\bf \Omega} (\theta_1, \theta_2) ,
\notag \\[1mm]
{\bf \Phi}_{sx}^{\circ\bullet} (\gamma_1, x_2) &= 
\int_{-\infty}^\infty d \theta_1 \, \int_{-\infty}^\infty d \theta_2 \, \frac{i}{2 \pi^2} \ \times
\\
&\qquad
\frac{\left(\cosh \left(\theta _1\right)-i \sinh \left(\gamma _1-\theta_1\right)\right) 
\left(x_2 \tanh \left(\theta_2\right)+1\right) }
{\cosh (\theta _1) \, \cosh (\theta _2) \, \sinh (\gamma _1-\theta _1) \,
\left(2 x_2 \tanh \left(\theta _2\right)+x_2^2+1\right)} \,
{\bf \Omega} (\theta_1, \theta_2) .
\notag
\end{align}
The $\Psi$-function in the mirror-mirror, string-mirror and mirror-string regions become
\begin{align}
{\bf \Psi}_{mx}^{\circ\bullet} (\gamma_1, x_2) &= 
\int_{-\infty}^\infty d \theta \, \frac{1}{\pi} \,
\frac{\left(x_2 \tanh (\theta )+1\right)}{\cosh \theta \, \left(2 x_2  \tanh (\theta )+x_2^2+1\right)} \,
{\bf \Omega} ( \gamma_1 + i \pi/2 , \theta) ,
\\[1mm]
{\bf \Psi}_{xm}^{\bullet\circ} (\gamma_1, x_2) &= 
\int_{-\infty}^\infty d \theta \, \frac{i}{2 \pi i} \,
\left( \frac{1}{\cosh (\theta -\gamma _2)} + \frac{1}{\cosh (\theta )} \right) \ \times
\\
&\hspace{-20mm} \left[ \log \Gamma \left(1-\frac{i  h \left( 1+x_1^2+2 \tanh (\theta ) x_1 \right)}{2 x_1}\right)
-  \log\Gamma \left( 1+ \frac{i h \left( 1+x_1^2+2 \tanh (\theta ) x_1 \right)}{2 x_1} \right) \right ] ,
\notag \\[1mm]
{\bf \Psi}_{sx}^{\circ\bullet} (\gamma_1, x_2) &= 
\int_{-\infty}^\infty d \theta \, \frac{1}{\pi} \,
\frac{\left(x_2 \tanh (\theta )+1\right)}{\cosh \theta \, \left(2 x_2  \tanh (\theta )+x_2^2+1\right)} \,
{\bf \Omega} ( \gamma_1 , \theta) ,
\\[1mm]
{\bf \Psi}_{xs}^{\bullet\circ} (x_1, \gamma_2) &= 
\int_{-\infty}^\infty d \theta \, \frac{1}{2 \pi } \,
\left( \frac{1}{\sinh (\theta -\gamma _2) } + \frac{i}{\cosh (\theta )} \right) \ \times
\\
&\hspace{-20mm} \left[ \log \Gamma \left( 1 + \frac{i h \left( 1+x_1^2+2 \tanh (\theta )  x_1\right)}{2 x_1} \right)
- \log \Gamma \left( 1 - \frac{i h \left(1+x_1^2+2 \tanh (\theta ) x_1\right)}{2 x_1}\right)\right] .
\notag 
\end{align}
Note that some of these functions should be regularised as in Appendix \ref{app:partial reg}.

\subsection{Expansion around the origin}\label{app:expand origin}

Recall that the dressing factor $\log \Sigma (x_1, x_2)$ is a sum over the $\Phi$-, $\Psi$- and $\Omega$-functions.
Some functions diverge when one of the arguments approaches the origin in the $x$-variable.
The divergence must be at most logarithmic, since otherwise the dressing factor $\Sigma (x_1, x_2)$ is not analytic around the origin.
We impose the analyticity of the Y-functions at the origin, and thus $\Sigma (x_1, x_2)$ must be analytic at that point.

We use the $\gamma$-rapidity to inspect the singular behaviour.
The origin in the $x$-variable is equal to the origin in the $\gamma$-rapidity, as can be seen from \eqref{def:tilde-gamma-mirror},
\begin{equation}
x (\gamma) = - \tanh \frac{\gamma}{2} \,, \qquad
\gamma = - 2x - \frac{2}{3} x^3 + O(x^5).
\end{equation}
It turns out that the following functions are potentially dangerous,
\begin{equation}
\begin{aligned}
\Psi_{mm} (\gamma_1, \gamma_2) &= \scr{G} + O(\log \gamma_1)
\\[1mm]
\Omega_{mm} (\gamma_1, \gamma_2) &= - \Omega_{mm} (\gamma_2, \gamma_1) 
= \Omega_{ms} (\gamma_1, \gamma_2) 
= - \Omega_{sm} (\gamma_2, \gamma_1) = \scr{G} + O(\log \gamma_1),
\end{aligned}
\label{app:dangerous functions}
\end{equation}
where 
\begin{equation}
\scr{G} \equiv \frac{2 h}{\gamma _1} \(  \log \abs{ \frac{h}{\gamma_1} } -1 \).
\label{def:dangerous G}
\end{equation}
This expansion is valid for any values of $\gamma_2$ in the mirror region, and all the other functions are at most logarithmic as $\gamma_1 \to 0$.\footnote{The function $\Psi_{ms} (\gamma_1, \gamma_2)$ is also singular, but this singularity disappears if we use $\Psi_{\rm reg} (x_1, x_2)$ in \eqref{def:Psi-epsilon}.}

This remark also applies to the function $\Psi (x_1, 0)$. Since the origin $x=0$ is in the mirror region of the $\gamma$-rapidity, $\Psi (x_1, 0)$ should be regarded as $\Psi_{mm} (x_1, 0)$ or $\Psi_{sm} (x_1, 0)$. Both are potentially dangerous around $x_1=0$.

\subsection{Massive-massive BES}

Following \cite{Arutyunov:2009kf}, the improved BES factor is defined by
\begin{equation}
\Sigma^{QQ'} = \sigma^{QQ'}\,\prod_{j=1}^Q\prod_{k=1}^{Q'} 
\frac{1-\frac{1}{x_j^+ z_k^-}}{1-\frac{1}{x^-_j z^+_k}} \,, 
\end{equation}
where $\theta_{jk} = - i \log \sigma_{jk}$ is called the dressing phase.

\paragraph{String-string region.}

The massive-massive BES phase is given by
\begin{equation}
\theta^{QQ'} (x_1, x_2) =
\Phi(x_1^+,x_2^+) -\Phi(x_1^+,x_2^-)-\Phi(x_1^-,x_2^+)+\Phi(x_1^-,x_2^-) ,
\label{mass^2 str^2 BES}
\end{equation}
where $x_1^\pm \,, x_2^\pm$ represent the bound state of $Q, Q'$ particles, respectively.

\paragraph{String-mirror region.}

The improved BES factor in the string-mirror region is \cite{Frolov:2010wt}
\begin{equation}
\begin{aligned}
\frac{1}{i}\log \Sigma_{1_*Q}(x_1,x_2) &=
\Phi(x_1^+,x_2^+) -\Phi(x_1^+,x_2^-)-\Phi(x_1^-,x_2^+)+\Phi(x_1^-,x_2^-)
\\
&\quad + \frac{1}{2}\Big[ \Psi(x_2^+, x_1^+) + \Psi( x_2^-, x_1^+)
- \Psi( x_2^+, x_1^-) - \Psi( x_2^-, x_1^-)\Big] 
\\
&\quad + \frac{1}{2i} \,
\log \frac{(x_1^- - x_2^+) (x_1^- - 1/x_2^-) (x_1^+ - 1/x_2^-)}{(x_1^+ - x_2^+) ( x_1^- - 1/x_2^+)^2 }
\end{aligned}
\label{mass^2 str-mir BES}
\end{equation}
where $x_1^\pm$ is in the string region, and $x_2^\pm$ is in the mirror region.

\paragraph{Mirror-string region.}

The improved BES factor is given by the unitarity
\begin{equation}
\frac{1}{i}\log \Sigma_{Q1_*}(x_2,x_1) = -\frac{1}{i}\log \Sigma_{1_*Q}(x_1,x_2)\,,
\end{equation}
which should agree with (6.12) of \cite{Arutyunov:2009kf}.

\paragraph{Mirror-mirror region.}

The improved BES factor is given by
\begin{equation}
\begin{aligned}
\frac{1}{i}\log\Sigma^{QQ'}(y_1,y_2)
 &=
\Phi(y_1^+,y_2^+)-\Phi(y_1^+,y_2^-)-\Phi(y_1^-,y_2^+)+\Phi(y_1^-,y_2^-)
\\ &-\frac{1}{2}\left(\Psi(y_1^+,y_2^+)+\Psi(y_1^-,y_2^+)-\Psi(y_1^+,y_2^-)-\Psi(y_1^-,y_2^-)\right) \\
&+\frac{1}{2}\left(\Psi(y_{2}^+,y_1^+)+\Psi(y_{2}^-,y_1^+)-\Psi(y_{2}^+,y_1^-)
-\Psi(y_{2}^-,y_1^-) \right)
 \\
&+\frac{1}{i}\log\frac{ i^{Q}\,\Gamma\big[Q'-\frac{i}{2} h \big(y_1^++\frac{1}{y_1^+}-y_2^+-\frac{1}{y_2^+}\big)\big]} {
i^{Q'}\Gamma\big[Q+\frac{i}{2} h \big(y_1^++\frac{1}{y_1^+}-y_2^+-\frac{1}{y_2^+}\big)\big]}
\frac{1-\frac{1}{y_1^+y_2^-}}{1-\frac{1}{y_1^-y_2^+}}\sqrt{\frac{y_1^+y_2^-}{y_1^-y_2^+}} \,,
\end{aligned}
\label{mass^2 mir^2 BES-2}
\end{equation}
as in \eqref{mass^2 mir^2 BES}.

\subsection{Massless-massless BES}

The massless BES phase in the string region is ambiguous. The phase is given in terms of the contour integrals over the unit circle, but $x^\pm \to (x)^{\pm 1}$ for a massless particle hit the integration contour.

The massless BES phase is {\it defined} as follows \cite{Frolov:2021fmj}. 
Before taking the massless limit for particles in the string region, we should first impose the condition $x^+ x^- = 1$. 
Our convention for the massless limit in the string region is $\abs{x^+}>1$ and $\abs{x^-}<1$.
In the anti-string region we take $\abs{x^+}<1$ and $\abs{x^-}>1$.
As shown in Figure \ref{fig:massless BES reg}, the $x^\pm$ in the string region is identical to the $x^\mp$ in the anti-string region. 
Since the BES phase is anti-symmetric with respect to the interchange of $x_1^+ \leftrightarrow x_1^-$, we find that the crossing relation for the massless BES phase must be trivial.
Given the analytic continuation procedure in Appendix \ref{app:analytic continuation}, it is straightforward to take the massless limit in the string region, which gives the massless BES phase.

The $x$ variable of massless particles in the mirror region stays on the real axis, and generally does not hit the integration contour. Therefore, there is no ambiguity in taking the massless limit. 
However, the BES phase diverges around $x=0$, and the degree of divergence depends on the regularisation scheme.
We will justify our regularisation scheme in Figure \ref{fig:massless BES reg} by studying the asymptotic behaviour of the Y-functions for excited states.\footnote{In \cite{Frolov:2021fmj}, the massless BES phase in the string-string region is compared with the results of the semiclassical string theory on \AdSxT, finding partial agreement. In fact, this comparison is somewhat subtle due to the IR and UV divergences in the perturbative computation.}

\tikzset{cross/.style={draw, thick, cross out, minimum size=2*(#1-\pgflinewidth), inner sep=0pt, outer sep=0pt},
cross/.default={1mm}}
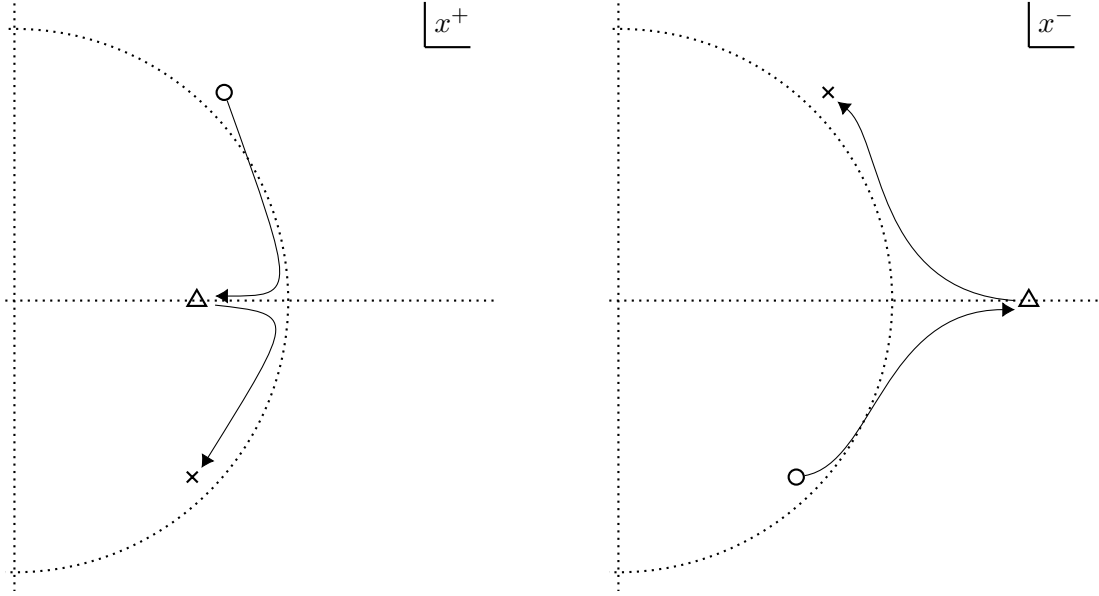
\begin{figure}[t]
\begin{center}
\begin{tikzpicture}[scale=1.2]
\clip (-0.1,-3.2) rectangle (5.31, 3.31);
\draw[-, thick, dotted] (-0.1,0) -- (5.3, 0);
\draw[-, thick,  dotted] (0,-3.3) -- (0, 3.3);
\draw[-, thick] (4.5, 3.3) -- (4.5, 2.8);
\draw[-, thick] (4.5, 2.8) -- (5, 2.8);
\node at (4.8, 3.1) {$x^+$};
\node [circle, draw, thick, dotted, minimum size=205] at (0,0) {};
\node [circle, draw, thick, minimum size=2mm, inner sep=0pt] (outup) at (2.3, 2.3) {};
\node[draw, thick, regular polygon, regular polygon sides=3, inner sep=.5mm] at (2,0) {};
\node [cross, draw] (indown) at (1.95, -1.95) {};
\draw [-{Latex[width=2mm]}] (outup) .. controls (3.1, 0.05) .. (2.2, 0.05);
\draw [-{Latex[width=2mm]}] (2.2, -0.05) .. controls (3.1, -0.15) .. (2.05, -1.85);
\end{tikzpicture}
 \hspace{12mm} 
\begin{tikzpicture}[scale=1.2]
\clip (-0.1,-3.2) rectangle (5.31, 3.31);
\draw[-, thick, dotted] (-0.1,0) -- (5.3, 0);
\draw[-, thick,  dotted] (0,-3.3) -- (0, 3.3);
\draw[-, thick] (4.5, 3.3) -- (4.5, 2.8);
\draw[-, thick] (4.5, 2.8) -- (5, 2.8);
\node at (4.8, 3.1) {$x^-$};
\node [circle, draw, thick, dotted, minimum size=205] at (0,0) {};
\node [cross, draw] (outup) at (2.3, 2.3) {};
\node[draw, thick, regular polygon, regular polygon sides=3, inner sep=.5mm] at (4.5,0) {};
\node [circle, draw, thick, minimum size=2mm, inner sep=0pt] (indown) at (1.95, -1.95) {};
\draw [-{Latex[width=2mm]}] (indown) .. controls (2.8, -1.8) and (2.9, -0.1) .. (4.35, -0.1);
\draw [-{Latex[width=2mm]}] (4.35,0) .. controls (2.8, 0.1) and (2.9,1.8) .. (2.4, 2.2);
\end{tikzpicture}
\end{center}
\caption{The location of $x^\pm = (x)^{\pm 1}$ for the massless particle in the string region ($\circ$), mirror region ($\triangle$) and anti-string region $(\times)$. We need to take the massless limit keeping the condition $x^+ x^- =1$. The location of $x^\pm$ in the string region is identical to the location of $x^\mp$ in the anti-string region. We can analytically continue $x^\pm$ from the string region to the anti-string region through the mirror region without crossing the branch cuts of the BES phase.}
\label{fig:massless BES reg}
\end{figure}

\bigskip
Let us compute the explicit massless-massless BES phase in various kinematical regions.

\paragraph{String-string region.}

As shown in \eqref{mass^2 str^2 BES}, the massive-massive BES phase in the string-string region is given by
\begin{equation}
\theta^{Q_* Q'_*} (x_1, x_2) =
\tilde \Phi(x_1^+,x_2^+) - \tilde \Phi(x_1^+,x_2^-)-\tilde \Phi(x_1^-,x_2^+)+\tilde \Phi(x_1^-,x_2^-) ,
\label{BES tilde Phi}
\end{equation}
where we replaced $\Phi$ with $\tilde \Phi$.
We evaluate $\tilde \Phi$ in the region $\abs{x_k^+}>1, \abs{x_k^-}<1$ for $k=1,2$ using \eqref{def:tilde Phi}.
The result is
\begin{multline}
\theta^{Q_* Q'_*} (x_1, x_2) =
\Phi(x_1^+,x_2^+) 
- \Bigl\{ \Phi(x_1^+, x_2^-) + \Psi (x_2^-, x_1^+) \Bigr\}
- \Bigl\{ \Phi(x_1^-, x_2^+) - \Psi (x_1^-, x_2^+)  \Bigr\}
\\
+ \Phi(x_1^-, x_2^-) - \Psi (x_1^-, x_2^-) + \Psi (x_2^-, x_1^-) + \Omega (x_1^-, x_2^-) .
\end{multline}
Then we take the massless limit $x_k^\pm = (x_k)^{\pm 1}$, which gives
\begin{multline}
\theta^{0_* 0_*} (x_1, x_2) =
\Phi(x_1,x_2) - \Phi(x_1, 1/x_2) - \Phi(1/x_1, x_2) + \Phi(1/x_1, 1/x_2)
\\
- \Psi (1/x_2, x_1) + \Psi (1/x_1, x_2) 
- \Psi (1/x_1, 1/x_2) + \Psi (1/x_2, 1/x_1) + \Omega (1/x_1, 1/x_2) .
\end{multline}
By using \eqref{identities Omega}, \eqref{identities Phi0} and \eqref{identity Phi00}, we get
\begin{align}
\theta^{0_* 0_*} (x_1, x_2) &=
\Phi(x_1,x_2) - \Phi(x_1, 1/x_2) - \Phi(1/x_1, x_2) + \Phi(1/x_1, 1/x_2)
\notag \\
&\qquad
- \Psi (x_2, x_1) + \Psi (x_1, x_2) 
- \Psi (x_1, 1/x_2) + \Psi (x_2, 1/x_1) + \Omega (x_1, x_2),
\notag \\[2mm]
&= 
2 \Phi(x_1,x_2) - 2 \Phi(x_1, 1/x_2) - \Phi(0, x_2) + \Phi(0, 1/x_2)
\notag \\
&\qquad
- 2 \Psi (x_2, x_1) + 2 \Psi (x_1, x_2) 
- \Psi (x_1, 0) + \Psi (x_2, 0) + \Omega (x_1, x_2),
\notag \\[2mm]
&= 
4 \Phi(x_1,x_2) - 2 \Phi(x_1, 0) - 2 \Phi(0, x_2) 
\notag \\
&\qquad
- 2 \Psi (x_2, x_1) + 2 \Psi (x_1, x_2) 
- \Psi (x_1, 0) + \Psi (x_2, 0) + \Omega (x_1, x_2),
\label{massless^2 str^2 BES}
\end{align}
which is (B.10) of \cite{Frolov:2021fmj}.

Numerically, the $\Phi$- and $\Psi$-functions in \eqref{massless^2 str^2 BES} are ambiguous because $x_{1}$ and $x_{2}$ hit the unit circle.\footnote{We do not have such a problem if we use $\tilde \Phi$ in \eqref{def:tilde Psi}, but everything must be integrated twice.}
We regularise them by using the functions  $\Phi_{\epsilon_1,\epsilon_2} (x_1, x_2)$, $\Phi_\epsilon (x_1, x_2)$ and $\Psi_\epsilon (x_1, x_2)$ as in \eqref{def:Phi-epsilon12}, \eqref{def:Phi-epsilon} and \eqref{def:Psi-epsilon}, to obtain
\begin{align}
\theta^{0_* 0_*} (x_1, x_2) &=
4 \Big( \Phi_{\rm reg,reg} (x_1,x_2) 
- \epsilon_2 \, \Psi_{\rm reg} (x_2, x_1) 
+ \epsilon_1 \, \Psi_{\rm reg} (x_2, x_1)
+ \epsilon_1 \, \epsilon_2 \, \Omega (x_2, x_1) \Big)
\notag \\
&\quad
+ 2 \Big( \Phi_{\rm reg} (0, x_1) - \epsilon_1 \, \Psi (x_1, 0) \Big)
- 2 \Big( \Phi_{\rm reg} (0, x_2) - \epsilon_2 \, \Psi (x_2, 0) \Big)
\notag \\
&\quad
- 2 \Big( \Psi_{\rm reg} (x_2, x_1) + \epsilon_1 \, \Omega(x_2, x_1) \Big)
+ 2 \Big( \Psi_{\rm reg} (x_1, x_2) + \epsilon_2 \, \Omega(x_1, x_2) \Big)
\notag \\
&\qquad
- \Psi (x_1, 0) + \Psi (x_2, 0) + \Omega (x_1, x_2),
\notag \\[1mm]
&=
4 \Phi_{\rm reg,reg} (x_1,x_2) 
+ 2 \Phi_{\rm reg} (0, x_1) 
- 2 \Phi_{\rm reg} (0, x_2)
\notag \\
&\quad 
- 2 \Psi_{\rm reg} (x_2, x_1) 
+ 2  \Psi_{\rm reg} (x_1, x_2)
- \Psi (x_1, 0) + \Psi (x_2, 0) + \Omega (x_1, x_2)
\notag \\
&\quad 
+ \epsilon_1 \Big( 
4 \Psi_{\rm reg} (x_2, x_1) - 2 \Psi (x_1, 0) 
\Big)
+ \epsilon_2 \Big( 
- 4 \Psi_{\rm reg} (x_2, x_1) 
+ 2 \Psi (x_2, 0) 
\Big)
\notag \\
&\quad
+ 4 \epsilon_1 \, \epsilon_2 \, \Omega (x_2, x_1) 
- 2 \epsilon_1 \, \Omega(x_2, x_1) 
 + 2 \epsilon_2 \, \Omega(x_1, x_2) .
\label{massless^2 str^2 BES-2}
\end{align}
According to our prescription in Figure \ref{fig:massless BES reg}, the variables $x_k$ of massless particles in the string region should lie outside the unit circle, which means $\epsilon_1 = \epsilon_2 = 0$.
The weak coupling expansion of $\theta^{0_* 0_*}$ was discussed in Appendix \ref{app:K exactBethe}.

\paragraph{Mirror-mirror region.}

The massless limit of \eqref{mass^2 mir^2 BES-2} is
\begin{align*}
\frac{1}{i}\log\Sigma^{00}(x_1, x_2) &=
\Phi(x_1, x_2)-\Phi(x_1,1/x_2)-\Phi(1/x_1, x_2)+\Phi(1/x_1,1/x_2)
\notag \\ 
&\quad -\frac{1}{2} \Bigl\{ \Psi(x_1,x_2)+\Psi(1/x_1,x_2)-\Psi(x_1,1/x_2)-\Psi(1/x_1,1/x_2) \Bigr\}
\notag \\[1mm]
&\quad +\frac{1}{2} \Bigr\{ \Psi(x_2,x_1)+\Psi(1/x_2,x_1)-\Psi(x_2,1/x_1)
-\Psi(1/x_2,1/x_1) \Bigr\}
\notag \\ 
&\quad +\frac{1}{i} \log \( - \frac{ \Gamma\big[ -\frac{i}{2} h \big(x_1+\frac{1}{x_1}-x_2-\frac{1}{x_2}\big)\big]} 
{ \Gamma\big[ +\frac{i}{2} h \big(x_1+\frac{1}{x_1}-x_2-\frac{1}{x_2}\big)\big]} \)
\notag \\[2mm]
&= \Phi(x_1, x_2)-\Phi(x_1,1/x_2)-\Phi(1/x_1, x_2)+\Phi(1/x_1,1/x_2)
\\
&\quad - \Psi(x_1,x_2) + \Psi(x_1,1/x_2) + \Psi(x_2,x_1) -\Psi(x_2,1/x_1)
+ \Omega (x_1, x_2),
\notag
\end{align*}
where we used $\Gamma (z) = \Gamma(1+z)/z$.
We can rewrite this further as
\begin{align}
\frac{1}{i}\log\Sigma^{00}(x_1, x_2) &=
4\Phi(x_1,x_2) - 2\Phi(0,x_1)-2\Phi(0,x_2)
\notag\\
&\quad
-2\Psi( x_1, x_2) +2\Psi(x_2,x_1)
+\Psi(x_1,0)-\Psi(x_2,0) + \Omega (x_1, x_2) ,
\label{massless^2 mir^2 improved BES}
\end{align}
which agrees with (A.15) of \cite{Frolov:2021fmj}.\footnote{There is a typo in (A.15).}

We can check the regularity of the improved dressing factor around the origin as in Appendix \ref{app:expand origin}.
The potentially dangerous terms are
\begin{equation}
\frac{1}{i}\log\Sigma^{00}(x_1, x_2) \simeq
\begin{cases}
-2\Psi( x_1, x_2) + \Psi(x_1,0) + \Omega (x_1, x_2) &\qquad (x_1 \to 0) 
\\[1mm]
+2\Psi(x_2,x_1) - \Psi(x_2,0) + \Omega (x_1, x_2) &\qquad (x_2 \to 0) .
\end{cases}
\end{equation}
The polynomial divergences cancel out in both cases.

\paragraph{Mirror-string region.}

We follow the same strategy as in the string-string region. Consider the massive-massive BES phase \eqref{BES tilde Phi}, and evaluate the $\tilde \Phi$ function using \eqref{def:tilde Phi} at
\begin{equation}
\abs{x_1^+} < 1, \quad \abs{x_1^-} >1, \qquad
\abs{x_2^+} > 1, \quad \abs{x_2^-} <1,
\end{equation}
where $x_1^\pm$ is in the mirror region and $x_2^\pm$ is in the string region. The result is
\begin{equation}
\begin{aligned}
\theta^{QQ'_*} (x_1, x_2) =
\Phi(x_1^+, x_2^+) - \Psi (x_1^+, x_2^+)
- \Phi(x_1^-, x_2^+)
+ \Phi(x_1^-, x_2^-) + \Psi (x_2^-, x_1^-)
\\
- \Bigl[ \Phi(x_1^+, x_2^-) - \Psi (x_1^+, x_2^-) + \Psi (x_2^-, x_1^+) + \Omega (x_1^+, x_2^-) \Bigr].
\end{aligned}
\end{equation}
By taking the massless limit, we obtain
\begin{align}
\theta^{00_*} (x_1, x_2) &=
\Phi(x_1, x_2) - \Phi(x_1, 1/x_2) - \Phi(1/x_1, x_2) + \Phi(1/x_1, 1/x_2)
\notag \\
&\qquad
- \Psi (x_1, x_2) + \Psi (x_1, 1/x_2) - \Psi (1/x_2, x_1)  + \Psi (1/x_2, 1/x_1) 
 - \Omega (x_1, 1/x_2) 
\notag \\[1mm]
&= 4 \Phi (x_1, x_2) - 2 \Phi (x_1,0) - 2 \Phi (0, x_2)
\notag \\
&\qquad
- 2 \Psi (x_1, x_2) - 2 \Psi (x_2, x_1)
+ \Psi (x_1, 0) + \Psi (x_2, 0) - \Omega (x_1, x_2) .
\label{massless^2 mir-str BES}
\end{align}

Here $\Phi (x_1, x_2)$ and $\Psi (x_1, x_2)$ in this expression are ambiguous because $x_{2}$ hits the unit circle.
By using $\Phi_\epsilon (x_1, x_2)$ in \eqref{def:Phi-epsilon} and $\Psi_\epsilon (x_1, x_2)$ in \eqref{def:Psi-epsilon}, we obtain
\begin{multline}
\theta^{00_*} (x_1, x_2) = 4 \Phi_{\rm reg} (x_1, x_2) - 2 \Phi (x_1,0) - 2 \Phi_{\rm reg} (0, x_2)
- \epsilon_2 \, \Bigl\{ 4 \Psi (x_2, x_1) - 2 \Psi (x_2, 0) \Bigr\}
\\
- 2 \Psi_{\rm reg} (x_1, x_2) - 2 \Psi (x_2, x_1)
+ \Psi (x_1, 0) + \Psi (x_2, 0) - (1 +2 \epsilon_2 ) \, \Omega (x_1, x_2) .
\label{massless^2 mir-str BES-2} 
\end{multline}
The potentially dangerous terms at the origin are
\begin{equation}
\theta^{00_*} (x_1, x_2) = \Psi (x_1, 0) - (1 +2 \epsilon_2) \Omega (x_1, x_2), \qquad
(x_1 \to 0).
\end{equation}
The polynomial divergences cancel out if $\epsilon_2=0$. This means $\abs{x_2}>1$, and is consistent with our prescription in Figure \ref{fig:massless BES reg}.

\paragraph{String-mirror region.}

In this region, we have
\begin{equation}
\abs{x_1^+} > 1, \quad \abs{x_1^-} < 1, \qquad
\abs{x_2^+} < 1, \quad \abs{x_2^-} > 1.
\end{equation}
Following the same procedures, we find
\begin{multline}
\theta^{0_* 0} (x_1, x_2) 
= 4 \Phi (x_1, x_2) - 2 \Phi (x_1,0) - 2 \Phi (0, x_2)
\\
+ 2 \Psi (x_1, x_2) + 2 \Psi (x_2, x_1)
- \Psi (x_1, 0) - \Psi (x_2, 0) - \Omega (x_1, x_2) .
\label{massless^2 str-mir BES}
\end{multline}
We regularise $\Psi (x_2, x_1)$ for $\abs{x_1}=1$ by using $\Psi_\epsilon (x_2, x_1)$ in \eqref{def:Psi-epsilon}, and find
\begin{multline}
\theta^{0_* 0} (x_1, x_2) 
= 4 \Phi (x_1, x_2) - 2 \Phi (x_1,0) - 2 \Phi (0, x_2)
\\
+ 2 \Psi (x_1, x_2) + 2 \Psi_{\rm reg} (x_2, x_1)
- \Psi (x_1, 0) - \Psi (x_2, 0) - (1 - 2\epsilon) \, \Omega (x_1, x_2) .
\label{massless^2 str-mir BES-2}
\end{multline}
The potentially dangerous terms at the origin are
\begin{equation}
\theta^{0_* 0} (x_1, x_2) = - \Psi (x_2, 0) - (1 - 2\epsilon_1) \, \Omega (x_1, x_2), \qquad
(x_2 \to 0).
\end{equation}
The polynomial divergences cancel out if $\epsilon_1=0$. This means $\abs{x_{1}}>1$, as in Figure \ref{fig:massless BES reg}.
With $\epsilon_1=0$, we consider the subleading divergence in the limit $x_2 \to 0$, which is logarithmic.
Let us rewrite the BES phase using the $\gamma$-rapidities as in Appendix \ref{app:BES gamma rap}, and take the limit $\gamma_2 \to 0$. The result is
\begin{equation}
\begin{aligned}
\theta^{0_* 0} \Big( x_s (\gamma_1), x (\gamma_2) \Big) =
\frac{2 i h}{\cosh \gamma_1} \log \abs{\frac{h}{\gamma_2}}
+ O(\gamma_2^0)
= i \cE_0 (x_1) \log \abs{\frac{h}{\gamma_2}}
+ O(\gamma_2^0).
\end{aligned}
\label{massless^2 str-mir BES asymp}
\end{equation}

\subsection{Massless-massive BES}\label{app:mixed mass BES}

We discuss the massless-massive BES phase, where the massive particle is a $Q$-particle bound state.
The massive-massless BES phase can be obtained by unitarity.
Since we only excite massless particles, we do not need to put a massive $Q$-particle in the string region.

Our basic strategy is to start from the massive-massive BES phase in the string-string region \eqref{BES tilde Phi}, take the massless limit for the first argument, and simplify the result using the identities in Appendix \ref{app:BES basic}.
This procedure should be in principle correct, but the final expression looks rather complicated.
It could be better to start from $\theta_\bes^{11}$, take the massless limit, simplify it and fuse them.

\paragraph{String-string region.}

We analytically continue $x_1^\pm$ of the massive-massive BES phase \eqref{BES tilde Phi}, to the region
\begin{equation}
\abs{x_1^+} > 1, \quad \abs{x_1^-}<1, \quad
\abs{x_2^\pm} > 1.
\end{equation}
The result is
\begin{multline}
\theta^{1_* Q'_*} (x_1, x_2) =
\Phi(x_1^+,x_2^+) 
-  \Phi(x_1^+,x_2^-)
\\
- \Phi(x_1^-, x_2^+) + \Psi (x_1^-, x_2^+)
+ \Phi(x_1^-, x_2^-) - \Psi (x_1^-, x_2^-) .
\end{multline}
By taking the massless limit of the first particle, we obtain
\begin{align}
\theta^{0_* Q'_*} (x_1, x_2) &=
\Phi(x_1,x_2^+) 
-  \Phi(x_1,x_2^-)
- \Phi(1/x_1, x_2^+) 
+ \Phi(1/x_1, x_2^-) 
\notag \\
&\quad + \Psi (1/x_1, x_2^+) - \Psi (1/x_1, x_2^-) ,
\notag \\[1mm]
&= 2 \Phi(x_1,x_2^+) - 2 \Phi(x_1,x_2^-)
- \Phi(0, x_2^+) + \Phi(0, x_2^-) 
\notag \\
&\quad + \Psi (x_1, x_2^+) - \Psi (x_1, x_2^-) ,
\end{align}
which is (4.13) of \cite{Frolov:2021fmj}.

\paragraph{Mirror-mirror region.}

We first consider the improved dressing factor \eqref{mass^2 mir^2 BES-2}. 
By taking the massless limit for the first argument, we obtain
\begin{align}
\frac{1}{i}\log\Sigma^{0 Q}(y_1,y_2)
 =&\;
\Phi(y_1,y_2^+)-\Phi(y_1,y_2^-)-\Phi(1/y_1,y_2^+)+\Phi(1/y_1,y_2^-)
\notag \\ 
&-\frac{1}{2}\left(\Psi(y_1,y_2^+)+\Psi(1/y_1,y_2^+)-\Psi(y_1,y_2^-)-\Psi(1/y_1,y_2^-)\right) 
\notag \\
&+\frac{1}{2}\left(\Psi(y_{2}^+,y_1)+\Psi(y_{2}^-,y_1)-\Psi(y_{2}^+, 1/y_1) -\Psi(y_{2}^-, 1/y_1) \right)
\notag \\
&+\frac{1}{i} \log
\frac{ \Gamma\big[Q-\frac{i}{2}h \big(y_1 +\frac{1}{y_1} -y_2^+-\frac{1}{y_2^+}\big)\big]} 
{i^{Q}\Gamma\big[+\frac{i}{2}h \big(y_1 +\frac{1}{y_1} -y_2^+-\frac{1}{y_2^+}\big)\big]}
\( - \frac{y_1 - \frac{1}{y_2^-}}{y_1 - y_2^+} \sqrt{  y_2^- \, y_2^+}  \)
\notag \\[1mm]
=&\;
2 \Phi(y_1,y_2^+) - 2 \Phi(y_1,y_2^-)-\Phi(0,y_2^+)+\Phi(0,y_2^-)
- \Psi(y_1,y_2^+) + \Psi(y_1,y_2^-)
\notag \\ 
&
+ \Psi(y_{2}^+,y_1) + \Psi(y_{2}^-,y_1)
- \frac{1}{2}\left( \Psi(y_{2}^+, 0) + \Psi(y_{2}^-, 0) \right)
\notag \\
&+ \frac{1}{i} \log \Biggl[
\frac{ \Gamma\big[ \frac{Q}{2} -\frac{i}{2}h \big(y_1 +\frac{1}{y_1} - u_2 \big)\big]} 
{\Gamma\big[ \frac{Q}{2} +\frac{i}{2}h \big(y_1 +\frac{1}{y_1} - u_2\big)\big]}
\( - i^{-Q} \frac{y_1 - \frac{1}{y_2^-}}{y_1 - y_2^+} \sqrt{  y_2^- \, y_2^+}  \) \Biggr] ,
\label{improved BES massless-massive mir^2}
\end{align}
where $y_2  = x( u_2)$.

Let us study the small $h$ expansion of the kernel
\begin{equation}
\cK_\bes^{0Q} \Big(x (\gamma_1) , u_2 \Big) = \frac{1}{2\pi i} \, \frac{\partial}{\partial \gamma_1} 
\log \Sigma_{\bes}^{0 Q} (x (\gamma_1) , x^\pm (u_2) ) ,
\label{def:cK0Q_bes}
\end{equation}
which showed up as the convolution $\log (1+Y_0) * K_\bes^{0Q}$ in Appendix \ref{app:weak mixed mass dressing}. 
If we take the limit $h \to 0$ with $\gamma_1$ fixed, we get
\begin{multline}
\cK_\bes^{0Q} \Big( \gamma_1, \frac{\tilde u_2}{h}  \Big) =
\frac{e^{\gamma _1} h}{2 \pi  Q (e^{2 \gamma _1}-1)^2} \Bigg\{
-4 \left(e^{\gamma _1}+1\right){}^2
-8 e^{\gamma _1} Q \left[\psi \left(\tfrac{Q}{2}\right)+\gamma_E \right]
- Q \left( e^{\gamma _1}-1\right){}^2 \times
\\
\left[
4 \gamma_E
+ \psi \left( 1-\tfrac{Q+i \tilde u_2}{2} \right)
+\psi \left( 1-\tfrac{Q-i \tilde u_2}{2} \right)
+\psi \left( 1+\tfrac{Q-i \tilde u_2}{2} \right)
+\psi \left( 1+\tfrac{Q+i \tilde u_2}{2} \right) \right ]
\Bigg\}
\\
+ O(h^2),
\end{multline}
where $\psi (x)$ is the digamma function and $\gamma_E$ is the Euler-Mascheroni constant.
We rescaled $u_2$ to ensure that the leading term in this expansion matches numerical evaluation.
This quantity is small unless $\gamma_1$ or $\gamma_2 = O(h)$.
If we take the limit $\gamma_1 \to 0$ with $h$ fixed, we get
\begin{align}
\cK_\bes^{0Q} (\gamma_1, \gamma_2) =
\frac{h  \log \left(\frac{h}{\gamma _1}\right)}{i \pi ^2 \gamma _1^2} \, \left\{
\log \left(\frac{i - x_2^-}{i + x_2^-}\right) - \log \left(\frac{i - x_2^+}{i + x_2^+}\right)
- \pi i \right\}
+ \cO( \gamma_1^{-1} )
\label{KBES0Q around origin}
\end{align}
which is potentially singular. With $x(\gamma_1) \simeq - \gamma_1/2$ around $\gamma_1=0$, this behaviour agrees with our earlier discussion at \eqref{psi-integrand-strongly-weak}.

\paragraph{String-mirror region.}

First, consider the improved dressing factor \eqref{mass^2 str-mir BES}.
If we analytically continue $x_1^-$ into the region $\abs{x_1^-}<1$, we obtain
\begin{align}
\frac{1}{i}\log \Sigma_{1_*Q}(x_1,x_2) &=
\Phi(x_1^+,x_2^+) -\Phi(x_1^+,x_2^-)- \Phi(x_1^-,x_2^+) 
+\Phi(x_1^-,x_2^-) 
+ \Psi (x_1^-, x_2^+)  
\notag \\
&\quad 
- \Psi (x_1^-, x_2^-)+ \frac{1}{2}\Big[ \Psi(x_2^+, x_1^+) + \Psi( x_2^-, x_1^+)
- \Psi( x_2^+, x_1^-) - \Psi( x_2^-, x_1^-) \Big]
\notag
\\
&\quad 
+ \frac{1}{2}\Big[ \Omega (x_2^+, x_1^-) + \Omega (x_2^-, x_1^-) \Big] \\
&\quad+ \frac{1}{2i} \,
\log \frac{(x_1^- - x_2^+) (x_1^- - 1/x_2^-) (x_1^+ - 1/x_2^-)}{(x_1^+ - x_2^+) ( x_1^- - 1/x_2^+)^2 }.
\notag 
\end{align}
By taking the massless limit for the first particle, we obtain
\begin{align}
&\frac{1}{i}\log \Sigma_{0_*Q}(x_1,x_2) =
\Phi(x_1,x_2^+) -\Phi(x_1,x_2^-)- \Phi(\frac{1}{x_1},x_2^+) 
+\Phi(\tfrac{1}{x_1},x_2^-) 
\notag \\
&\quad 
+ \Psi (\tfrac{1}{x_1}, x_2^+) - \Psi (\tfrac{1}{x_1}, x_2^-) 
+ \frac{1}{2}\Big[ \Psi(x_2^+, x_1) + \Psi( x_2^-, x_1)
- \Psi( x_2^+, \tfrac{1}{x_1}) - \Psi( x_2^-, \tfrac{1}{x_1}) \Big] 
\notag \\
&\quad 
+ \frac{1}{2}\Big[ \Omega (x_2^+, x_1) + \Omega (x_2^-, x_1) \Big] 
+ \frac{1}{2i} \,
\log \frac{(1/x_1 - x_2^+) (1/x_1 - 1/x_2^-) (x_1 - 1/x_2^-)}{(x_1 - x_2^+) ( 1/x_1 - 1/x_2^+)^2 } \,.
\notag \\[1mm]
&= 2 \Phi(x_1,x_2^+) - 2 \Phi(x_1,x_2^-)- \Phi(0,x_2^+) +\Phi(0,x_2^-) 
\notag \\
&\quad 
+ \Psi (x_1, x_2^+) - \Psi (x_1, x_2^-) 
+ \Psi(x_2^+, x_1) +  \Psi( x_2^-, x_1)
- \frac{1}{2}\Big[ \Psi( x_2^+, 0) + \Psi( x_2^-, 0) \Big] 
\notag \\
&\quad 
+ \frac{1}{2}\Big[ \Omega (x_2^+, x_1) + \Omega (x_2^-, x_1) \Big] 
+ \frac{1}{2i} \,
\log \frac{(1/x_1 - x_2^+) (1/x_1 - 1/x_2^-) (x_1 - 1/x_2^-)}{(x_1 - x_2^+) ( 1/x_1 - 1/x_2^+)^2 } \,.
\end{align}

In the string-mirror region, we need to regularise $\Phi_{sx}^{\circ \bullet}$ and $\Psi_{xs}^{\bullet\circ}$ as in
\begin{alignat}{9}
\Phi_{sx}^{\circ \bullet} \ &\to &\ \Phi_{\epsilon_1} (x_1, x_2^\pm) 
&= - \Phi_{\rm reg} (x_2^\pm, x_1) + \epsilon_1 \Psi (x_1, x_2^\pm)
\\[1mm]
\Psi_{xs}^{\bullet\circ} \ &\to &\ \Psi_{\epsilon_2} (x_2^\pm, x_1) 
&= \Psi_{\rm reg} (x_2^\pm, x_1) + \epsilon_2 \, \Omega (x_2^\pm, x_1) ,
\end{alignat}
where \eqref{def:Phi-epsilon1} and \eqref{def:Psi-epsilon} are used. It follows that
\begin{align}
&\frac{1}{i}\log \Sigma_{0_*Q}(x_1,x_2) =
- 2 \Phi_{\rm reg} (x_2^+, x_1) 
+ 2 \Phi_{\rm reg} (x_2^-, x_1) 
- \Phi(0,x_2^+) +\Phi(0,x_2^-) 
\notag \\
&\quad 
+ (1 + 2 \epsilon_1) \Psi (x_1, x_2^+) 
- (1 + 2 \epsilon_1) \Psi (x_1, x_2^-) 
+ \Psi_{\rm reg} (x_2^+, x_1) 
+ \Psi_{\rm reg} (x_2^-, x_1) 
\notag \\
&\quad 
- \frac{1}{2}\Big[ \Psi_{\rm reg} ( x_2^+, 0) + \Psi_{\rm reg} ( x_2^-, 0)  \Big] 
- \frac{\epsilon_2}{2}\Big[ \Omega (x_2^+, 0) + \Omega (x_2^-, 0) \Big] 
\notag \\
&\quad + \frac{i ( 1 + 2 \epsilon_2) }{2} \, 
\log \Biggl(
\frac{\Gamma\big[1 - \frac{Q}{2} +\frac{i}{2}h \big( u_2 - x_1 - \frac{1}{x_1} \big)\big]}
{\Gamma\big[1 + \frac{Q}{2} -\frac{i}{2}h \big( u_2 - x_1 - \frac{1}{x_1} \big)\big]} \,
\frac{\Gamma\big[1 + \frac{Q}{2} +\frac{i}{2}h \big( u_2 - x_1 - \frac{1}{x_1} \big)\big]}
{\Gamma\big[1 - \frac{Q}{2} -\frac{i}{2}h \big( u_2 - x_1 - \frac{1}{x_1} \big)\big]}
\Biggr)
\notag \\
&\quad
+ \frac{1}{2i} \,
\log \frac{(1/x_1 - x_2^+) (1/x_1 - 1/x_2^-) (x_1 - 1/x_2^-)}{(x_1 - x_2^+) ( 1/x_1 - 1/x_2^+)^2 } \,.
\label{improved massless-massive str-mir}
\end{align}
Below we set $\epsilon_1 = \epsilon_2 = 0$ as before.

Let us study the small $h$ expansion of the improved dressing factor.
If we take the limit $h \to 0$ with $\gamma_1$ and $\tilde u_2$ fixed, we get
\begin{multline}
\frac{1}{i}\log \Sigma_{0_*Q}(x_1,x_2) \Big( \gamma_1, \frac{\tilde u_2}{h}  \Big) =
\frac{1}{2i}  \Biggr\{ \log \left(\frac{\left(e^{\gamma _1}+i\right) h \left(Q+i \tilde{u}_2\right)}{\left(-1-i e^{\gamma _1}\right) \left(Q-i \tilde{u}_2\right){}^2}\right)
+\log\Gamma\left( 1 -\frac{Q + i \tilde{u}_2}{2} \right)
\\
-\log\Gamma\left( 1 -\frac{Q - i \tilde{u}_2}{2} \right)
+\log\Gamma\left( 1 + \frac{Q-i \tilde{u}_2}{2} \right)
-\log\Gamma\left( 1 + \frac{Q+i \tilde{u}_2}{2} \right) \Biggr\}
+ \cO(h).
\end{multline}
We can obtain the kernel by taking the derivative with respect to $\gamma_1$\,.
The phase diverges logarithmically as $h \to 0$, which comes from the last line of \eqref{improved massless-massive str-mir}.
We can combine this divergence with the driving term in the TBA \eqref{ex-TBA for YQ} and \eqref{ex-TBA for YbQ} as
\begin{equation}
L \tilde{\mathcal{E}}^Q (u) - 2 \sum_{j=1}^{2M} \, \log \Sigma_{0_*Q}(x_{*j} , x (u) ) 
= L \log (\tilde u^2 + Q^2) - (2L+2M) \log h + \cO(1),
\end{equation}
where we used \eqref{def:tilde EQ}.

The kernel is
\begin{equation}
K^\Sigma_{0_*Q}(x_1,x_2) \Big( \gamma_1, \frac{\tilde u_2}{h}  \Big) =
-\frac{1}{4 \pi \cosh (\gamma_1)} + \cO(h),
\end{equation}
which is regular at this order.

\section{Large-\texorpdfstring{$L$}{L} analysis}
\label{app:largeL}

The numerical solution of the small-tension TBA suggests that the exact and the asymptotic energy spectra converge to the energy spectrum of a free particle in the large $L$ limit. In this appendix we explain how to derive this conclusion from a semi-analytic solution of TBA.

The momentum $p$ of a free particle is quantised as
\begin{equation}
p = \frac{2 \pi \nu}{L} = -i\log\left(\frac{e^{\gamma_s} -i}{e^{\gamma_s} +i}\right)^2 , 
\qquad \(\nu =0 ,1, \dots, \frac{L}{2} \)
\qquad \Leftrightarrow \qquad
\gamma_s = \log \tan \frac{p}{4} \,.
\label{free p and gamma}
\end{equation}
As before, we assume that all momenta come in pairs $(\gamma_j \,, - \gamma_j)$ with $j=1,2, \dots, M$.
The energy is given by
\begin{equation}
E_{\rm free} (L) = \sum_{j=1}^M \frac{4h}{\cosh \gamma_j} \,.
\label{app:free energy}
\end{equation}

\subsection{Intuitive explanation}\label{app:intuitive}

We consider excited states whose mode numbers come in pairs, $(\nu_j \,, -\nu_j)$ with $j=1,\dots M$, and assume that the length $L$ and the mode numbers $\{ \nu_j \}$ are large. Below we set $N_0=2$ for simplicity.

Consider the first term in the exact energy \eqref{weak exact energy}. The Jacobian $\frac{d \tilde{p}}{d \gamma}$ is exponentially small outside the origin $\gamma=0$. However, the massless Y-function $Y_0 (\gamma)$ vanishes at $\gamma=0$ owing to the driving term \eqref{app: exponentially small at gamma=0}. This means that the second term gives the dominant contribution to the exact energy,
\begin{equation}
\mathcal{E}(L) \simeq \sum_{j=1}^{2M} \frac{2h}{\cosh (\gamma_j)} \,.
\label{approximately free energy}
\end{equation}

Consider the exact Bethe equations \eqref{weak exact Bethe eqs}. 
When the mode number $\nu_k$ is large, then $2 \pi i \nu_k$ gives a large imaginary number.
However, a logarithm can bring at most $\pm \pi i$ because ${\rm Im} \, \log z \in (-\pi, \pi)$ for any $z \in \mathbb{C}$.
This means that only the term $- i L p(\gamma_k)$ can compensate the large imaginary number, giving us
\begin{equation}
2 \pi i \nu_k \simeq - i L p(\gamma_k),
\label{approximately free momentum}
\end{equation}
which is the quantisation condition of the free particle momenta \eqref{free p and gamma}.

From \eqref{approximately free energy} and \eqref{approximately free momentum}, we find that the energy of the states with large $L$ and large mode number $\nu_j$ should be approximated by the free particle spectrum.

\subsection{Large-\texorpdfstring{$L$}{L} ansatz}\label{app:large L ansatz}

In order to demonstrate the argument in Appendix \ref{app:intuitive}, we study ``approximate'' solutions of the mirror TBA equations at small $h$, and compute the energy spectrum of the states at $M=1,2$ for various $L$ and mode numbers.
For simplicity, we set $N_0=2$ below.

\subsubsection*{TBA at $\gamma=\pm \infty$}

We assume that the Y-functions $Y_0 (\gamma), Y(\gamma)$ approach constant values as $\gamma \to \pm \infty$.\footnote{Recall that $u (\gamma)=-2\coth \gamma$ from \eqref{eq:ugammarelation}.}
Then, the convolutions $\log (1+Y_0) * s, \log(1-Y)*s$ effectively become integration over a constant, because the Cauchy kernel $s (\gamma)$ is concentrated around the origin $\gamma=0$. It follows that
\begin{alignat}{9}
\lim \limits_{\gamma \to \pm \infty} \log(1+Y_0) \star s (\gamma) \ &\to \ 
\frac12 \, \log(1+y_0) , &\qquad
y_0 &\equiv Y_0 (\pm \infty)
\\
\lim \limits_{\gamma \to \pm \infty} \log(1-Y) \star s (\gamma) \ &\to \ 
\frac12 \, \log(1-y) , &\qquad
y &\equiv Y (\pm \infty).
\end{alignat}
The source terms become
\begin{equation}
\tE_0 (\pm \infty) = 0, \qquad
S_* (\pm \infty)= \mp i , \qquad
S (\pm \infty)= \mp i .
\label{source S infty}
\end{equation}
Under these assumptions, the small-tension TBA \eqref{weak TBA Y0Yy} in the region $\gamma \to \pm \infty$ becomes
\begin{equation}
\begin{aligned}
-\log y_0 &= - \log(1+y_0 )
- M \log (-1)
- 2 \log (1-y) \,,
\\
\log y =& \log(1+y_0 )
+ M \log (-1) .
\end{aligned}    
\label{weak TBA Y0Yy pminfty}
\end{equation}
If we impose further the following conditions
\begin{equation}
y_0 \ge -1, \quad {\rm and} \quad
1 \ge y \ge -1,
\end{equation}
we find the solution
\begin{equation}
( y_0 , y ) = \begin{cases}
(0,1) &\qquad ({\rm even} \ M), 
\\
( -1 - \omega, \omega ) &\qquad ({\rm odd} \ M),
\end{cases} 
\label{def:omega as}
\end{equation}
where $\omega \simeq - 0.353$ is the real root of $\omega^3 - 2 \omega^2 + 2 \omega + 1  = 0$.

By subtracting the constant equations \eqref{weak TBA Y0Yy pminfty} from the small-tension TBA \eqref{weak TBA Y0Yy}, we obtain
\begin{equation}
\begin{aligned}
-\log \frac{Y_0}{y_0} &= L \tilde{\mathcal{E}_0}
- \sum_{j=1}^{M} \log ( -S_{*}(-\gamma_j^{\dot{\alpha}_j}-\gamma)\,
S_{*}(\gamma_j^{\dot{\alpha}_j}-\gamma))\\
&\qquad 
-  2 \, \log \Big( \frac{1+Y_0}{1+y_0} \Big) * s
- 4 \log \Big( \frac{1-Y}{1-y} \Big)  * s \,,\\
\log \frac{Y}{y} &= 2 \, \log \Big( \frac{1+Y_0}{1+y_0} \Big) *s
+ \sum_{j=1}^{M} \log \left( -S_{*}(-\gamma_j^{\dot{\alpha}_j}-\gamma)S_{*}(\gamma_j^{\dot{\alpha}_j}-\gamma)\right) .
\end{aligned}    
\label{weak TBA Y0Yy-sub}
\end{equation}
Note that these equations are singular when $(y_0 , y) = (0,1)$.
The exact Bethe equations \eqref{weak exact Bethe eqs} should be modified in the same way, although the convolution with a constant is real-valued; $1 * s_* = 1/2$.

\subsubsection*{Approximate TBA}

We introduce {\it approximate} TBA equations and Bethe equations by neglecting the convolutions 
\begin{equation}
I_0 (\gamma) \equiv \( \log \frac{1+Y_0}{1+y_0} * s \) (\gamma), \qquad
I_{0*} (\gamma) \equiv \( \log \frac{1+Y_0}{1+y_0} * s_* \) (\gamma)
\label{def:I0 I0*}
\end{equation}
as\footnote{We omit the $\alg{su}(2)_\circ$ index from the Bethe roots.}
\begin{align}
-\log \frac{Y_0^\blacktriangle}{y_0} &= L \tilde{\mathcal{E}_0}
- \sum_{j=1}^{M} \log ( - S_{*}(-\gamma_j -\gamma)\,
S_{*}(\gamma_j -\gamma))
- 4 \log \Big( \frac{1-Y^\blacktriangle}{1-y} \Big)  * s ,
\notag \\
\log \frac{Y^\blacktriangle}{y} &= 
\sum_{j=1}^{M} \log \left( -S_{*}(-\gamma_j -\gamma)S_{*}(\gamma_j -\gamma)\right) ,
\label{weak TBA Y0Yy-triangle} \\
L \, \log\left(\frac{e^{\gamma_k} -i}{e^{\gamma_k} +i}\right)^2
&=
- i \pi (2 \nu_k + 1) 
+ \sum_{j=1}^{2M} \log S(\gamma_j -\gamma_k )
- 4 \(\log (1-Y^\blacktriangle)  * s_*\)(\gamma_k ) \,.
\notag
\end{align}    
The first two equations can be solved as
\begin{equation}
\begin{aligned}
Y_0^\blacktriangle (\gamma) &= y_0 \, F (\gamma)^4 \, e^{- L \tE_0} 
\prod_{j=1}^M  \Big( - S_{*}(-\gamma_j -\gamma) S_{*}(\gamma_j -\gamma) \Big) ,
 \\
\log F (\gamma) &= \log \Big( \frac{1-Y^\blacktriangle}{1-y} \Big) *s ,
\\
Y^\blacktriangle (\gamma) &= y \, 
\prod_{j=1}^M  \Big( - S_{*}(-\gamma_j -\gamma) S_{*}(\gamma_j -\gamma) \Big) .
\end{aligned}
\label{def:Yas triangle}
\end{equation}
We determine the Bethe roots $\gamma_j \in \mathbb{R}$ by solving the last line of \eqref{weak TBA Y0Yy-triangle}.

The reliability of this approximation depends on $M$.

\begin{figure}[ht]
\includegraphics[width=\linewidth]{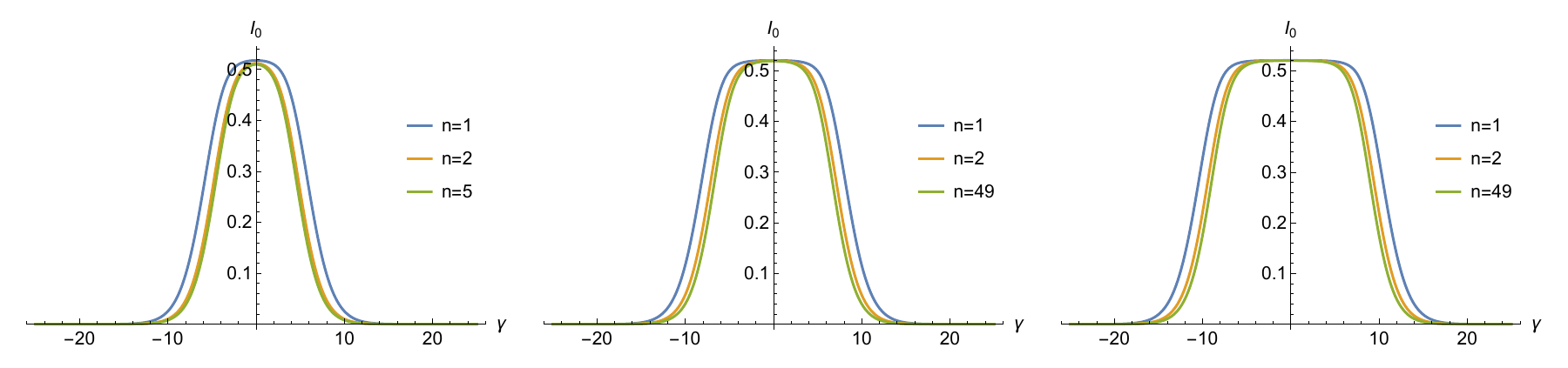}
\caption{Plot of $I_0 (\gamma)$ for two-particle states, at $L=10$ (left), $L=100$ (middle) and $L=1000$ (right). If $I_0 (\gamma) \approx 0$ for all $\gamma \in \bb{R}$, the function $Y_0^\blacktriangle (\gamma)$ is a good approximation of the genuine solution of TBA equations. The height of the plateau is close to $- \frac12 \, \log (1+y_0)$ in the left figure.}
\label{fig: approxY02conv}
\end{figure}

In Figure \ref{fig: approxY02conv}, we plotted the function $I_0 (\gamma)$ for two-particle states at $L=10, 100, 1000$ with the mode numbers $(\nu_1, \nu_2) = (\nu,-\nu)$. 
At $M=1$ we find $I_0 (\gamma)$ is non-vanishing around the origin. Thus $Y_0^\blacktriangle$ is quite different from the genuine solution of TBA.

\begin{figure}[ht]
\includegraphics[width=\linewidth]{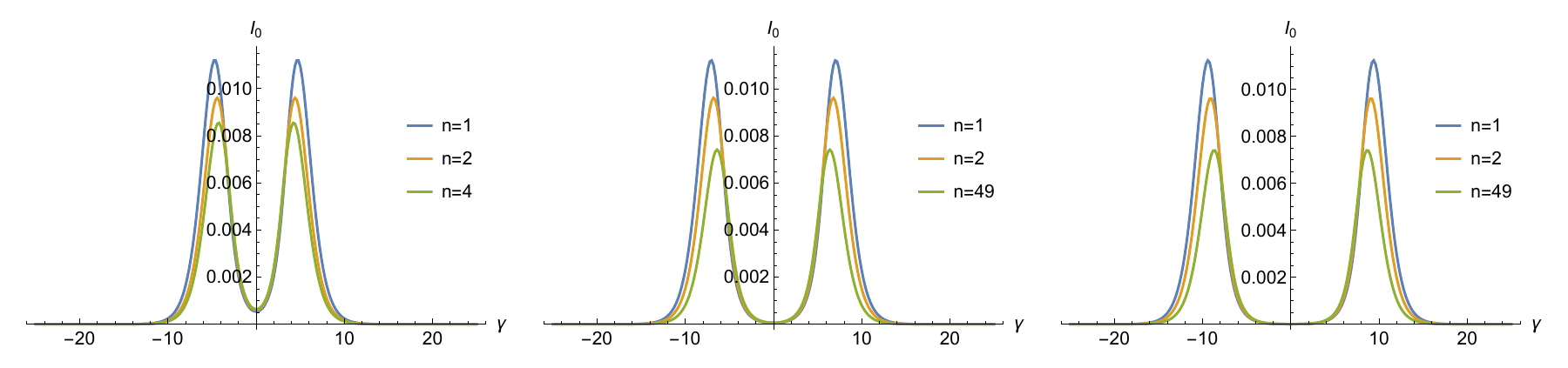}
\caption{Plot of $I_0 (\gamma)$ for four-particle states, at $L=10$ (left), $L=100$ (middle) and $L=1000$ (right).}
\label{fig: approxY04conv}
\end{figure}

In Figure \ref{fig: approxY04conv}, we plotted the same function for four-particle states at $L=10$, 100 and 1000. We fixed a pair of mode numbers at $(\nu_1, \nu_2) = (1,-1)$ for the four-particle states with the mode numbers $(\nu_1, \nu_2, \nu_3, \nu_4) = (1,-1,\nu,-\nu)$.
At $M=2$ we find $I_0 (\gamma)$ has two small peaks. As $L$ grows large, these peaks move towards $\gamma = \pm \infty$ while maintaining height.
Even though our ansatz $(Y_0^\blacktriangle , Y^\blacktriangle)$ does not solve TBA around these peaks, we may still think of it as an approximate solution of the original TBA, in the sense that the corrections would change the Y-functions only in the region far away from the origin.
Such corrections are not important when we compute the contribution to the exact energy \eqref{weak exact energy}, because the Jacobian $(d \tilde p_0/d \gamma)$ is exponentially suppressed as $\gamma \to \pm \infty$.

\begin{figure}[ht]
\centering
\includegraphics[width=0.4\linewidth]{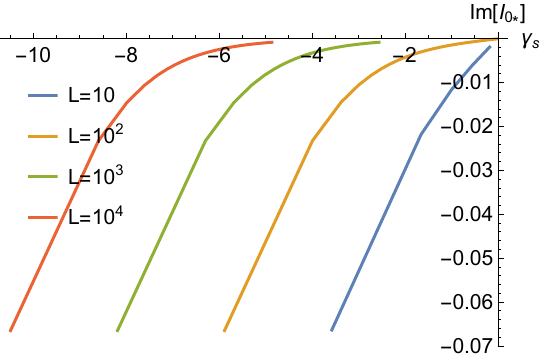}
\hspace{6mm}
\includegraphics[width=0.4\linewidth]{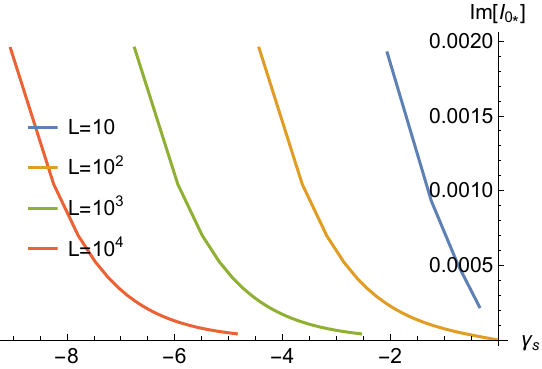}
\caption{Plot of $I_{0*} (\gamma)$ for two-particle states (left) and four-particle states (right). 
If $I_{0*} (\gamma) \approx 0$ at $\gamma=\gamma_s$\, then $\gamma_s$ is a good approximation of the genuine Bethe root that solves the exact Bethe equations. We use the same setup as in Figure \ref{fig: approxY02conv} and \ref{fig: approxY04conv}, respectively.}
\label{fig: approxI0sconv}
\end{figure}

In Figure \ref{fig: approxI0sconv} we plotted the function $I_{0*} (\gamma)$ for two-particle and four-particle states. In both cases $I_{0*} (\gamma)$ remains small. This means that the position of the Bethe roots determined by the approximate Bethe equations should be reliable, at least for four-particle states.


\paragraph{Corrections to the exact energy.}
The exact energy \eqref{weak exact energy} is approximated by
\begin{equation}
\mathcal{E} (L) \simeq
\mathcal{E}^\blacktriangle (L) = 2h \Bigg( 
\sum_{j=1}^{2M} \frac{1}{\cosh (\gamma_j)}
- \int\limits_{-\infty}^{+\infty} \frac{\de \gamma}{\pi} \, \frac{\cosh\gamma}{\sinh^2\gamma}
\log\left(1+Y^\blacktriangle _0(\gamma)\right) \Bigg).
\label{weak large L energy}
\end{equation}
It is straightforward to check that the approximate energy becomes close to the energy of free particles when $L \gg 1$ and $n \gg 1$.

\clearpage

\section{Numerical results}
\label{app:numerical}
Here we collect some numerical results for the readers' convenience.

\subsection{\texorpdfstring{$M=1$}{M=1} energies and Bethe roots}
\begin{table}[ht]
\centering
\begin{minipage}{0.48\linewidth}\centering
\begin{tabular}{l||l|l}
  		$\nu_1$& $\gamma_1$ & $H_{(1)}^{\nu,L}$\\
  		\hline
		1	&1.3698653240	&2.0371546985	\\
		2	&0.3412311709	&3.9521799805	
  		\end{tabular}
\caption{$M=1$, $L=4$, $N_0=2$}
\medskip
  \begin{tabular}{l||l|l}
  $\nu_1$& $\gamma_1$ & $H_{(1)}^{\nu,L}$\\
  \hline
1	&2.1248622820	&1.0028559276	\\
2	&1.1089105186	&2.4707908032	\\
3	&0.5792737159	&3.5117487126	\\
4	&0.1817026465	&4.0386428712	
  \end{tabular}
\caption{$M=1$, $L=8$, $N_0=2$}
\medskip
  \begin{tabular}{l||l|l}
  $\nu_1$& $\gamma_1$ & $H_{(1)}^{\nu,L}$\\
  \hline
1	&2.8618207790	&0.4849717915	\\
2	&1.8438211035	&1.2794684349	\\
3	&1.3390651829	&2.0126766051	\\
4	&0.9929598227	&2.6600005573	\\
5	&0.7219619376	&3.2000441129	\\
6	&0.4924934000	&3.6163564964	\\
7	&0.2870226887	&3.8981848212	\\
8	&0.0943269199	&4.0400428155	\\
\hline
  \end{tabular}
  \medskip
\caption{$M=1$, $L=16$, $N_0=2$}
\end{minipage}
\hfill
\begin{minipage}{0.48\linewidth}\centering
  \begin{tabular}{l||l|l}
  $\nu_1$& $\gamma_1$ & $H_{(1)}^{\nu,L}$\\
  \hline
1	&3.5807368927	&0.2368807583	\\
2	&2.5603603627	&0.6365227133	\\
3	&2.0603600910	&1.0281132980	\\
4	&1.7251311148	&1.4082316357	\\
5	&1.4710119498	&1.7734401398	\\
6	&1.2648112263	&2.1204079213	\\
7	&1.0899414525	&2.4459763296	\\
8	&0.9368974686	&2.7472074554	\\
9	&0.7997021734	&3.0214244963	\\
10	&0.6743276527	&3.2662459101	\\
11	&0.5579036274	&3.4796125862	\\
12	&0.4482825396	&3.6598056719	\\
13	&0.3437821208	&3.8054518369	\\
14	&0.2430230319	&3.9155132154	\\
15	&0.1448207412	&3.9892619250	\\
16	&0.0481100780	&4.0262441073	
  \end{tabular}
\caption{$M=1$, $L=32$, $N_0=2$} 
\end{minipage}
\end{table}

\clearpage

\begin{table}[ht]
\centering
  \begin{tabular}{l||l|l|||l||l|l}
  $\nu_1$& $\gamma_1$ & $H_{(1)}^{\nu,L}$ &$\nu_1$& $\gamma_1$ & $H_{(1)}^{\nu,L}$   \\
  \hline
1	&4.2879096161	&0.1168564842	&17	&0.8401019031	&2.9267028133	\\
2	&3.2664446921	&0.3155568408	&18	&0.7742505012	&3.0576318067	\\
3	&2.7671383824	&0.5132072504	&19	&0.7111470670	&3.1811630271	\\
4	&2.4341433236	&0.7093766819	&20	&0.6504336814	&3.2970173851	\\
5	&2.1837123783	&0.9036191403	&21	&0.5918022443	&3.4049356209	\\
6	&1.9826258225	&1.0954854863	&22	&0.5349845242	&3.5046788549	\\
7	&1.8143008697	&1.2845282941	&23	&0.4797444118	&3.5960290077	\\
8	&1.6692618496	&1.4703047162	&24	&0.4258717961	&3.6787890734	\\
9	&1.5415834217	&1.6523785504	&25	&0.3731776535	&3.7527832329	\\
10	&1.4273122065	&1.8303219179	&26	&0.3214900533	&3.8178568013	\\
11	&1.3236779082	&2.0037167259	&27	&0.2706508662	&3.8738760127	\\
12	&1.2286629854	&2.1721559956	&28	&0.2205130145	&3.9207276592	\\
13	&1.1407515544	&2.3352450966	&29	&0.1709381451	&3.9583186158	\\
14	&1.0587748047	&2.4926029101	&30	&0.1217946312	&3.9865753002	\\
15	&0.9818115704	&2.6438629291	&31	&0.0729558279	&4.0054431247	\\
16	&0.9091219985	&2.7886743016	&32	&0.0242985168	&4.0148860079	
  \end{tabular}
  \caption{$M=1$, $L=64$, $N_0=2$}
\end{table}

\subsection{\texorpdfstring{$M=2$}{M=2} energies and Bethe roots}

\begin{table}[ht]
\centering
  \begin{tabular}{l|l||l|l|l}
  $\nu_1$&$\nu_3$& $\gamma_1$ & $\gamma_3$ & $H_{(1)}^{\nu,L}$\\
  \hline
1	&1	&2.9376378954	&1.8375583252	&1.6623510390	\\
1	&2	&2.6967090197	&1.0391156116	&3.0495341815	\\
1	&3	&2.6170173068	&0.5539483099	&4.0342421895	\\
1	&4	&2.5887794648	&0.1749771889	&4.5341663005	\\
2	&1	&1.0391156116	&2.6967090197	&3.0495341815	\\
2	&2	&0.9518480877	&1.4877186973	&4.4041253580	\\
2	&3	&1.4462042179	&0.5189041649	&5.2978941757	\\
2	&4	&1.4299341712	&0.1653301442	&5.7541077048	\\
3	&1	&0.5539483099	&2.6170173068	&4.0342421895	\\
3	&2	&0.5189041649	&1.4462042179	&5.2978941757	\\
3	&3	&0.4980953042	&0.8831783606	&6.3729119753	\\
3	&4	&0.8730916450	&0.1592766021	&6.7925692823	\\
4	&1	&0.1749771889	&2.5887794648	&4.5341663005	\\
4	&2	&0.1653301442	&1.4299341712	&5.7541077048	\\
4	&3	&0.1592766021	&0.8730916450	&6.7925692823	\\
4	&4	&0.1555881589	&0.4805085942	&7.5289149203	
  \end{tabular}

  \caption{$M=2$, $L=8$, $N_0=2$}
\end{table}

\clearpage

\begin{table}[ht]
\centering
  \begin{tabular}{l|l||l|l|l}
  $\nu_1$&$\nu_3$& $\gamma_1$ & $\gamma_3$ & $H_{(1)}^{\nu,L}$\\
  \hline
1	&1	&3.6537610671	&2.5602970649	&0.8203534117	\\
1	&2	&3.3876886449	&1.7628137640	&1.6018607650	\\
1	&3	&3.2857192526	&1.3024011789	&2.3227799421	\\
1	&4	&3.2336036393	&0.9729210873	&2.9591944500	\\
1	&5	&3.2035863552	&0.7101339967	&3.4901834518	\\
1	&6	&3.1856343126	&0.4854953019	&3.8996058876	\\
1	&7	&3.1753421098	&0.2832967956	&4.1768538170	\\
1	&8	&3.1706273572	&0.0931556025	&4.3164455616	\\
4	&1	&0.9729210873	&3.2336036393	&2.9591944500	\\
4	&2	&0.9423478941	&2.0822853425	&3.6872964642	\\
4	&3	&0.9212795353	&1.5406661679	&4.3864271809	\\
4	&4	&0.9059731295	&1.1760363379	&5.0319943849	\\
4	&5	&1.1652460981	&0.6682476284	&5.5204633427	\\
4	&6	&1.1581088866	&0.4598471864	&5.8977454073	\\
4	&7	&1.1537644634	&0.2693710960	&6.1536717395	\\
4	&8	&1.1517095215	&0.0887380979	&6.2827075858	\\
8	&1	&0.0931556025	&3.1706273572	&4.3164455616	\\
8	&2	&0.0912746810	&2.0401356906	&5.0050489064	\\
8	&3	&0.0898643130	&1.5089325981	&5.6697697168	\\
8	&4	&0.0887380979	&1.1517095215	&6.2827075858	\\
8	&5	&0.0878354445	&0.8764190409	&6.8223972096	\\
8	&6	&0.0871308216	&0.6471361855	&7.2716629274	\\
8	&7	&0.0866115471	&0.4457230371	&7.6176657203	\\
8	&8	&0.0862699163	&0.2612714949	&7.8519608180	
  \end{tabular}

  \caption{$M=2$, $L=16$, $N_0=2$}
\end{table}

\begin{table}[ht]
\centering
  \begin{tabular}{l|l||l|l|l}
  $\nu_1$&$\nu_3$& $\gamma_1$ & $\gamma_3$ & $H_{(1)}^{\nu,L}$\\
  \hline
1	&1	&4.3667760502	&3.2749047890	&0.4030448796	\\
1	&2	&4.0925190899	&2.4758876263	&0.8008463492	\\
1	&3	&3.9835601638	&2.0205102633	&1.1905732274	\\
1	&4	&3.9248851827	&1.7019564268	&1.5688541068	\\
1	&5	&3.8883291398	&1.4559402105	&1.9322832043	\\
1	&6	&3.8635441692	&1.2543177294	&2.2775537076	\\
1	&7	&3.8458143772	&1.0823054700	&2.6015269927	\\
1	&8	&3.8326777910	&0.9311781182	&2.9012824112	\\
1	&9	&3.8227233795	&0.7953422171	&3.1741580831	\\
1	&10	&3.8150863070	&0.6709773571	&3.4177851279	\\
1	&11	&3.8092099606	&0.5553341129	&3.6301147693	\\
1	&12	&3.8047236021	&0.4463397930	&3.8094360711	\\
1	&13	&3.8013752502	&0.3423611847	&3.9543811970	\\
1	&14	&3.7989931291	&0.2420529768	&4.0639155381	\\
1	&15	&3.7974629520	&0.1442557986	&4.1373126777	\\
1	&16	&3.7967145735	&0.0479245310	&4.1741191045	\\
8	&1	&0.9311781182	&3.8326777910	&2.9012824112	\\
8	&2	&0.9216319023	&2.7088951040	&3.2779452910	\\
8	&3	&0.9141849574	&2.1842219252	&3.6517239986	\\
8	&4	&0.9079885188	&1.8364873234	&4.0183658379	\\
8	&5	&0.9027267397	&1.5744210131	&4.3743620916	\\
8	&6	&0.8982298328	&1.3626539505	&4.7164383992	\\
8	&7	&0.8943850814	&1.1837177617	&5.0414825239	\\
8	&8	&0.8911070856	&1.0276784945	&5.3465517214	\\
8	&9	&1.0249251329	&0.7638168954	&5.6053929680	\\
8	&10	&1.0226557697	&0.6461621688	&5.8365914251	\\
8	&11	&1.0208101070	&0.5359482231	&6.0381767125	\\
8	&12	&1.0193396941	&0.4314761487	&6.2084956100	\\
8	&13	&1.0182063344	&0.3313762548	&6.3462183473	\\
8	&14	&1.0173808114	&0.2344982199	&6.4503309849	\\
8	&15	&1.0168419384	&0.1398351973	&6.5201144774	\\
8	&16	&1.0165759091	&0.0464692922	&6.5551150808	
  \end{tabular}
  \caption{$M=2$, $L=32$, $N_0=2$}
\end{table}

\clearpage

\subsection{\texorpdfstring{$N_0\neq2$}{N0!=2} energies and Bethe roots}

\begin{table}[ht]
\centering
  \begin{tabular}{l|l||l|l|||l|l||l|l}
  $N_0$ &$\nu_1$& $\gamma_1$ & $H_{(1)}^{\nu,L}$ &$N_0$ &$\nu_1$& $\gamma_1$ & $H_{(1)}^{\nu,L}$ \\
  \hline
1	&1	&2.1478952374	&0.9635873862	&4	&1	&2.0953342432	&1.0566264100	\\
1	&2	&1.1133031093	&2.4362266137	&4	&2	&1.1031296462	&2.5183340592	\\
1	&3	&0.5807795815	&3.4803921364	&4	&3	&0.5772837367	&3.5550539708	\\
1	&4	&0.1820951353	&4.0088079500	&4	&4	&0.1811833307	&4.0799379863	
  \end{tabular}
  \caption{$M=1$, $L=8$, $N_0=1$ or $N_0=4$}
\end{table}

\begin{table}[ht]
\centering
  \begin{tabular}{l|l||l|l|||l|l||l|l}
  $N_0$ &$\nu_1$& $\gamma_1$ & $H_{(1)}^{\nu,L}$ &$N_0$ &$\nu_1$& $\gamma_1$ & $H_{(1)}^{\nu,L}$ \\
  \hline
1	&1	&2.8857343672	&0.4654209354	&4	&1	&2.8311082468	&0.5117915477	\\
1	&2	&1.8488625594	&1.2606517243	&4	&2	&1.8371707568	&1.3053067737	\\
1	&3	&1.3412153995	&1.9945438917	&4	&3	&1.3362158215	&2.0376028122	\\
1	&4	&0.9941067817	&2.6424719775	&4	&4	&0.9914373805	&2.6841218621	\\
1	&5	&0.7226306849	&3.1830157305	&4	&5	&0.7210735368	&3.2234992687	\\
1	&6	&0.4928863986	&3.5997081486	&4	&6	&0.4919710848	&3.6393056282	\\
1	&7	&0.2872311467	&3.8817888137	&4	&7	&0.2867455680	&3.9207979770	\\
1	&8	&0.0943923424	&4.0237715433	&4	&8	&0.0942399382	&4.0624898925	
  \end{tabular}
  \caption{$M=1$, $L=16$, $N_0=1$ or $N_0=4$}
\end{table}

\newpage

\bibliography{refs}

\end{document}